\documentclass[10pt]{amsart}
\usepackage{geometry}       
\usepackage{subfig}
\usepackage[english]{babel}
\usepackage{authblk}
\usepackage{todonotes}
\usepackage{amsaddr}
\usepackage[latin1]{inputenc}
\usepackage{amsmath,amssymb,latexsym,amsthm}
\numberwithin{equation}{section}
\usepackage{float}
\usepackage{enumerate}
\usepackage{fancyhdr}
\usepackage{url}

\makeatletter
\g@addto@macro{\endabstract}{\@setabstract}
\newcommand{\authorfootnotes}{\renewcommand\thefootnote{\@fnsymbol\c@footnote}}%
\makeatother

\title{Predictive Index  for slope instabilities in  open pit mining}
\begin{document}

\maketitle

\begin{center}
  \normalsize
  \authorfootnotes
  J. H. Ortega\footnote{Email: jortega@dim.uchile.cl, mrapimanc@gmail.com, rlecaros@dim.uchile.cl, fmedel@dim.uchile.cl, fpadilla@dim.uchile.cl, agarcia@gesecology.com}\textsuperscript{2,3}, 
  M. Rapiman\textsuperscript{1}, 
  R. Lecaros\textsuperscript{3}, 
  F.  Medel\textsuperscript{3}, 
  F. Padilla\textsuperscript{3}, 
  A. Garc\'ia\textsuperscript{1} \par \bigskip
  \textsuperscript{1}Gesecology Ltda., Los Andes, Chile\par
  \textsuperscript{2}Departamento de Ingenier\'ia Matem\'atica, Universidad de Chile, Santiago, Chile \par
  \textsuperscript{3}Center for Mathematical Modeling (CMM), Universidad de Chile, Santiago, Chile\par 
   \bigskip
\end{center}

\begin{abstract}
In this paper we study the stability and deformation of structures, in particular the wall of an open pit mine is studied by using information obtained from a variety of remote sensors  and some extra data, with a novelty approach considering the use of mathematical models and data mining techniques. In particular we present two models to help the study the slope stability of pit and the possible occurrence of movements. Primarily we present an static model for slow movements, which will help us identify areas of possible risks areas with  time horizons of several months or years, depends on the available information,  before the wall start moving, and secondly a dynamic short-term model, which help us to determine risks of collapse zones  with several days in advance. We remark that this methodology can be a powerful tool  to plain future actions in order to simulate possible scenarios considering the production plans.
\end{abstract}

\setcounter{page}{2} 

\section{Introduction}
\
\par

The study of slope stability is an issue of great interest to mining companies worldwide, as well as for roads and slopes nearby communities and infrastructure, among others.
These phenomena usually committed significant financial resources for mining companies, and moreover, may even cause loss of life. 
We must note that these phenomena, and factors associated with the movement of the slopes and landslides, are extremely complex and highly nonlinear. Therefore, methods of data mining turned out to be ideal for discovering new information in the data, which are recorded periodically in the continuous monitoring.
 Monitoring the movement and deformation of the slope of open pit mine is a basic task to assess security conditions in the mine, so that the safety of workers and material resources involved is increased (Du et al., (2013) \cite{Du2013}). Economic losses caused by landslides reach several million dollars (Huang et al., (2014) \cite{Huang2014}).

The movement of the slopes is a common phenomenon in open-pit mines. These slopes  are designed with a safety factor to control the risk of injury and equipment damage due to slope failures and falling rocks.  Dubrovskiy and Sergeev (2006) \cite{Dubrovskiy2006}, indicate that there is the possibility of finding phenomena that predict the unstable state of a slope and which can be predicted in time and scale collapse thereof.
Slopes never fail spontaneously, before the failure, the slope indications delivered in the form of measurable movement and\/or the appearance of stress cracks. In contrast, the movements that have slopes, are the result of long-term movement thereof crawling for hundreds of years resulting in cumulative movement of tens of meters. Precursors and early indicators movements can warn of disaster landslide (Rai (2014) \cite{Rai2014}).
Meanwhile, there are different modes of deformation and failure that can happen on a steep slope. But common to them all, when a slope is excavated or exposed, there is an initial response period as a result of stress relaxation (Závodní (2000) \cite{Zavodni2000}), which it is common in mines with a rapid rate of excavation.

Reeves et al. (2000) \cite{Reeves2000} studied the detection of precursors slope movements using a system with actual radar interferometry, the SSR (Slope Stability Radar). 
The hypothesis is that a managed system itself would be able to detect these very small movements occurring precursors before the collapse of the slope, allowing estimate the slope stability study (Reeves et al. (2000) \cite{Reeves2000}). Faillttaz and Or (2013) \cite{Faillettaz2013}, notes that on to identify who is facing imminent failure by gravity in a natural environment (e.g. a bank), it is a complex and daunting task, especially because sudden ruptures are highly nonlinear, sensitive to unknown heterogeneities inherent in the natural environment processes.
These authors postulate that these mechanical failures measurable release elastic energy or noise emissions as microseisms. So that by monitoring these activities should provide information about the mechanical condition of faults. However, the occurrence of catastrophic failure in heterogeneous materials is not an instantaneous event, but typically is preceded by smaller internal faults before rupture.
 
Hartwig et al.  (2013) \cite{Hartwig2013} used PSI  (\textit{Persistent scatterer Interferometry}) to monitor an iron ore mine in Carajas, Brazil. 18 scenes using TerraSAR-X acquired during the dry season. While most of the study area was stable, high rates of strain were detected in dumps (312 mm\/year).
Although the geotechnical design can be improved to increase the safety factors and designs of banks, they can be improved to minimize the dangers of rockfalls. Even on slopes with conservative designs, however they may experience unexpected failures due to the presence of geological structures unknown, abnormal weather conditions or earthquakes (Osasan and Afeni (2012) \cite{Osasan2012}).
For the monitoring of banks, they have developed a number of systems capable of identifying and tracking the precursors slope collapses or older rock mass movement.
INSAR was used to confirm the move from an area of the Italian Alps, before and after some events appears, but with constant rates and relatively low motion (Notti et al., 2013 \cite{Notti2013}). In addition to detecting and tracking slope instabilities, due to its ability to detect small movements (Colombo, 2013 \cite{colombomeasuring}).
Rosser et al. (2007) \cite{NickRosser2007}  used 3D Laser Scanner High resolution monitor slopes of cliffs in the Philippines, where precursors movements of the slope gave information imminent collapse. The system provides real-time measurement of the rock under study. Solano et al. (2012) \cite{Solano2012} used Landsat and SqueeSAR images to analyze the temporal evolution of fault risk areas associated with subsidence. With this information they  constructed risk maps of the area under study.
The use of ground-based radar gives slope monitoring continuously and higher resolution than other systems of ground radars. The evolution of slope movements delivers indications of the collapse of the slope. Farina et al. (2011) \cite{farina2011ibis}, given as example, the displacement of a slope until its collapse, measured by the radar IBIS-M, where two changes in velocity (acceleration) are seen before triggering an exponential acceleration collapse.
Furthermore, prediction of landslides is a highly nonlinear and highly complex phenomenon. In most cases, the use of mathematical models is very difficult to clearly describe the processes involved. Thus, the technology of data mining or data mining that has the ability to find patterns hidden data and models can be used to predict movements.
With respect to collapses in mining, 
Pletey et al. (2005) \cite{Petleya2005} analyzed the movement of landslides on a seasonally hotspot slope collapses. They found different patterns of movement occurring at the same event slide, primarily as a result of changes in groundwater conditions. 
Ghasemi and Ataei (2013) \cite{Ghasemi2013} used fuzzy logic to predict failure rates in tunnels of underground mines. Meanwhile, Venkatesan et al. (2013) constructed a model of landslide susceptibility early warning based on classification techniques of data mining, also they showed that Bayesian classifiers were more accurate than classifiers SVM (\textit{Support Vector Machine}) in the analysis of landslides.
To estimate areas for these events, Venkatesan et al. (2012) \cite{Venkatesan2012}, used a spatial database and GIS (Geographic Information System) to process information. They also showed that the Bayesian approach was more appropriate and accurate SVM.
Some authors have also used data mining techniques to create maps of landslide risk, as the case of the work by Pradhan et al. (2009) \cite{Pradhan2009}, where they combined a model derived from the data (frequency ratio) and derived from the knowledge (fuzzy operators) for analyzing the risk of slipping slope in Penang Island in Malaysia model. 
This work is  divided in four sections. Section 2  corresponds to a literature review of some sensors used in the monitoring of mines such as SAR (terrestrial and satellite radars), prisms, piezometers, among others, as well as a review of the state of the art technologies precursors movements slopes and embankments, and predictive modeling of landslides and movements slope. Section 3, realizes data received from the sensors, their formats and the size of the database. Section 4  shows an analysis  of the data received, especially from a statistical standpoint and verification of the phenomenon studied. Finally, the results, which corresponds to two movements estimation models are presented. A long-term static model that is capable of generating a motion estimation six months in advance or more, depending on the available information, and a dynamic or short-term model, based on data generated by the GBR, and which can be predicted with a week in advance about a movement of greater magnitude.

\section{Data acquisition and preprocessing}

In this section we will present some ideas on the data used and the analysis of this information, which will be useful for the study of slope stability.
We note that the data used in the study are different sensors installed on the mine owned. Among the considered sensors we can cite ground-based radars, INSAR, prisms, piezometers, digital elevation models, meteorology  and geological data, etc. The reporting period covers three years, with the availability of data in each of the sensors variable, this due to various causes and nature of each sensor. Aimed at studying the stability of the mine was raised, so that given the availability of information, we focus on  a particular region, for which they had record of an event, but also  was studied briefly another sector of the mine  and finally risk maps were made for all mine.

\subsection{Total Stations - Prisms}

The prisms are used to monitor surface displacements in three directions. They consist of a robotic station that records the position of prisms installed in the mine.
The measurements are performed by installing the station in stable areas and prisms in the areas of interest. Usually, the measurements of vertical movements are simple. The average error in leveling height 1,000 m is estimated as 2.5 mm.

Information concerning the prisms is contained in a database comprised of 15 tables, each table contains different number of prisms.
Within each table, the prisms have different measurement periods, the sampling rate is irregular, there being no measurements and other days with concentrated at certain times of day measurements. The sampling rate can vary orders from one hour to several days.

\subsection{Ground Radar}
\
\par

Radars for monitoring of embankments and hillsides emerged in the last 10 years, as a cutting-edge tool for monitoring movements of slopes and embankments, largely because of its ability to  measure the movements of the slopes with a level of pinpoint accuracy over large areas, and regardless of weather conditions, avoiding the installation of artificial reflectors.
The first type of radar monitoring introduced in the mining market was based on radar satellite, which corresponds to Real Aperture Radar (RAR), which takes advantage of a thin beam radar illuminates the target along a small number of scanning areas.

\subsubsection{Synthetic Aperture Radar (SAR)}
\
\par

In 2000, Italian researchers developed the GB InSAR (\textit{Ground-Based Interferometric Synthetic Aperture Radar}), a system for measuring surface movements of slopes and embankments (Tarchi et al. (2003) \cite{Tarchi2003}). The main differences between SAR and RAR is that GBInSAR is faster, since it takes the whole scene at once, not with RAR systems that make sweeping the area. Furthermore GB InSAR has higher ranges of effectiveness (Atzeni et al. (2014) \cite{Atzeni2014})
The sensor records the movements in the area to monitor. Registration can be done in various environmental conditions, no matter whether it is day or night, with fog or dust. It measures remotely the simultaneous displacement of thousands of points of a surface in large areas (natural reflections of the slope), without the need for any access to the slope for observation.

\subsubsection{Previous experiences using this technology}
\
\par

The use of radar for slope monitoring is the standard in open pit mining. Radar units are used efficiently for safety monitoring in critical situations, providing alerts on events with progressive movement.
Monitoring over conventional radar technology presents notable advantages, high precision measurements, long-range capability, limited impact of atmospheric effects on measurements and the possibility of rapid measurement of a large number of points.
Technology such radars is based on radar interferometry, a technology originally developed for satellite applications, in order to recover the displacement of the surface of the Earth related to natural events \cite{Rosen2000}. However, there is a limitation related to satellite platforms, given the low coverage of satellites available and the consequent temporal correlation of the radar signal, plus the problem of geometric distortions induced by the almost vertical line of sight. To overcome these limitations, the same technology has been implemented in terrestrial systems. ground radar interferometry have been used for monitoring natural slopes \cite{Antonello2004,Corsini2006,Bozzano2011,Bozzano2008} and slopes  \cite{Reeves2001, Harries2006}.

In this work we consider  information of five projects IBIS,  from  October 2012 until January  2014, see Table \ref{tablaIBIS}.  These projects have been transferred through text files, separated by tab. The information is organized in a series of project files.
The files of these projects contain the following information:
\begin{itemize} 
\item The coordinates $X$ and  $Y$ in meters and in the mine local reference system, the axis $OX$ is from south to north and the axis  $OY$ is from west to east. 
\item The date and time correspond to the measurement time and we have is one record per hour
\item The displacement per hour is given in millimeters per hour.
\item  The total information available corresponds to 10,143 hours
\end{itemize}
\begin{table}[htp]
\caption{SAR projects available}
\begin{center}
\resizebox{0.8\textwidth}{!} {
\begin{tabular}{|c|c|c|c|c|c|c|c|c|c|}
\hline
Project&
Rate & 
Hours&
Points&
$x_{min}$ & 
$x_{max}$ &  
$y_{min}$ &
$y_{max}$ \\
&
[hora] & 
\#&
\#&
[m]& 
 [m]&  
[m]&
[m]\\
\hline
\hline
Project 1 & 
1 &
1.132&
146.202&
15.465,5&
17.388,5&
107.815,5&
109.543,5
\\
\hline
Project 2 &
1& 
 4.564& 
 137.684&
15.638,1& 
 17.327,1& 
 107.911,5&
 109,450,5
\\
\hline
Project 3 &
 1& 
 646& 
 156.348&
 15.371,5& 
 17.189,5& 
107.944,5&
 109.516,5
 \\
\hline
Project 4 &
 1& 
 2.784& 
 160.450&
 15.331,6& 
 17.036,8& 
107.949,4&
 109.519,4
 \\
\hline
Project 5 &
 1& 
 1.017& 
 155.372&
15.253,4& 
17.350,1& 
107.821,1&
109.499,7
 \\
\hline
\end{tabular}
}
\end{center}
\label{tablaIBIS}
\end{table}%
The number of points in a scene varies according to the distance from the radar to the surface to be monitored, in this case, the number of points is around 160,000 points.

\subsection{Piezometers}
\
\par

Piezometers are specially designed to measure  pore water pressure in soil or rock. They are very important in the research and evaluation of slope stability, dams and dikes.
The most common application in geotechnic is to determine  water pressure in the ground or water level in perforations \cite{piezometro03}.

 In our case, a total of 139 piezometers is recorded, with  measurements between June 2012 and December 2013, for a given area of the pit. Recording measurements piezometers are not regular, differences in measurement are observed from 2-to-20 days.

\subsection{ Interferometric Satellite Images (InSAR)}
\
\par

Since its development, the satellite interferometry has been used to detect changes in the Earth's surface, landslides and other applications, we refer to the works of Colombo (2013) \cite{colombomeasuring},  Ferreti et al. \cite{ferretti2011new},  Raspini et al. \cite{raspini2013landslide} and Notti et al.  (2013) \cite{Notti2013} for details.
With its ability to detect very small movements to millimeter level, large areas, SqueeSAR can be complementary to conventional geological and geomorphological studies on the detection and monitoring of slope instability.

\begin{figure}[h]
\begin{center}
 \includegraphics[width=0.8\textwidth]{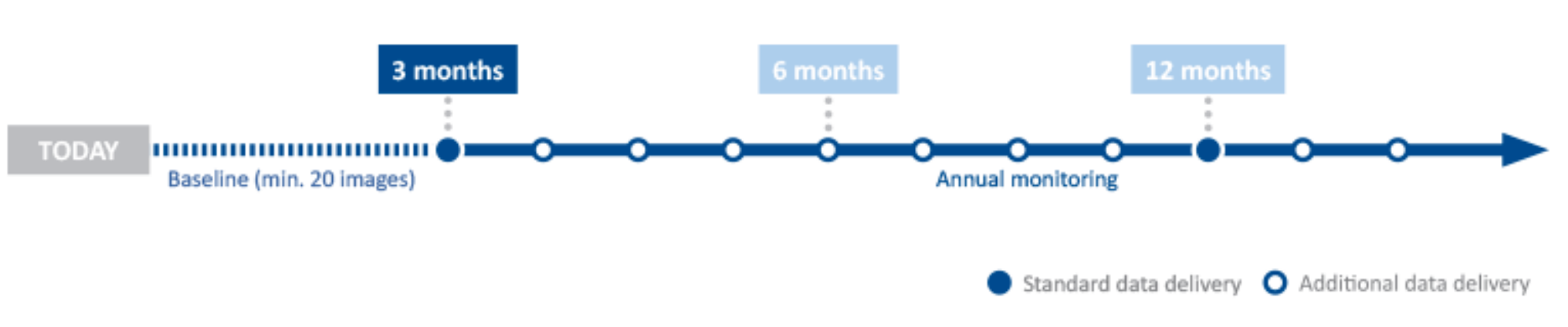}
\caption{Scheme of use of satellite information}
\label{imgsSAR}
\end{center}
\end{figure}

In our analysis we considered SqueeSAR images, 
which provides regular monitoring of the movement of the land surface using satellite SAR images. 
Thus, four maps or files SqueeSAR, with 6 and 12 months of movement accumulation  for the years 2012, 2013 and 2014 were used. We note that once a minimum number of satellite SAR images for an analysis of the base line (minimum 20) is acquired, regular updates of the surface can be  generated, with new images that are acquired over the area of interest (Figure \ref{imgsSAR}).

\begin{figure}[h]
\begin{center}
 \includegraphics[width=0.4\textwidth]{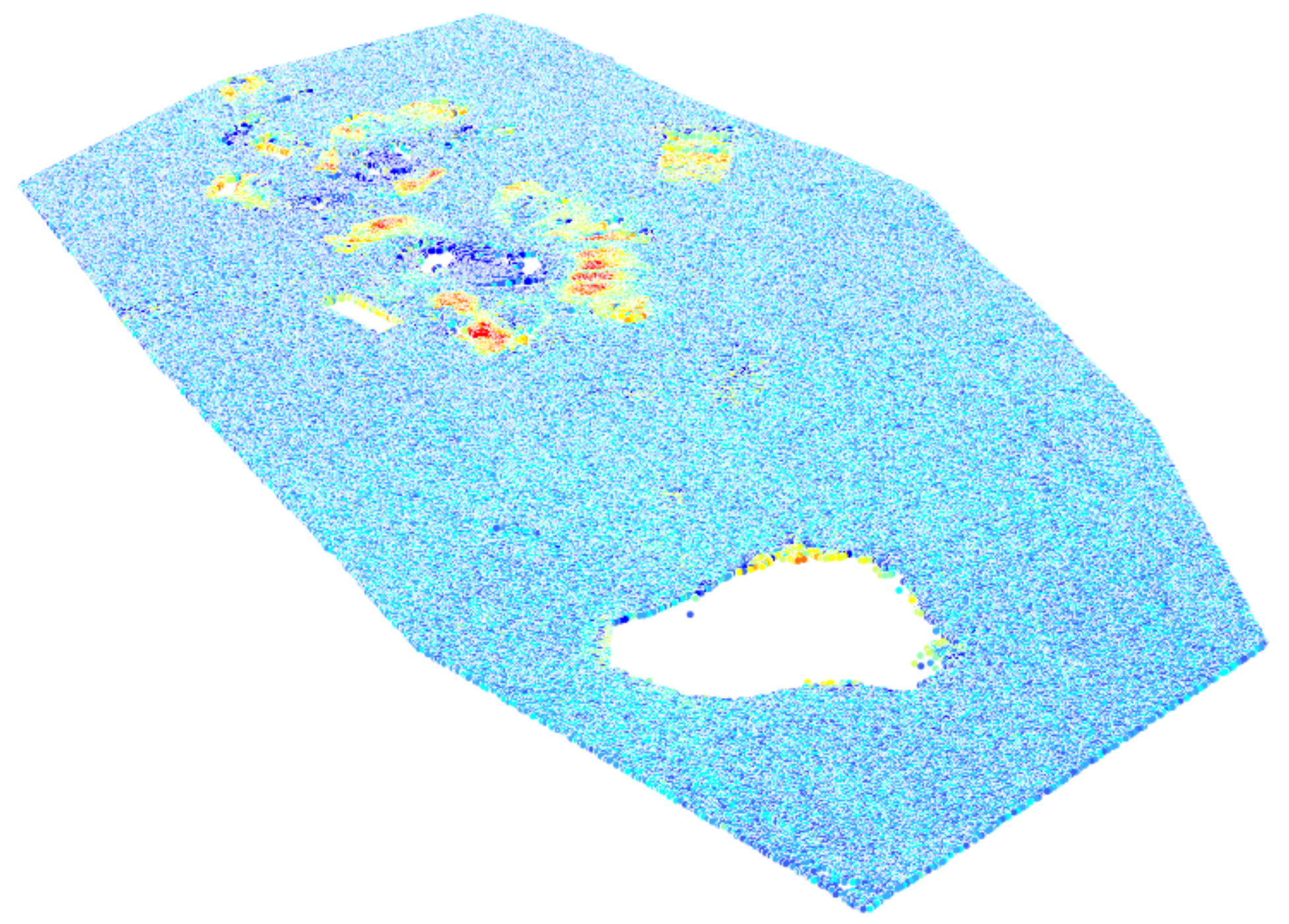}
\caption{SqueeSAR image with accumulated  movements}
\label{squeesarM}
\end{center}
\end{figure}

Figure \ref{squeesarM} shows a map of movements of the Earth's surface between September 4 to October 10, 2012.

\subsection{Other Sensors and Variables}
\
\par

\subsubsection{Digital Elevation Model  - DEM}
\
\par

\begin{figure}[h]
\begin{center}
  \subfloat{
     \includegraphics[width=0.7\textwidth]{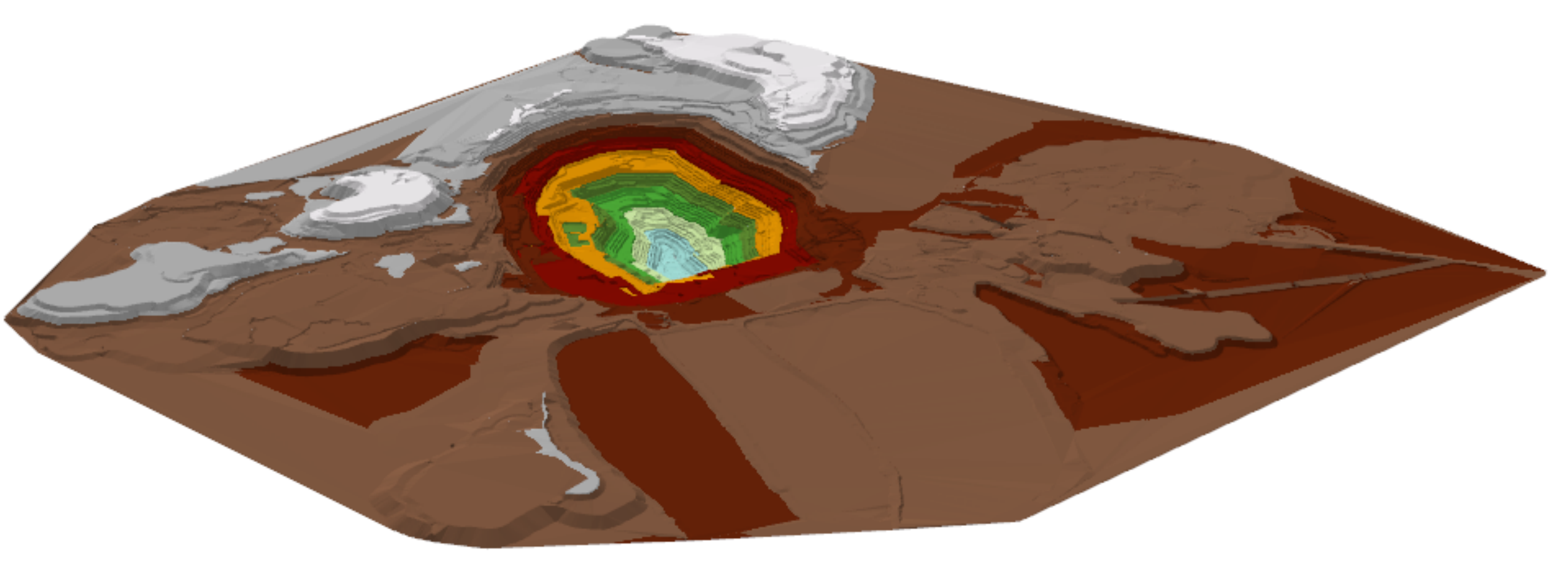}}
  \subfloat{
     \includegraphics[width=0.3\textwidth]{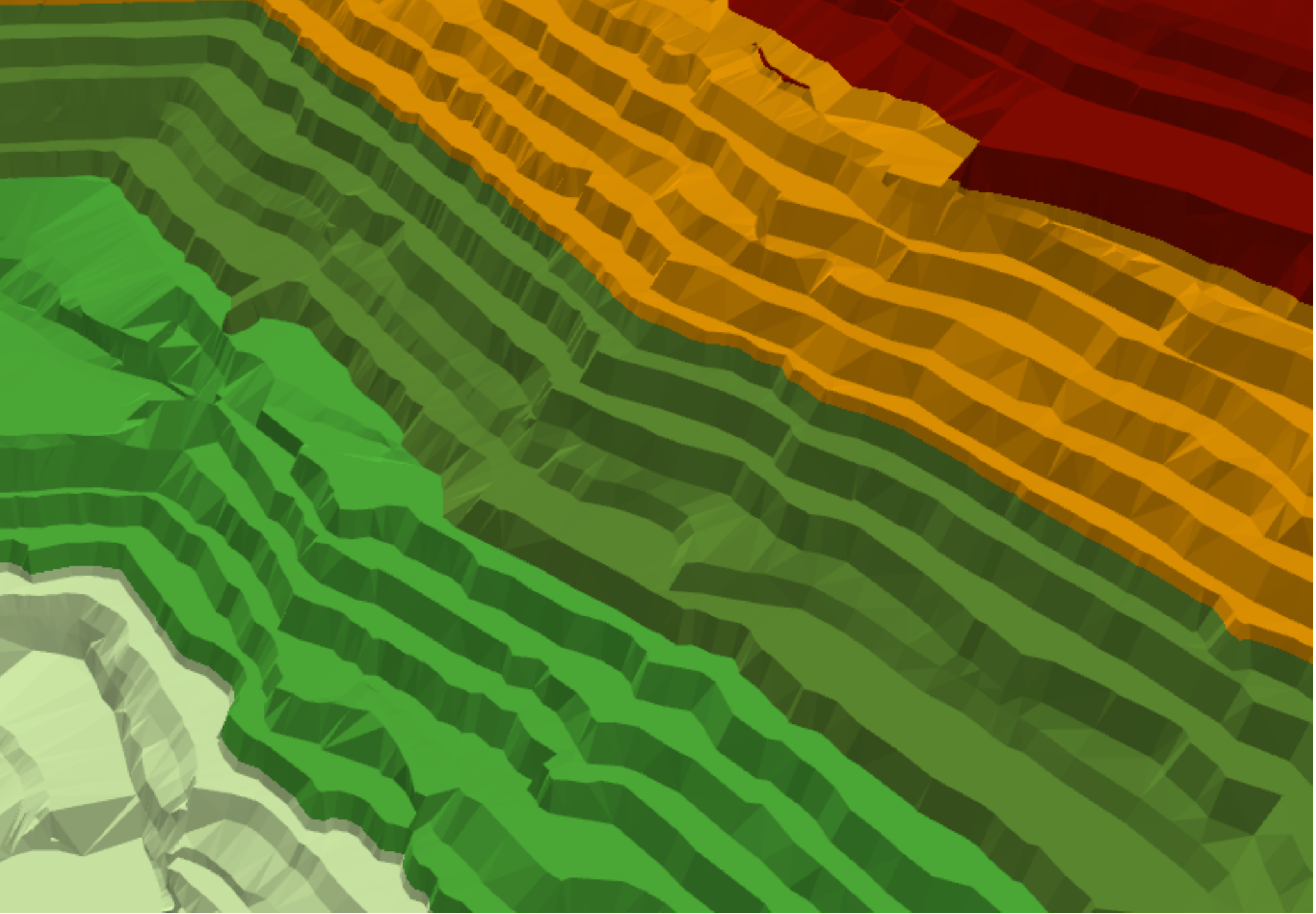}}
\caption{Overview and detail of a DEM}
\label{DEM1}
\end{center}
\end{figure}

In our study, 40 files DEMs were received on a monthly basis, from April 2011 to September 2013. In Figure \ref{DEM1} DEM overview  and below the detail of the  studied area is shown. The DEM realizes changes so that experiences the pit, mainly by exploitation and deformation thereof.

\subsubsection{Block Model}
\
\par

Some extra data of the geology of the mine is also received. This includes five tables containing: 
\begin{itemize} 
\item Lithology 
\item Geological alterations 
\item Mineralogy
\item Rock Mass Rating (RMR)
\item Unconfined Compression Strength (UCS)
\end{itemize}
This information comes from a 3D model of the entire  mine.

\subsubsection{Structural models}
\
\par

\begin{figure}[h]
\begin{center}
 \subfloat[Top view of the alterations of the mine]{
 \includegraphics[width=0.4\textwidth]{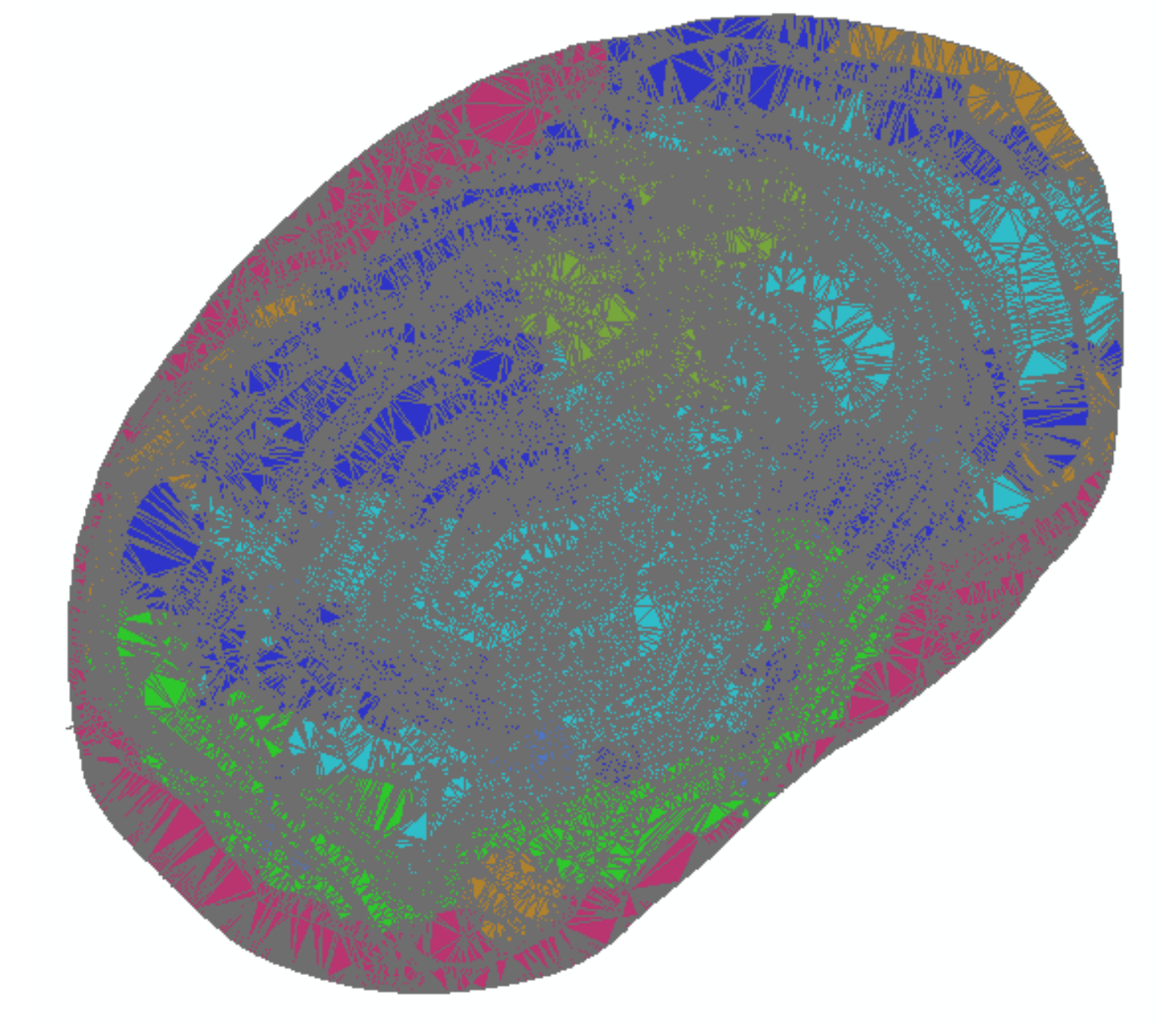}} \qquad
  \subfloat[Top view of the structural model of a geological fault]{
 \includegraphics[width=0.4\textwidth]{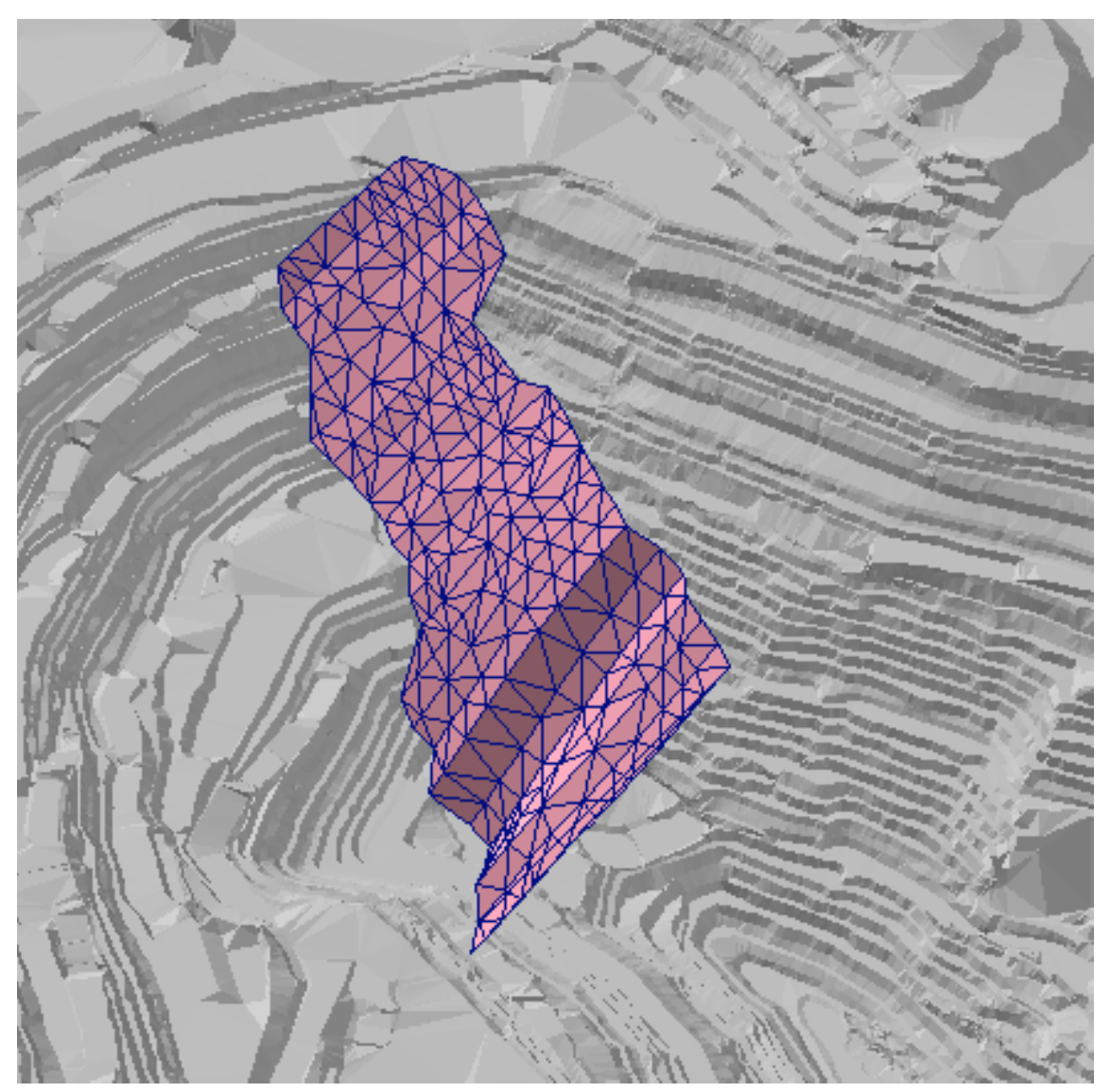}
 \label{falla}}
\caption{Structural Models}
\end{center}
\end{figure}

Additionally we obtained the structural models, i.e., models of geological faults present in the pit. A fault is a discontinuity in the earth's crust that occurs naturally by the propagation of a fracture in the rock. Figure \ref{falla} shows a structural model.
This variable was not considered in the models developed due to their difficult representation on the domain models. But it could be included in later development.

\section{Data Analysis}
\
\par

In this section we presents an analysis of the data provided, corresponding to the different instruments used for geotechnical management. This analysis is very important for the construction of the models under study.

\subsection[Ground-based Radar]{Ground-based Radar}
\
\par

The data acquired by the sensor were deeply analyzed to obtain the information necessary to build models. 
We note that in this case, the data delivered by the company corresponds to a ground synthetic aperture radar SAR.
This sensor has the particularity of being a sensor which delivers a very detailed information of the study area, with high spatial and temporal resolution. In addition to generating data in near real time. That is why, much of the analysis were used to this sensor, which has the potential to deliver the evolution of behavior pit, and therefore the events that are happening in it.

\begin{figure}[h]
\begin{center}
\subfloat[Average value of negative displacements per hour]{ \includegraphics[width=0.45\textwidth]{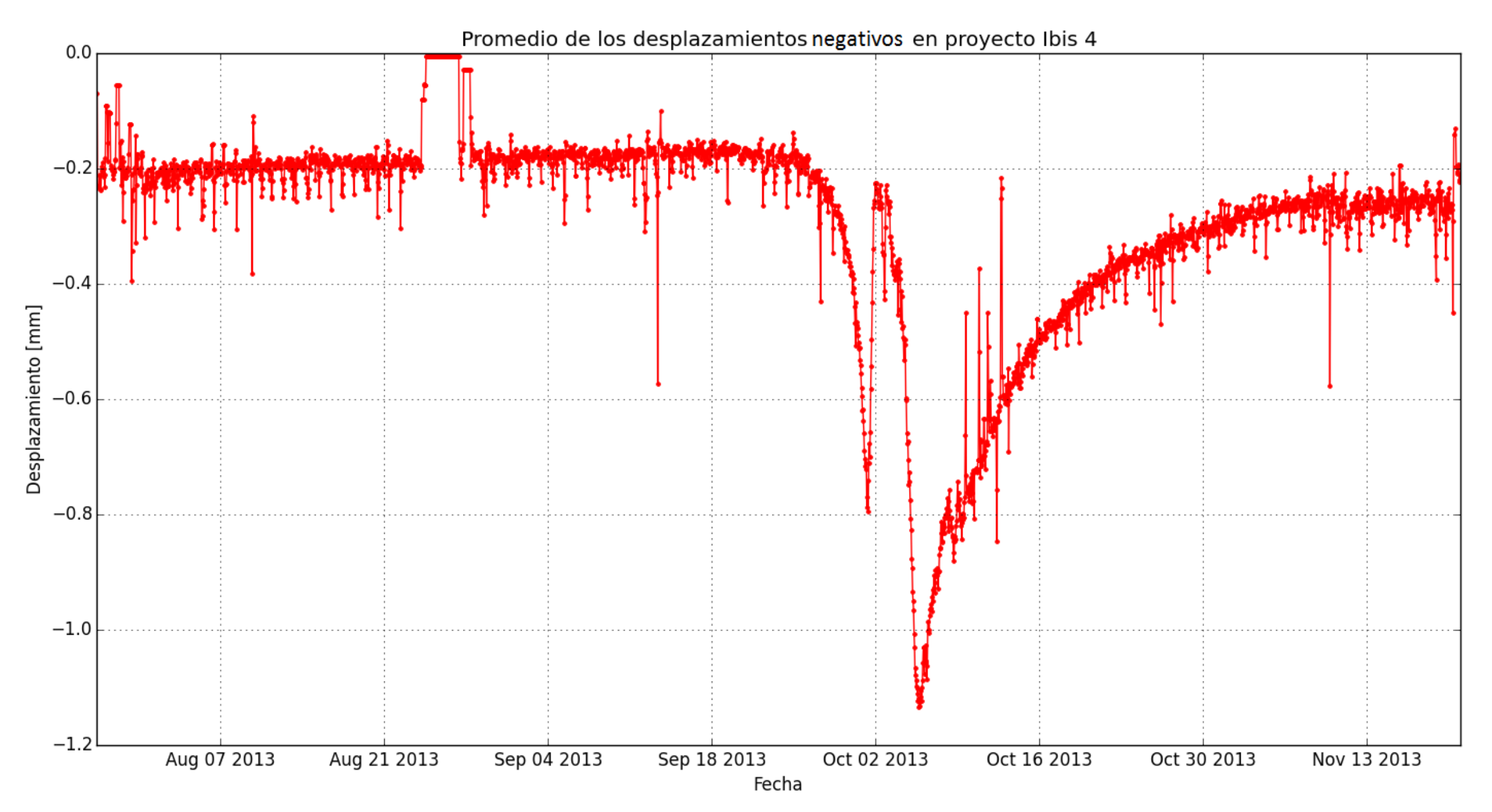}
\label{Ibis4_n}
}
\subfloat[Average value of positive displacements per hour]{
\includegraphics[width=0.45\textwidth]{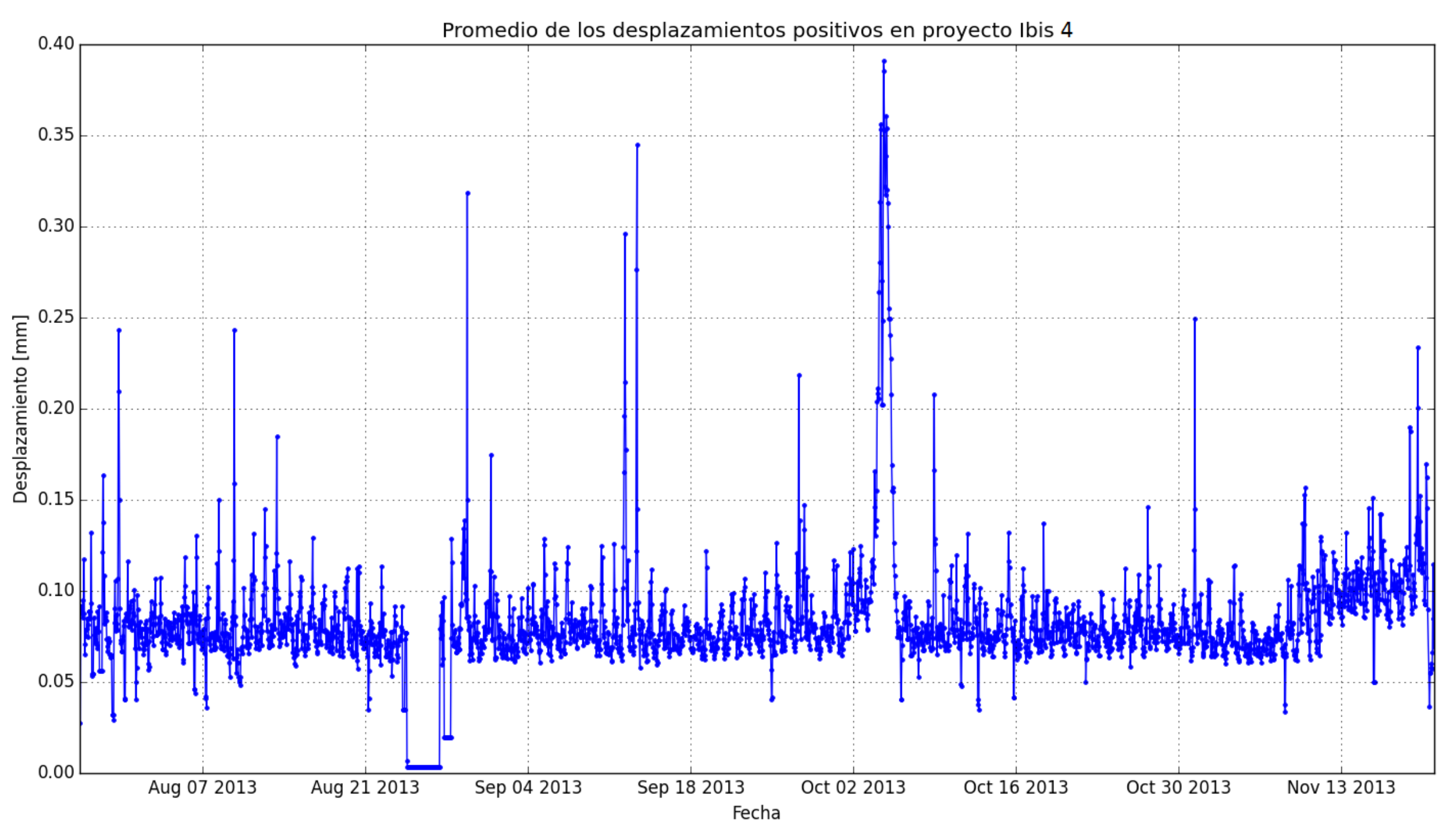}
\label{Ibis4_p}
}
\caption{Average value of displacements per hour between July and November 2013}
\end{center}
\end{figure}

Preliminary analyzes were performed with a dimensional reduction of the data, that is, the temporal resolution of the data, by leaving them in daily data with a decreasing measuring points. At this stage, thanks to the use of other data model such as HDF5 format, which allows the handling and manipulation of a large amount of information, among other advantages, we consider atemporal resolutions of hours. This allowed the one hand, capture a greater amount of information, but also a higher level of variability and noise in the data.
Also positive and negative data were separated due to a lack of correlation between the two sets of values, but only negative data will be consider for our study. To illustrate this fact we refer to Figures \ref{Ibis4_n} and \ref{Ibis4_nD}  where
average hourly shows negative displacements of all projects.

In Figures \ref{Ibis4_n} and \ref{Ibis4_p} we can appreciate the project 4, separating negative and positive displacements. In this period a significant displacement happens in a particular  area, which begins to move significantly from October 2013. A good characterization of the  event is achieved with the values of negative displacement (Figure \ref{Ibis4_n}) and not with the positive values (Figure \ref{Ibis4_p}). This is because the main displacement movement is toward the sensor.

\subsubsection[Statistics]{Statistics}
\
\par

\begin{figure}[h]
\label{prom_var01_n}
\begin{center}
\includegraphics[width=0.48\textwidth]{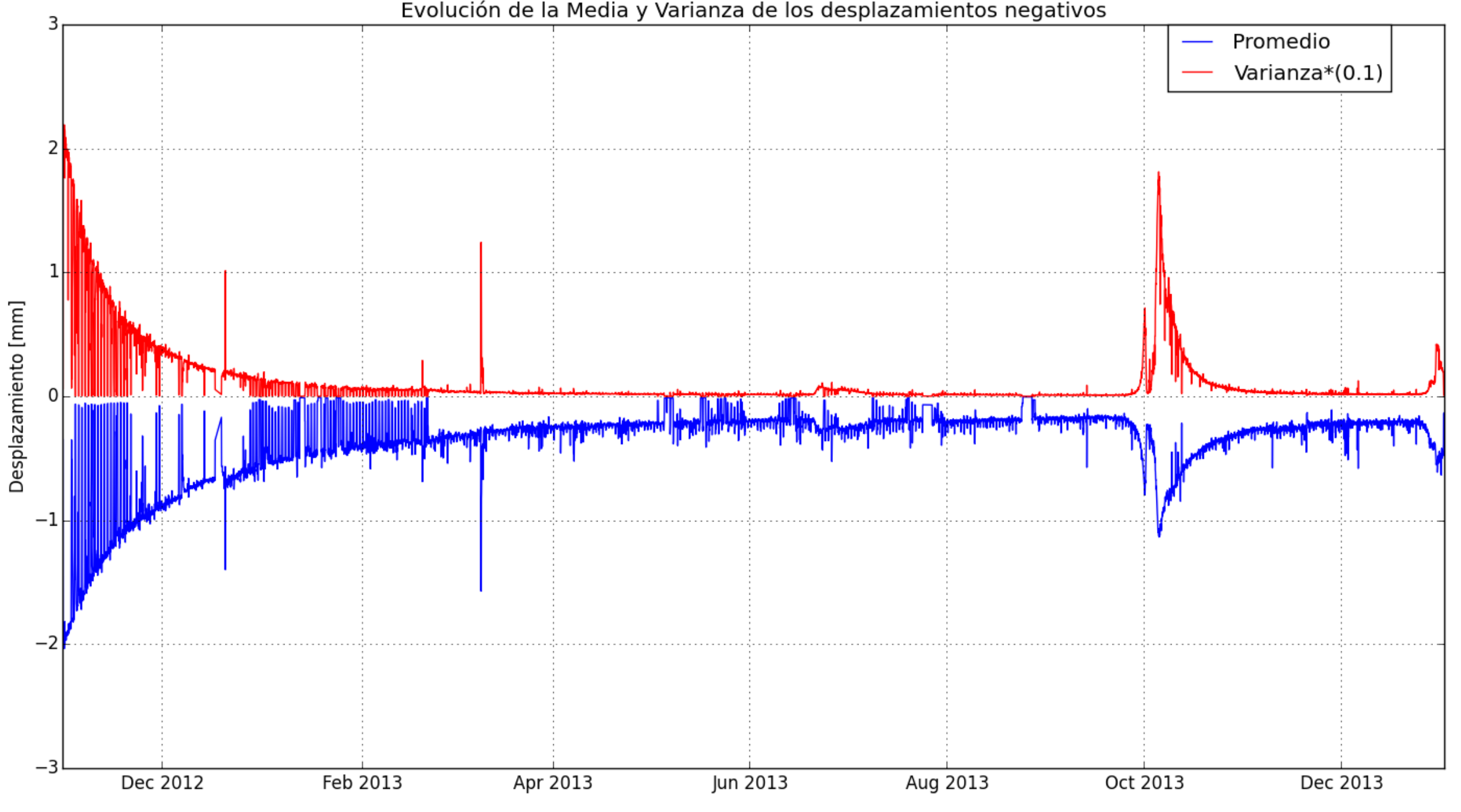}
\caption{Evolution of the mean value and variance of movements per hour.}
\end{center}
\end{figure}

Firstly we consider the five projects and we compute the mean value and variance hourly average. We separate positive and negative values due to differences in their behavior, which  was verified in previous analyzes separated. Both negative and positive data, the mean value and variance of the displacements of all points for each IBIS project was calculated, that is
\begin{equation*}
\sigma^2_t = \frac{1}{n} \sum_{i=1}^n  \left( x_{it}- \overline{x}_t \right)^2 =  \left( \frac{1}{n} \sum_{t=1}^n x^2_{it} \right) - \overline{x}_t^2 \qquad\textrm{where}\qquad \overline{x}_t = \frac{1}{n} \sum_{i=1}^n,   x_{it}
\end{equation*}
where $\sigma^2_t$ is the variance at time $t$ in all points, $\ x_{it}$ is the displacement of point $i$ at time $t>0$, $\overline{x}_t$ is the mean value at time $t$. Thus we obtain the results showed in Figures \ref{prom_var01_n}, \ref{prom_var_p}, 
In order to compare values in the graph of negative values (Figure \ref{prom_var01_n} 
the variance values weighted 0.1 were used.

Figure 8 shows the evolution of the mean value and variance of the displacement of the northern sector of the pit recorded by the IBIS sensor, between October 2012 and January 2014. The data include five projects studied. We note that the negative values have a clear tendency to describe the phenomena of sliding slope. 

\subsubsection[Coefficient of Variation]{Coefficient of Variation}
\
\par

\begin{figure}[h]
\centering
\subfloat[All values]{\includegraphics[width=0.49\textwidth]{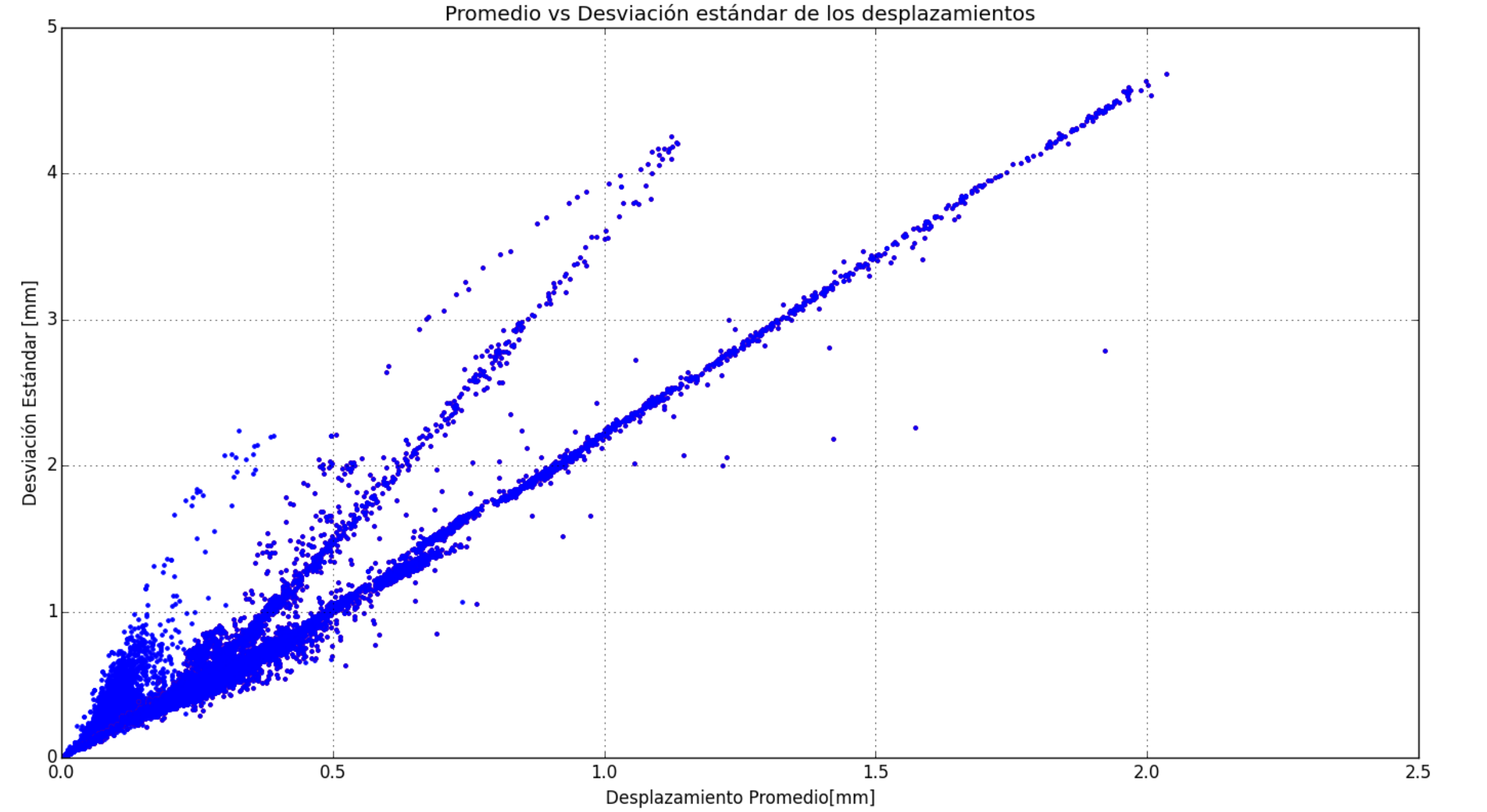}} \vspace{0.01 \linewidth}
\subfloat[Negativ Values]{\includegraphics[width=0.49\textwidth]{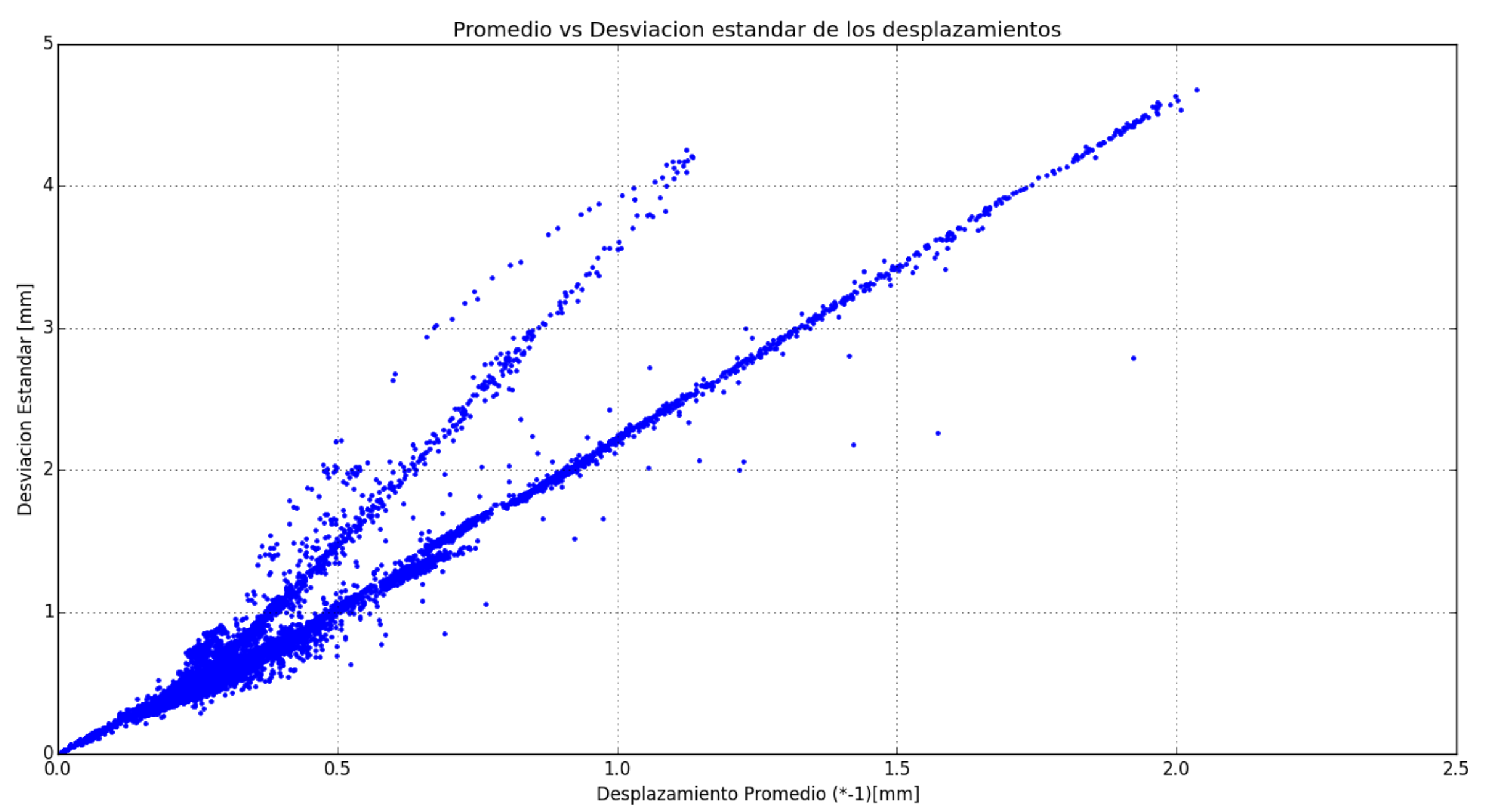}}
\caption{Dispersion graphs fro the mean values and variace of the displacements.} 
\label{med_desv_all}
\end{figure}

When analyzing the data presented above, a close relationship between the mean value and variance of movements recorded by the sensor IBIS is observed. Especially in negative values, i.e., those points of the pit which approach to the sensor.
With average displacement per hour a scatter plot of the absolute value of the mean value versus the standard deviation of such movement is built. This graph shows a high variability and low correlation between these values (see Figure \ref{med_desv_all}). However, if we separate the negative and positive values, we  can see that there are certain symmetries if we consider only the negative values.
Upon analysis it has verified that the increase in travel speed generates points that relate the mean value and standard deviation nonlinearly, however when the affected area starts to slow down gradually, the point cloud behaves more or less linearly.
The slope of this  cloud represents what is known as Pearson variation coefficient, or coefficient of variation, defined in Equation \eqref{CVeq}:
\begin{equation}
CV_t =  \frac{\sigma_t}{| \overline{x}_t|},
\label{CVeq}
\end{equation}
where $\sigma_t$ is the standard deviation and $\overline{x}_t$ is the mean value of the data.

\begin{figure}[h]
\begin{center}
 \includegraphics[width=0.6\textwidth]{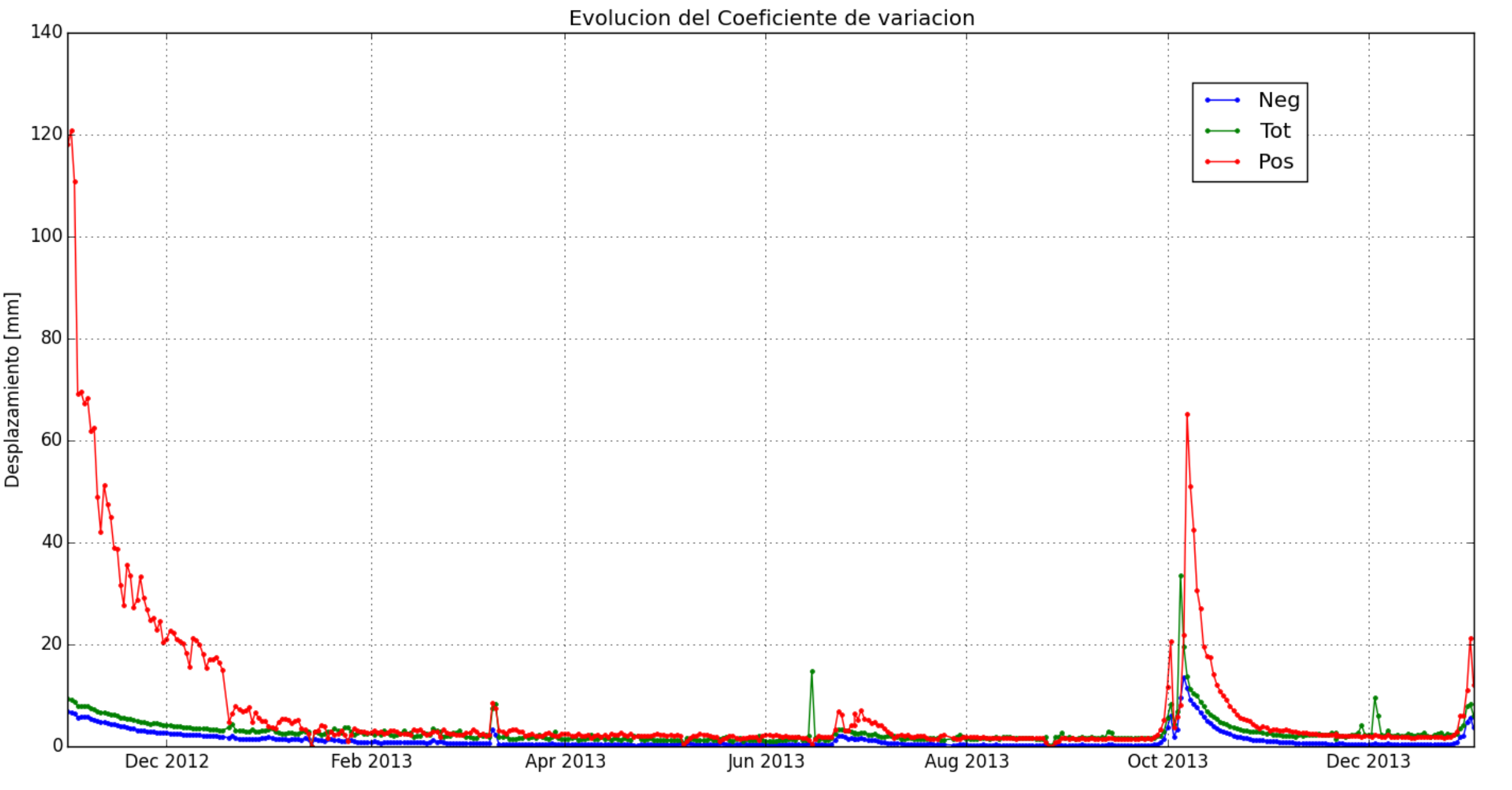}
\caption{Coefficient of Variation of the data across the five projects, for positive, negative and total data.}
\label{CV}
\end{center}
\end{figure}

\begin{figure}[h]
\begin{center}
 \includegraphics[width=0.6\textwidth]{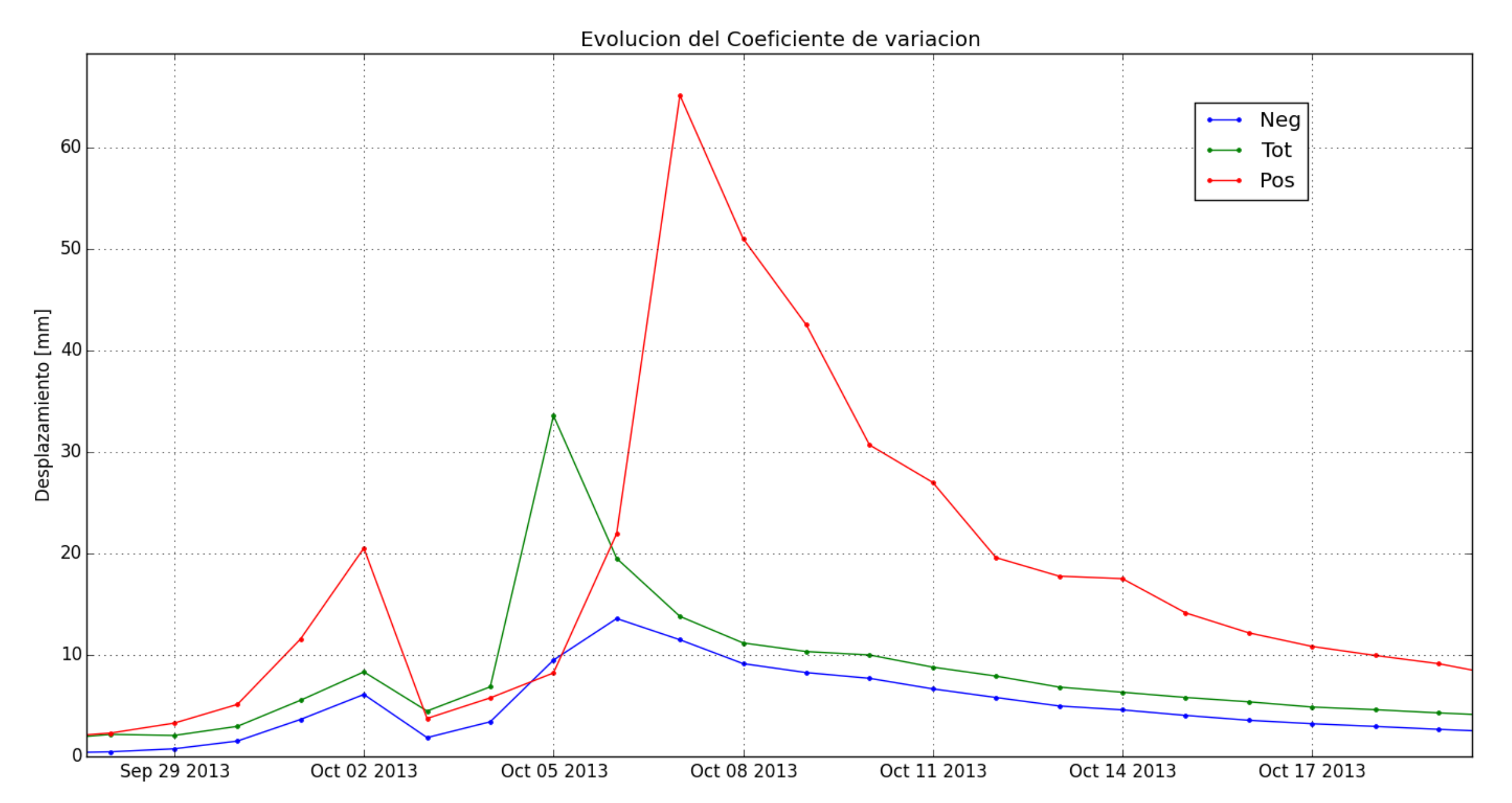}
\caption{Coefficient of Variation detail during the event.}
\label{CV_d}
\end{center}
\end{figure}

Figure \ref{CV} presents the variation experienced by the Coefficient of Variation from November 2012 to January 2014. The different curves represent the coefficient of variation, positive (red) and negative (blue)  values and all values (green). Three changes are observed in the trend of values, corresponding to March 2013, June 2013 and October 2013, what indicate the occurrence of a phenomenon related to the movement of the pit, or perhaps with the malfunction of the sensor.
The coefficient of variation ($CV_t$) represents the standard deviation as a percentage of the average. The more CV there is greater heterogeneity of data, and the average is a worse representative of the measured data. The estimate of this coefficient on the data reveals that the heterogeneity of data increases when events occur, and if Figure \ref{CV_d}  is observed, we can  see that its value is off shortly before the event  occur (October 1, 2013), considering all the (positive and negative together) displacements and the positive and negative values separately.

These analyzes show that it is possible to detect incipient changes in the monitored area. That is, it is possible to anticipate increasing movement, comparing said current behavior with the historical behavior of the same area.

\subsection{Prisms}
\
\par

 With this data coming from different prisms, we perform several exploratory and comparative analyzes in order to increase the sensitivity of the information.
 
 \subsubsection{Spectral Analysis}
 \
 \par

\begin{figure}[h]
 \centering
    \includegraphics[width=0.7\textwidth]{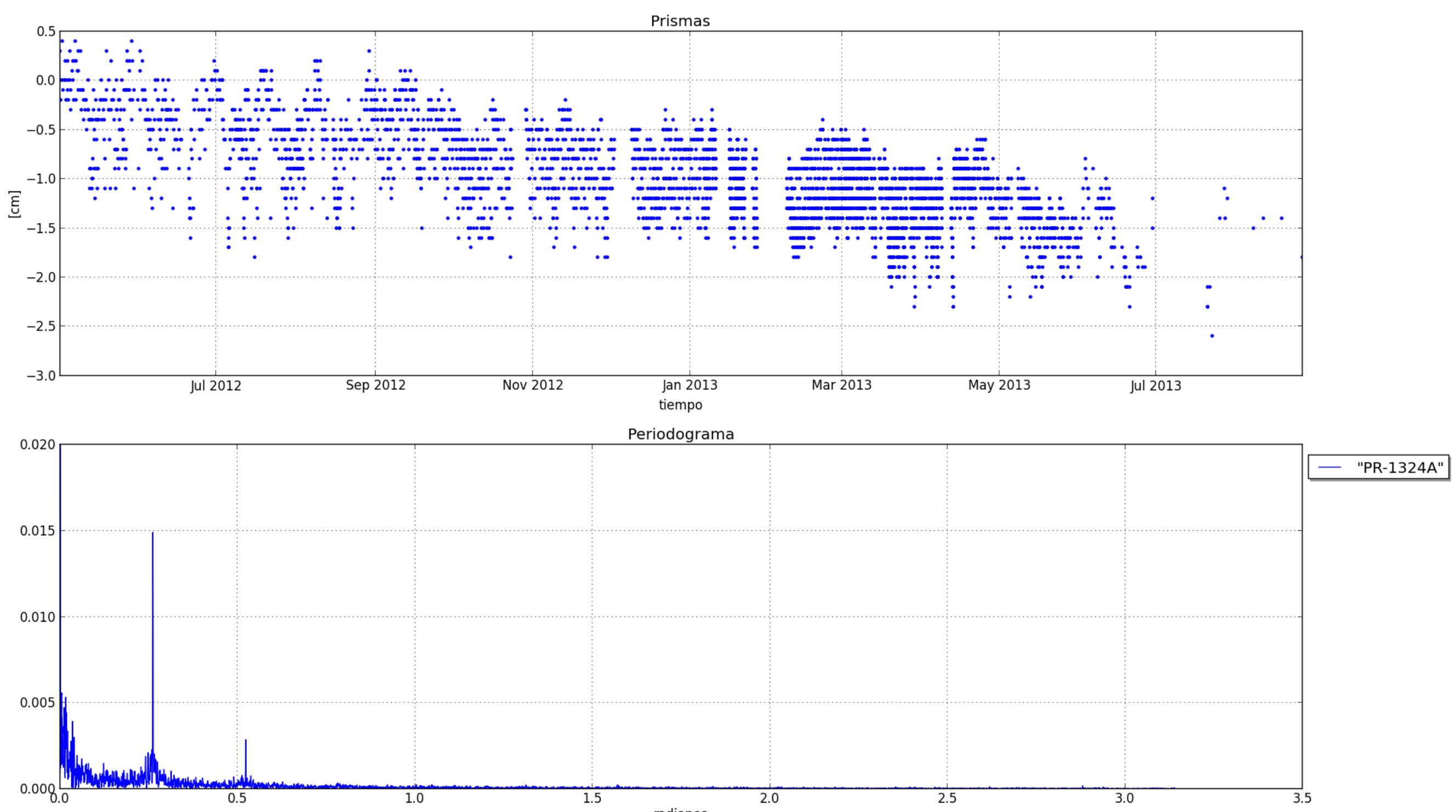}
 \caption{Periodogram corresponding to prism  ID PR-1324A}
\label{FourierEjemplo02}
\end{figure}

By using data from the prisms, a spectral analysis is performed, i.e., the Fourier transform of the data was calculated and the periodogram was obtained, see Figure \ref{FourierEjemplo02}.
Since the data have an irregular sampling rate, it was decided interpolated the data hourly, considering a linear interpolation. Moreover, The FFT (Fast Fourier Transform) was used to obtain the periodogram.
We denote by $ f: [a, b] \to \mathbb {R}$ the function representing the movements recorded by a prism, measured in centimeters and assuming that $ f (t_0) = 0, $, where $ t_0 $ it is the first time that information in the registration prism. Let  $ \overline {f} = \int_a^b f (t) dt, $ be the average of the data, thus  removing the average we obtain 
$$\hat{F}(s)=FFT(f-\overline{f})(s),$$
where $FFT(\cdot)$ is the Fast Fourier transform. Thus, normalizing we have 
$$s\in[0,\pi]\to\frac{|\hat{F}(s)|}{\int_0^\pi|\hat{F}(s)|ds},$$
which can be represented in the periodogram of Figure \ref{FourierEjemplo02}.

\begin{figure}[h]
 \centering
    \includegraphics[width=0.8\textwidth]{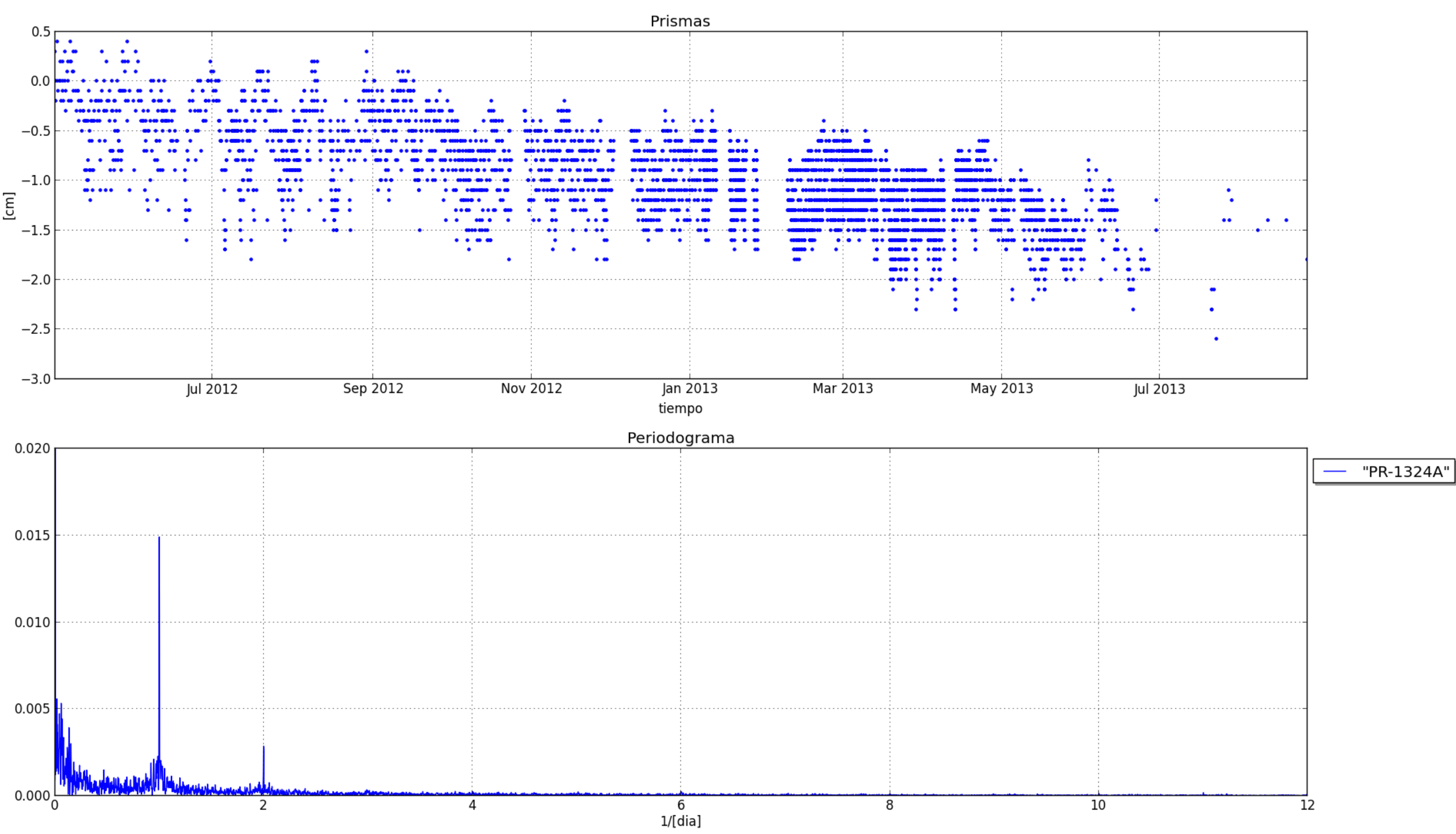}
 \caption{Periodogram of prism ID PR-1324A}
\label{FourierEjemplo01}
\end{figure}

Looking at the periodogram of the prism PR-1324a in Figure \ref{FourierEjemplo02}, we can notice representations of singular moments (Dirac masses).
Then, making a change of scale on the axis $ OX $ we can place these unique moments in whole numbers, specifically we consider the periodogram function
\begin{equation}\label{funcionPeriodograma}
\nu\in[0,12]\to \frac{|\hat{F}(\frac{\pi\nu}{12})|}{\int_0^\pi|\hat{F}(s)|ds},
\end{equation}
thus, using the \ref{funcionPeriodograma} we find that the unique observed frequencies are quantized, that is, we find the unique frequency values
$\nu_1=1,\nu_2=2,$ see Figure \ref{FourierEjemplo01}.
Thus, we find that the unique frequencies, are clearly identified in the values $\nu_1 = 1, \nu_2 = 2 $, and hence at frequencies $ \frac {\pi \nu_1}{12} = \frac {\pi}{12}, \frac {\pi \nu_2}{12} = \frac {2 \pi}{12}, $ with which the periods are given by
$$T_1=\frac{24}{\nu_1}=24,\;\;\; T_2=\frac{24}{\nu_2}=12,$$
This means that data presented this prism intervals every 24 hours and 12 hours. This phenomenon is observed in most of the prisms. Whereupon, we conclude that there is a clear swing in the day for each prism and smaller oscillations are observed, every half day.

\section{Data Correlation}
\
\par

We are interested in to  establish the degree of correlation between the data from the ground radar and the information provided by prisms.
The ideas presented in this section gives an interesting line of development, to be explored in future work, enhancing interaction and complementarity of information and standardization.

The main difficulties in this comparison are:
\begin{itemize}
\item Temporal regularity: both sensors do not have the same rate of acquisition time, while radar  have data per hour with high regularity, the prism records data per hour and with low regularity (4 or 5 data per day ).
\item While radar data, for each time step  register  displacements in a mesh of order $ 10^5 $ elements, the prism register  only one record for each time step.
\item The number of prisms is low compared to the  monitoring area  of the radar.
\end{itemize}

In order to establish a comparative framework between the two sensors, from the spatial point of view, all the data are considered ina fixed grid, i.e.,  radar data IBIS are taken to a screen and  prisms data are reduced to the same mesh. In our case we consider a mesh of  $90\times 90$ meters and with 22 elements on axis  $OX$ and 20 elements on axis $OY$.

\begin{figure}[h]
 \centering
    \includegraphics[width=0.7\textwidth]{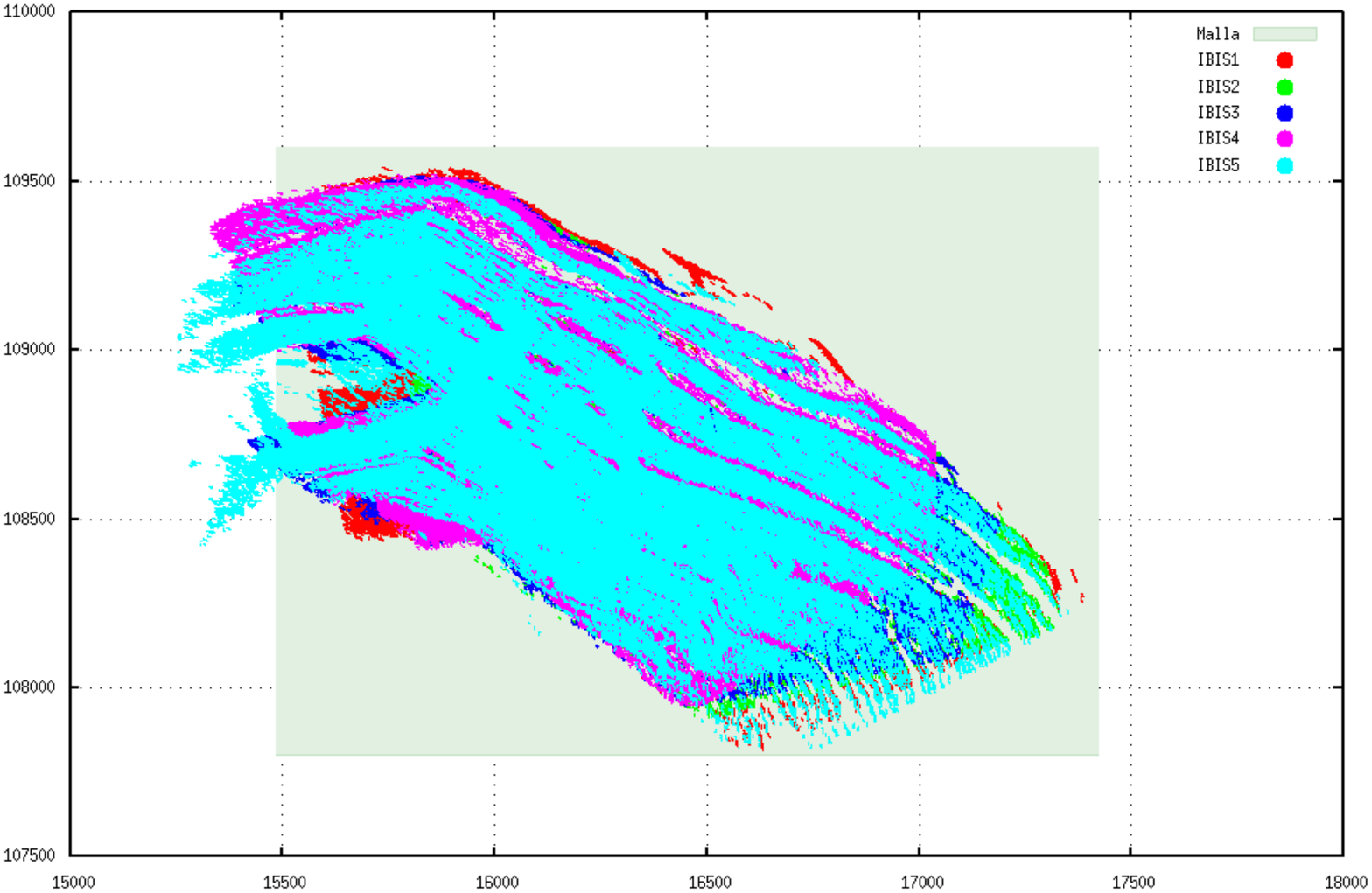}
 \caption{Mess $M$ 
 and the considered area for radar projects.}
\label{MallaIBISs}
\end{figure}

In Figure \ref{MallaIBISs} we can observe the regions involved in the different radar projects and our mesh.

To project the radar data in our mesh, we consider the following approximation.
Using the mesh above defined, we can consider the average of all the data contained in each cell and within the same day.
For the projection of the prisms data in the mesh, we must consider  a temporal interpolation, filling with linear interpolation (days that no data) where is necessary and averaging (when more than one entry per day) to carry all data to our  mesh. Once considered the data of the prisms with a temporal resolution of the data by day, we proceed to perform a spatial interpolation data for a single day that are contained in the region covering the mesh considered. Thus we have a non-uniform spatial interpolation for prisms, to address this problem we decided to use an exponential interpolation.

 We assume that the deformation is characterized by a function $ f:M \subseteq \mathbb{R}^2  \to \mathbb{R}$, thus, the deformation recorded by the prisms are on the mesh consists of an evaluation this feature on a number of points. Let $ \{p_n \} _ {n = 1}^m \subset M$ be the position of the $m$-prisms considered within the mesh $ M $. Thus we have the following interpolation problem:

To find 
$\{\alpha_n\}_{n=1}^m\subset \mathbb{R}$ 
such that
 \begin{equation}\label{iterpola01}
 f(x)\simeq \sum_{n=1}^m \alpha_n e^{-\|x-p_n\|^2/D},
 \end{equation}
where $D$ is a parameter that accounts  the influence of the Gaussian function used. We can note that this problem can be carried to a linear system of equations, but unfortunately we can assure that the corresponding matrix is not invertible, because the sampling is not regular. To solve this problem we can rewrite the interpolation problem in the following form:
To find $\{\alpha_n\}_{n=1}^m\subset \mathbb{R}$ such that 
 \begin{equation}\label{iterpola04}
 f(x)\simeq \sum_{n=1}^m \alpha_n e^{\frac{-\|x-p_n\|^2}{D_n}}1\!\!1_{\|x-p_n\|\leq R},
 \end{equation}
 where $D_n$ is an arbitrary parameter which can be choose later in order to assure the invertibility of the system. 
In this case the Gaussian function is centered  at  $p_n$ with a coefficient  $D_n$, moreover we multiply each component by the indicator function of ratio $R$
 $$1\!\!1_{\|x-p_n\|\leq R}=
 \left\{\begin{array}{cc}
 1 & \textnormal{Si }\|x-p_n\|\leq R\\ \\
 0 & \textnormal{Si }\|x-p_n\|> R.
 \end{array}\right.$$

 Thus we consider the interpolation problem associated to the function  \eqref{iterpola04}, which can be rewritten as a linear system of equations
 \begin{equation}\label{sistemaLineal}
 b=A\alpha,
 \end{equation}
where $b_i=f(p_i)$ and $\alpha$ is an unknown vector, while the matrix  $A\in \mathcal{M}_{m\times m}(\mathbb{R})$ is defined by 
 $$A_{i,n}= e^{\frac{-\|p_i-p_n\|^2}{D_n}}1\!\!1_{\|p_i-p_n\|\leq R},\;\;\; i,n=1,...,m.$$
 
 In order to guarantee the invertibility of the matrix $A$, we can consider the dominant diagonal condition, that is,
 $$A_{i,i}\geq \sum_{n\neq i}A_{n,i}, \;\;\; i=1,...,m,$$
thus, it is enough to satisfies the condition 
\begin{equation}\label{condi01}
\sum_{n\neq i}e^{\frac{-\|p_i-p_n\|^2}{D_n}}\leq 1.
\end{equation}

For that reason we can choose the parameter $D_n$ verifying  
\begin{equation}\label{condi02}
D_n\leq \frac{1}{\ln(m)}\min_{i\neq n}\|p_i-p_n\|^2.
\end{equation}

We remark that it is easy to prove that it is enough to consider the condition 
\begin{equation}\label{condi03}
D_n\leq \frac{1}{\ln(m)}\;\min_{
\tiny\begin{array}{c}
i\neq n,\\
 \|p_i-p_n\|\leq R
 \end{array}
 }\|p_i-p_n\|^2,
\end{equation}
to guarantee the invertibility of the matrix and the existence and uniqueness of solution for our linear system \eqref{sistemaLineal}.

\subsubsection{Data correlation}
\
\par

In order to study the data correlation  of different sensors, in particular  the spatial and temporal interpolation of data prisms, can be consider in the same mesh spatiotemporal data from the ground radar. Later  the correlation between these two data vectors is studied,  that is, a temporal correlation is calculated, where there is no data of ground radar data IBIS, these cells can be  not considered in the calculation of the correlation. Thus we obtain a 
\begin{equation}
\textnormal{Correlation}=-0.00238684593651.
\end{equation}

Clearly this shows a low correlation between these two variables, 
but we must mention factors that might influence this fact, 
among these factors we consider:
\begin{itemize} 
\item There are many time-slots in which indistinctly, both variables do not have recorded data, this clearly affect the correlation. 
\item The sparseness of prisms in the area and high concentration (location of nearby prisms). 
\item The interpolation model is not responsible for present failures in the geological structures. 
\item Recorded data from the ground radar consider  displacement variation per unit time, whereas the prisms consider  total displacement. 
\end {itemize}

\subsection{Piezometers}
\
\par

In this case we consider 31 piezometers from a total of 139. We choose these sensors considering their position with respect to the region of study and the availability of the data during the analyzed period.

\begin{figure}[h]
\begin{center}
 \includegraphics[width=.7\textwidth]{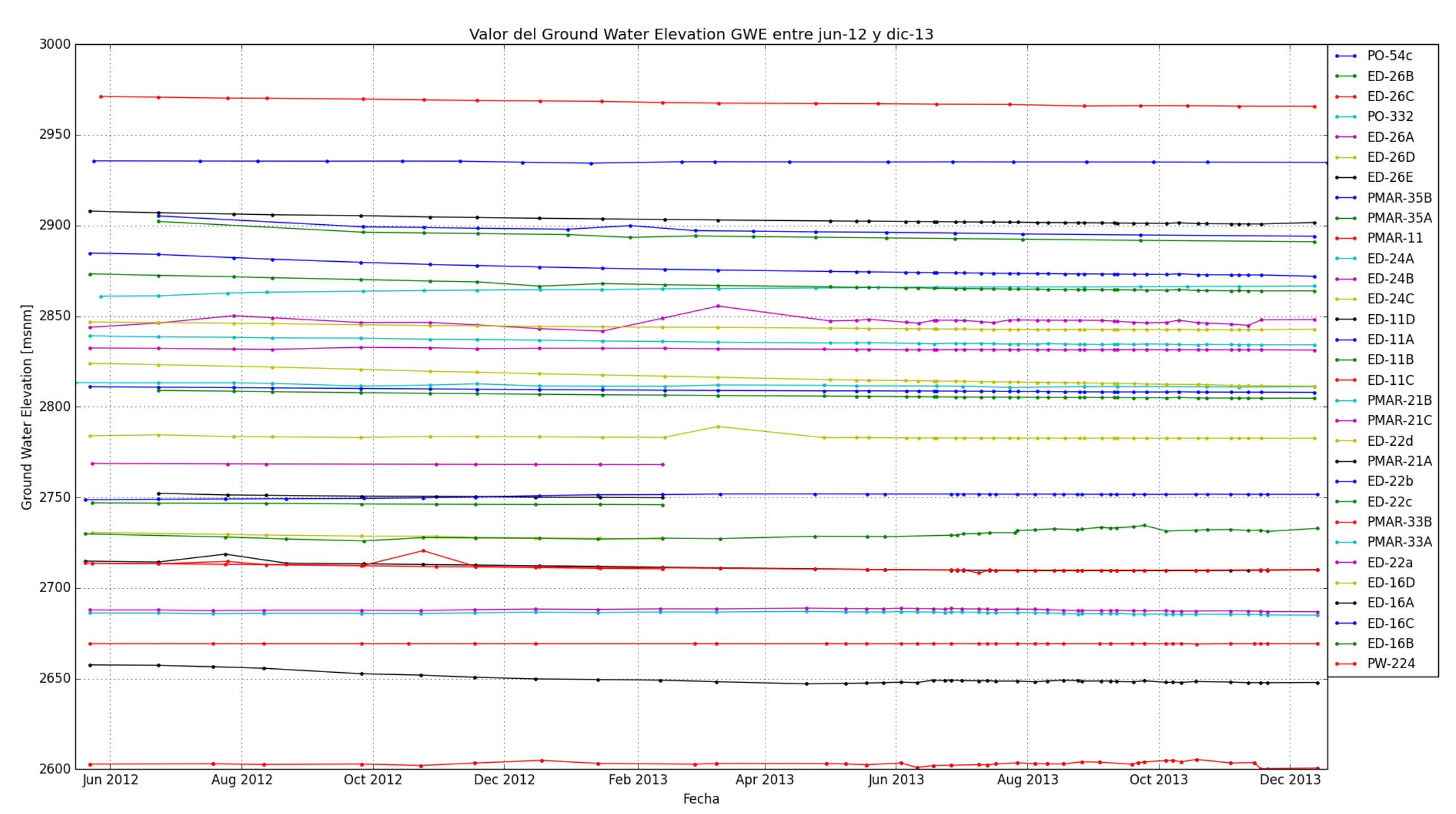}
\caption{Measurement of Groundwater Elevation (GE) for selected piezometers}
\label{GWE_ALL}
\end{center}
\end{figure}

The graph in Figure \ref{GWE_ALL} shows the measurement of GE in selected piezometers. It can be seen that, in general, maintain their values around the start of records period, nevertheless, there are more noticeable changes in measurements of GWE for PMAR35A, PMAR35B, ED24B, ED11D, ED11C and ED22d piezometers.

\subsubsection{Statistics for \textit {Groundwater Elevation}}
\
\par

We calculated some basic statistics of the data studied, in particular the mean, standard deviation, maximum values and the minimum values of measurements of GE. When analyzing the records obtained we can see that GE levels remain more or less stable over the period considered. The maximum values of standard deviation are around 3 meters, for 16 piezometers located in position. The graphics of the maximum and minimum for each of the piezometers values are very similar due to the narrow range of variation of the data, compared to the order of magnitude of GE recorded.

\subsubsection{Conclusions}
\
\par

While data analyzed from piezometers show some changes in their measurements, these variations were not related directly in spatial and temporal manner, with the analyzed event. Particularly in the study area there are no piezometers to verify the relationship between events and measurements of \textit {Groundwater Elevation}.

\section{Results}
\
\par

The results of the analysis of the processed data and information generated were two models for predicting occurrence of phenomena of deformation of the wall of the pit.
The first is a static model which estimates a long-term risk zone of likely future movements, creating a risk map of the study area with an horizon of time of six months before the estimate. The second model corresponds to a dynamic model that predicts the time of occurrence of the next significant movement of the wall of the pit.

 Static long-term model analyzes  the data obtained from the studied region, with  a grid of 50 x 50 meters originally and then a grid of 30 x 30 meters was used to reduce the dimensionality of information. Data feeding the model corresponds to records of 6 to 12 months before the event occurred.
The dynamic model consists of calculating an index motion and a motion vector. They are built based on information recorded by the ground radar. These indexes are based on the historical behavior of the study area, which is compared with the behavior of the area at the time of the prediction or estimate.

\subsection{Construction of Analytic Basis}
\
\par

To build the analytical basis we need three basic elements:

\subsubsection{Mesh}
\
\par

To reduce the dimensionality of the problem, the inputs to the models, i.e., the set of variables generated from the data  were taken to a georeferenced grid extending over the wall of the pit. Several dimensions to the grid were tested. Firstly a grid of 50x50 meters was used, the total number of grid cells, in this case, was  35x39 elements. Subsequently, other measures for grid cells were used, these measures considered were: 90, 60, 30 and 15 meters. Trying to establish a relationship with a bank length (15 meters).

\subsubsection{Geographic proximity}
\
\par

\begin{figure}[h]
 \includegraphics[width=0.75\textwidth]{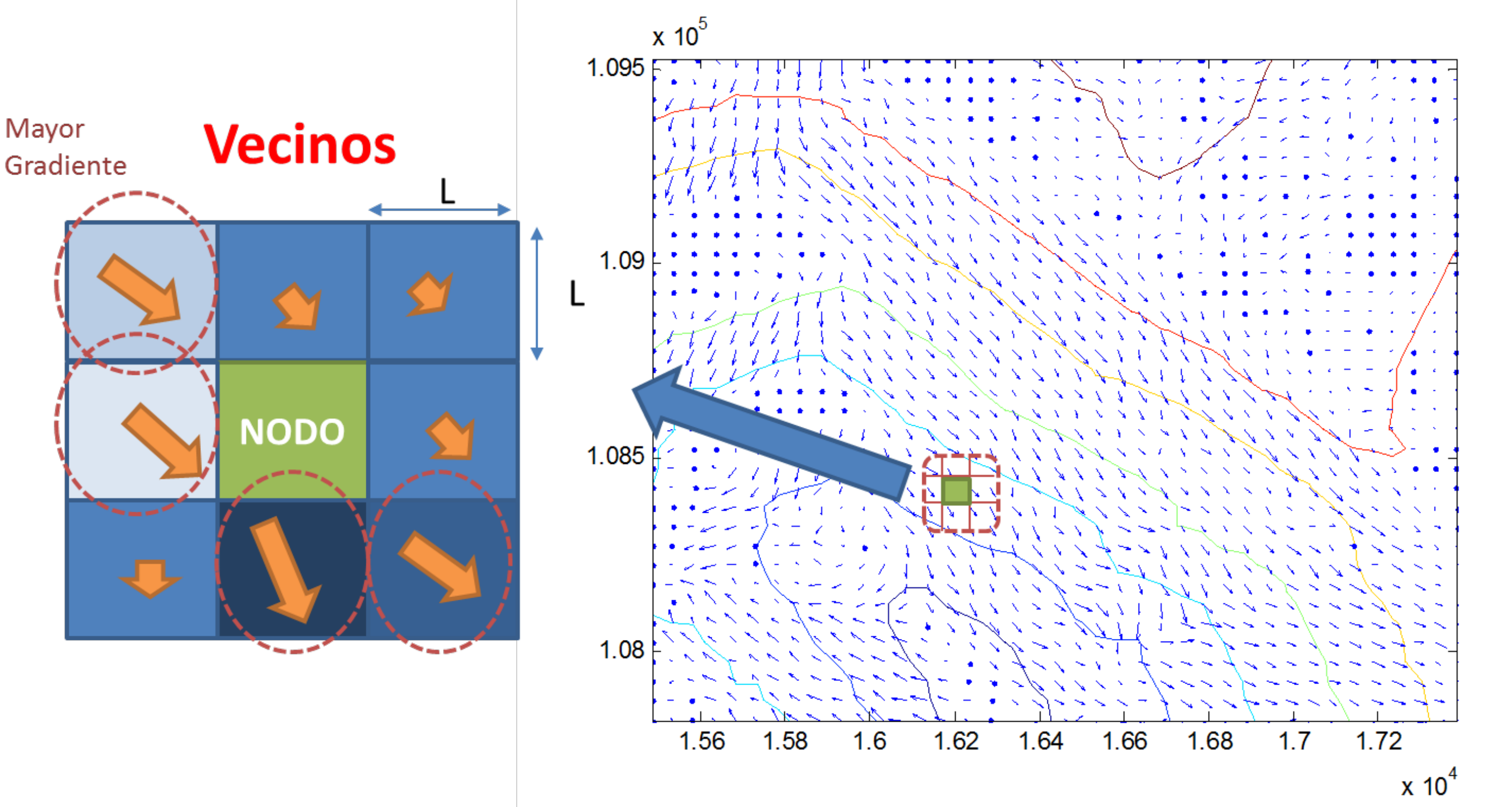}
\caption{Choice of neighbors}
\label{FigBaseAna_01}
\end{figure}

To handle the geographical proximity of the points we consider 4 relevant neighbors, they were chosen from the adjacent grid cells with the greatest gradient. Figure \ref{FigBaseAna_01} shows a cell (node) with its eight closest neighbors, and four of them are elected with the biggest  gradient (two above and two below).

\subsubsection{Allocation of Lost Data}
\
\par

For the imputation of missing or non measured data, radial functions for interpolation were used in the interior areas of the grid, in the case of no data exists. The adjustment of the radial functions are made through neural networks based on radial basis functions (RBF-\textit{Radial Basis Function}).

According to the type of  data,  a Gaussian function RBF were used in order to control the influence of the other elements of the grid.  The neural network RBF has the advantage for the particular case of INSAR, which delivers higher or lower values than those contained in the sample, unlike the IDW values. This means that in the case of data that have been lost, and we assume that they are high (negative and positive high speed), the approximation is better. In the case of comparison with kriging type models, these can deliver similar results, if a  kriging Gaussian is concerned and if the number of neurons in the number of samples is left fixed in the RBF. The advantage is that RBF a simpler (less power) so that the kriging also delivers more consistent results and soft. Nevertheless, the interpolation and effectiveness in the formulation are very similar.

\begin{figure}[h]
\centering
\subfloat[Gaussiana.]{\includegraphics[width=0.4\textwidth]{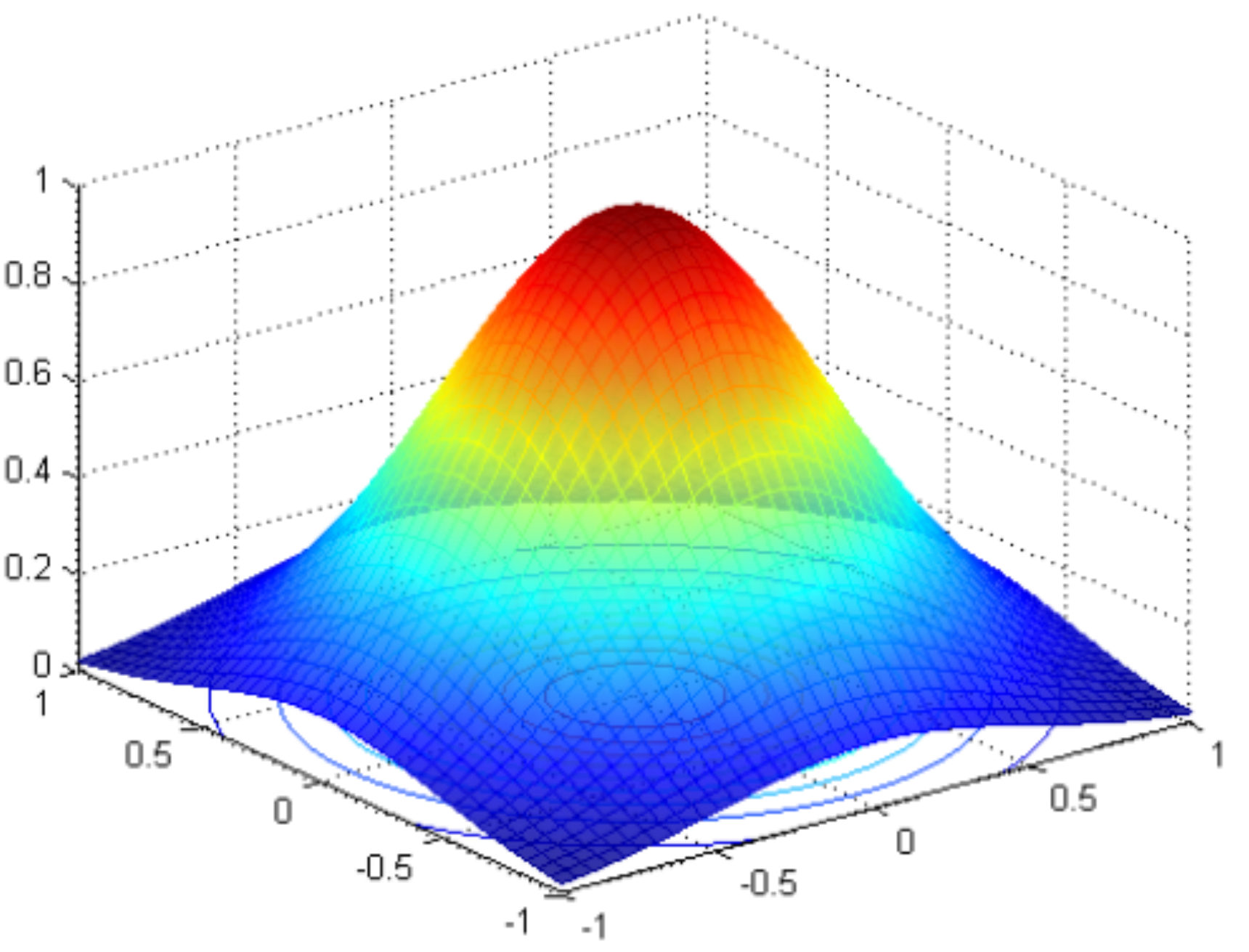}} 
\subfloat[Big]{\includegraphics[width=0.40\textwidth]{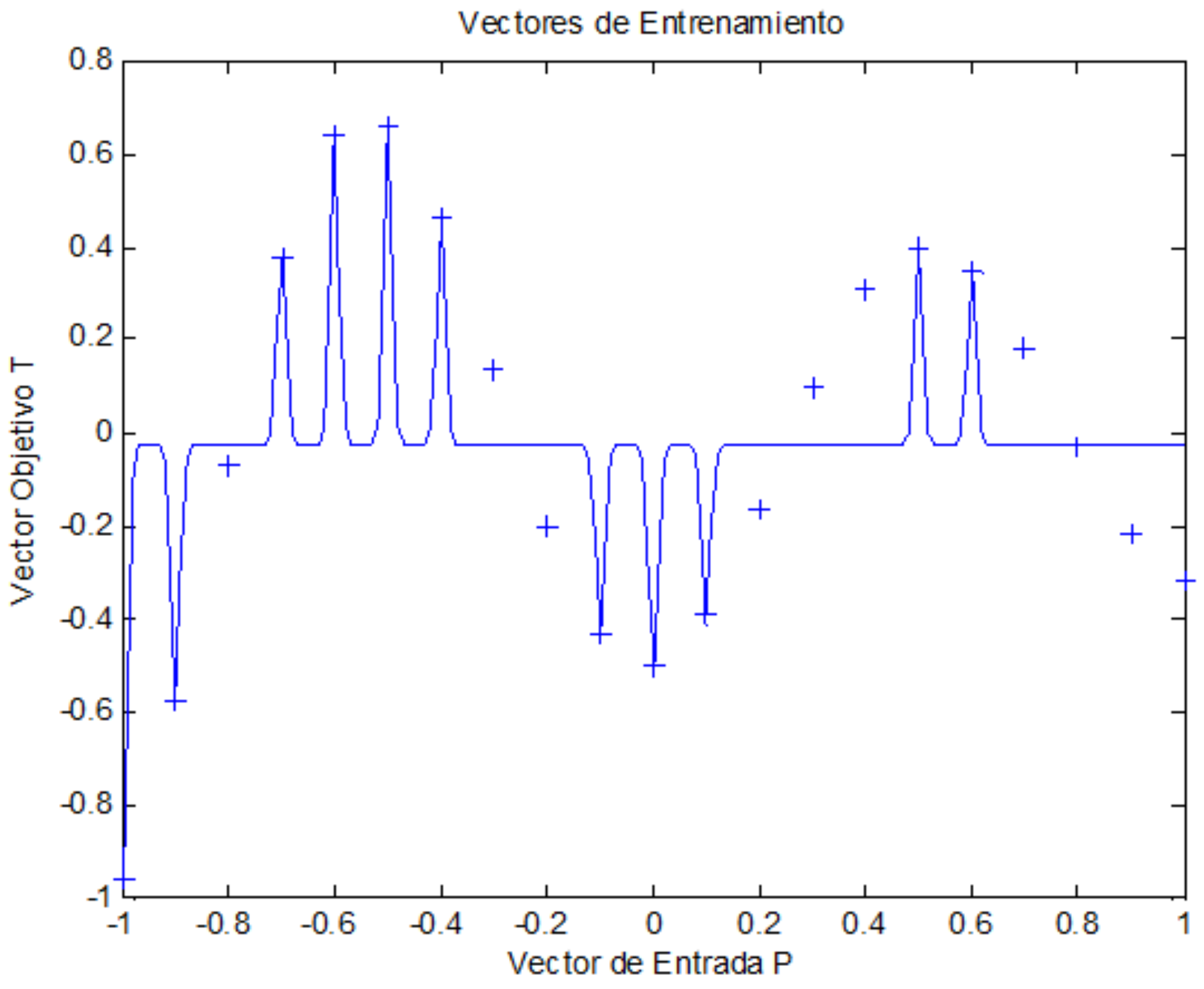}}\vspace{0.01 \linewidth}
\subfloat[Small]{\includegraphics[width=0.40\textwidth]{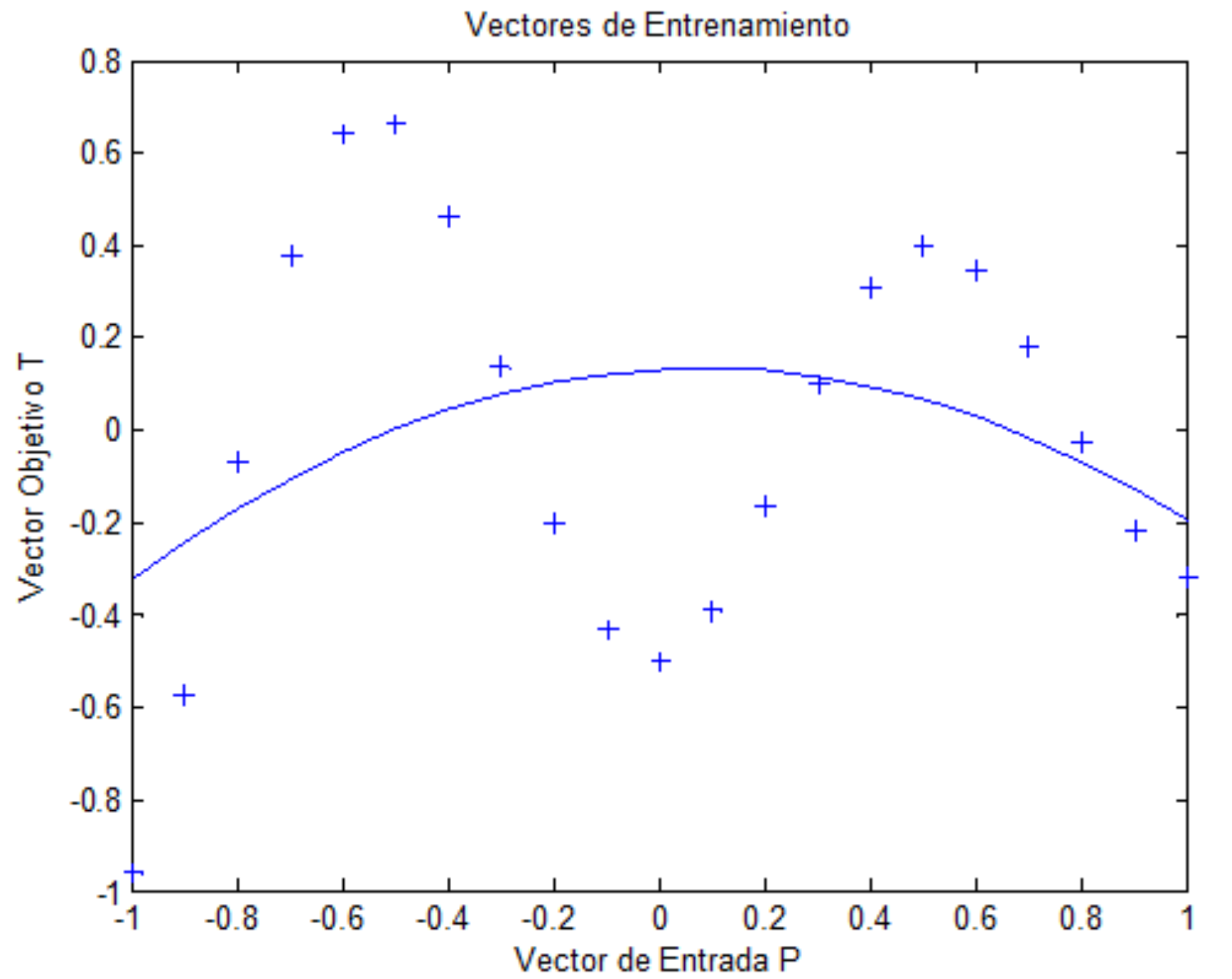}} 
\subfloat[Good]{\includegraphics[width=0.40\textwidth]{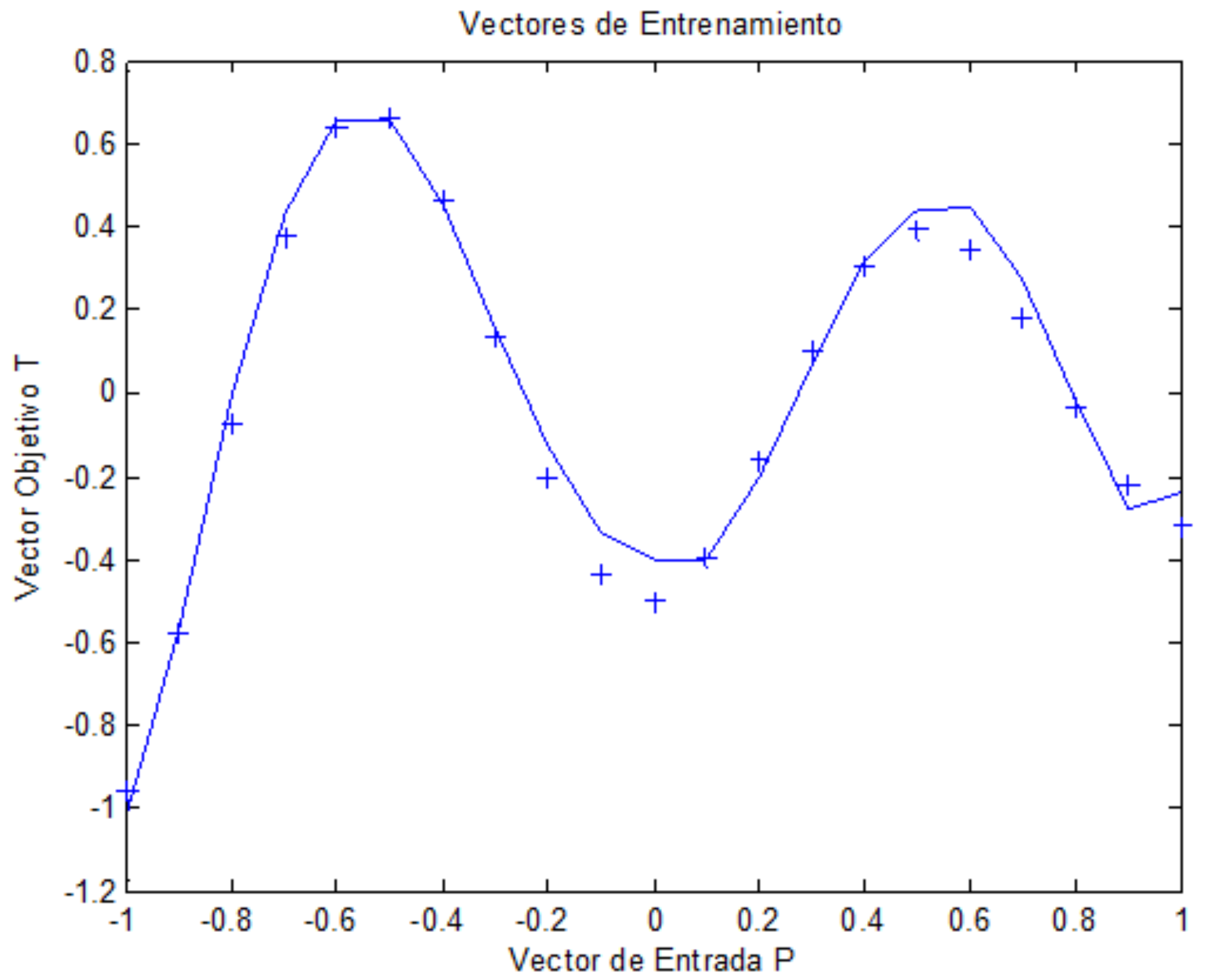}}
\caption{Example of the use of  RBF \textit{radial basis function network}}
\label{interp}
\end{figure}

Interpolation is performed using a Gaussian function at each grid point and is used to interpolate $ n $ dimensions, having more influence the closer you are to the center of the function. Figure \ref{interp} you can see the representation of a 3D Gaussian function, and the training vectors. Within the parameters of training of a RBF, plus the maximum number of neurons is the \emph{entertainment} of the Gaussian, usually if you want to build a probabilistic network this is equal to 1, but if you want to interpolate is better achieve higher values. Figure \ref{interp} can appreciate the use of different values on a fixed set of points, for example, for the same number of points and neurons low overlap is chosen and the result of (b) is, while a high overlap provides a very soft and exaggerated generalization and (c) while adequate considered can be shown in (d). In the case of a geographical problem like this, the \emph{sparsity} is much more interpretable. 

\subsection{Correlations}
\
\par

\begin{figure}[h]
\begin{center}
\subfloat[Linear Correlation (Pearson) of the variables outside the event zone]{\includegraphics[width=.5\textwidth]{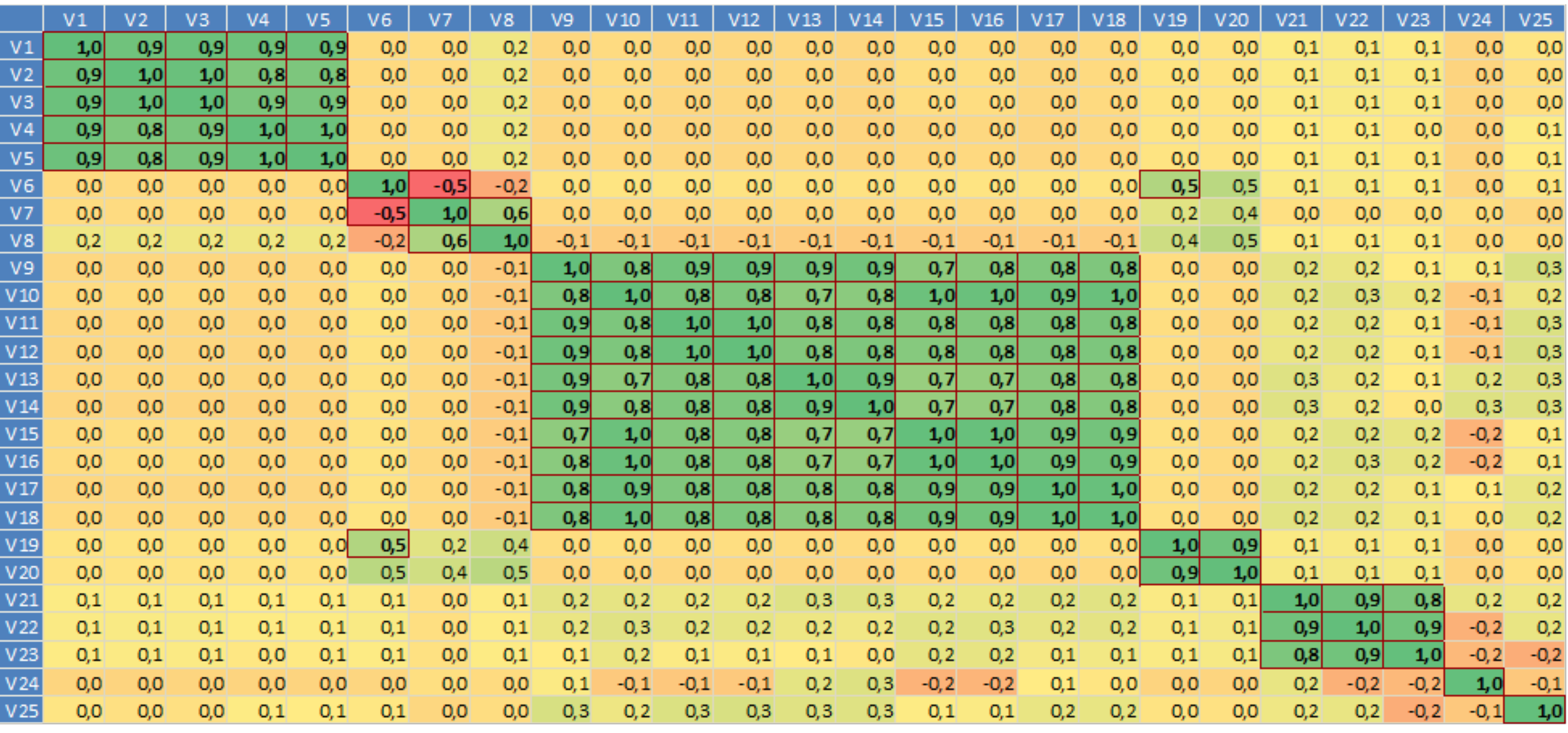}
\label{pearson1}
}
\subfloat[Linear Correlation (Pearson) of the variables inside of the event zone]{\includegraphics[width=.5\textwidth]{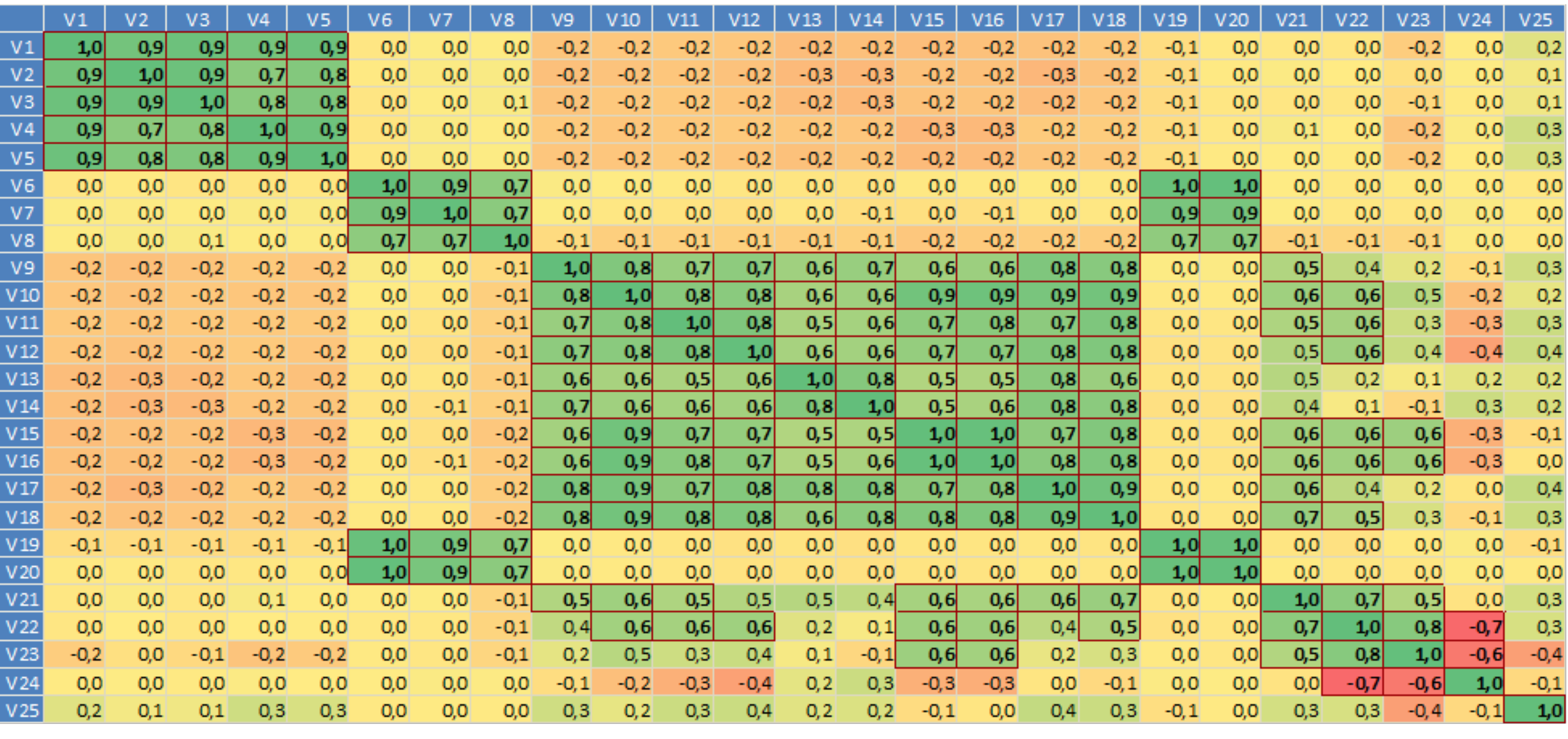}
\label{pearson2}
}
\caption{}
\end{center}
\end{figure}

\begin{figure}[h]
\subfloat[Kendall Correlation of the variables outside of the event zone]{\includegraphics[width=.50\textwidth]{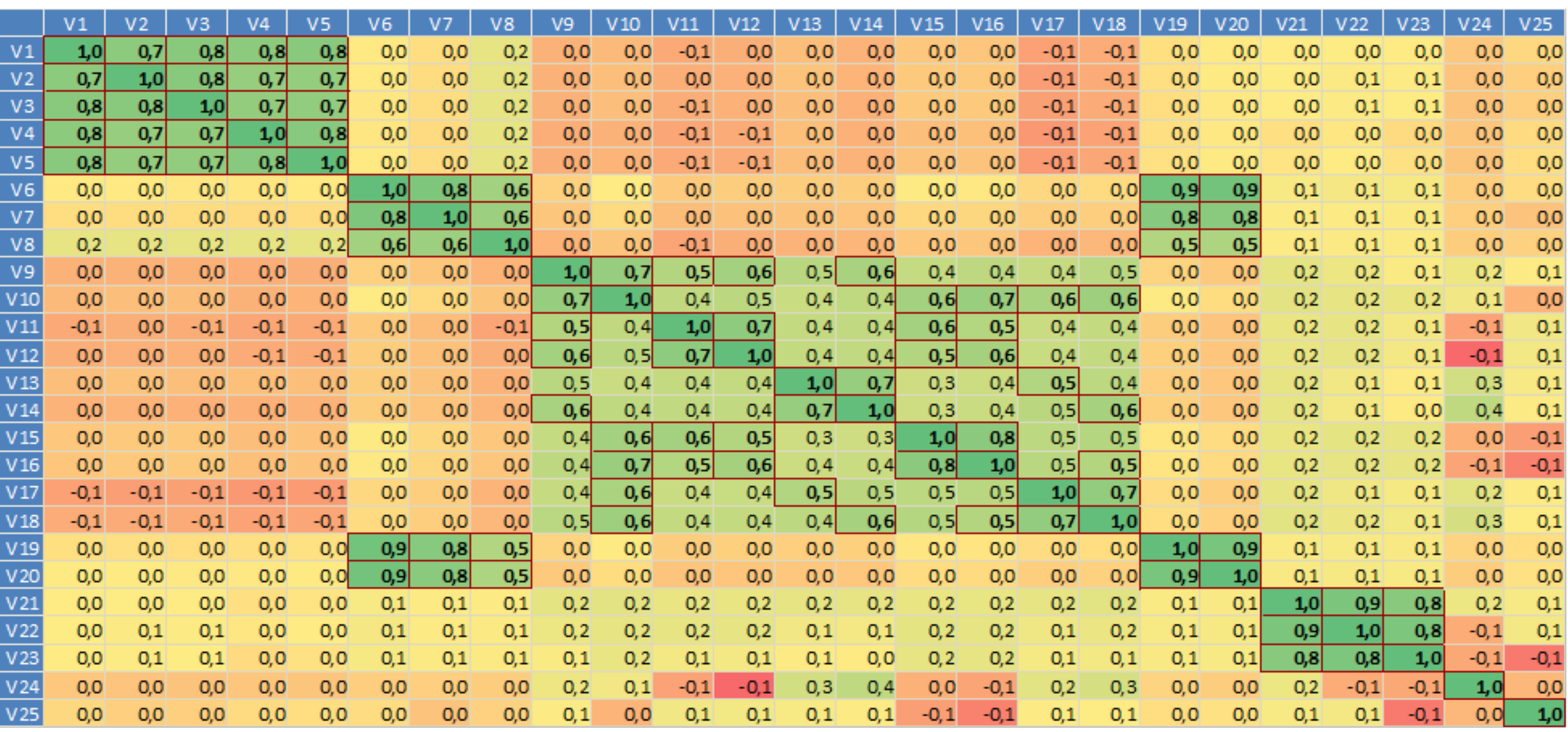}
\label{kendall1}
}
\subfloat[Kendall Correlation of the variables inside of the event zone]{\includegraphics[width=.5\textwidth]{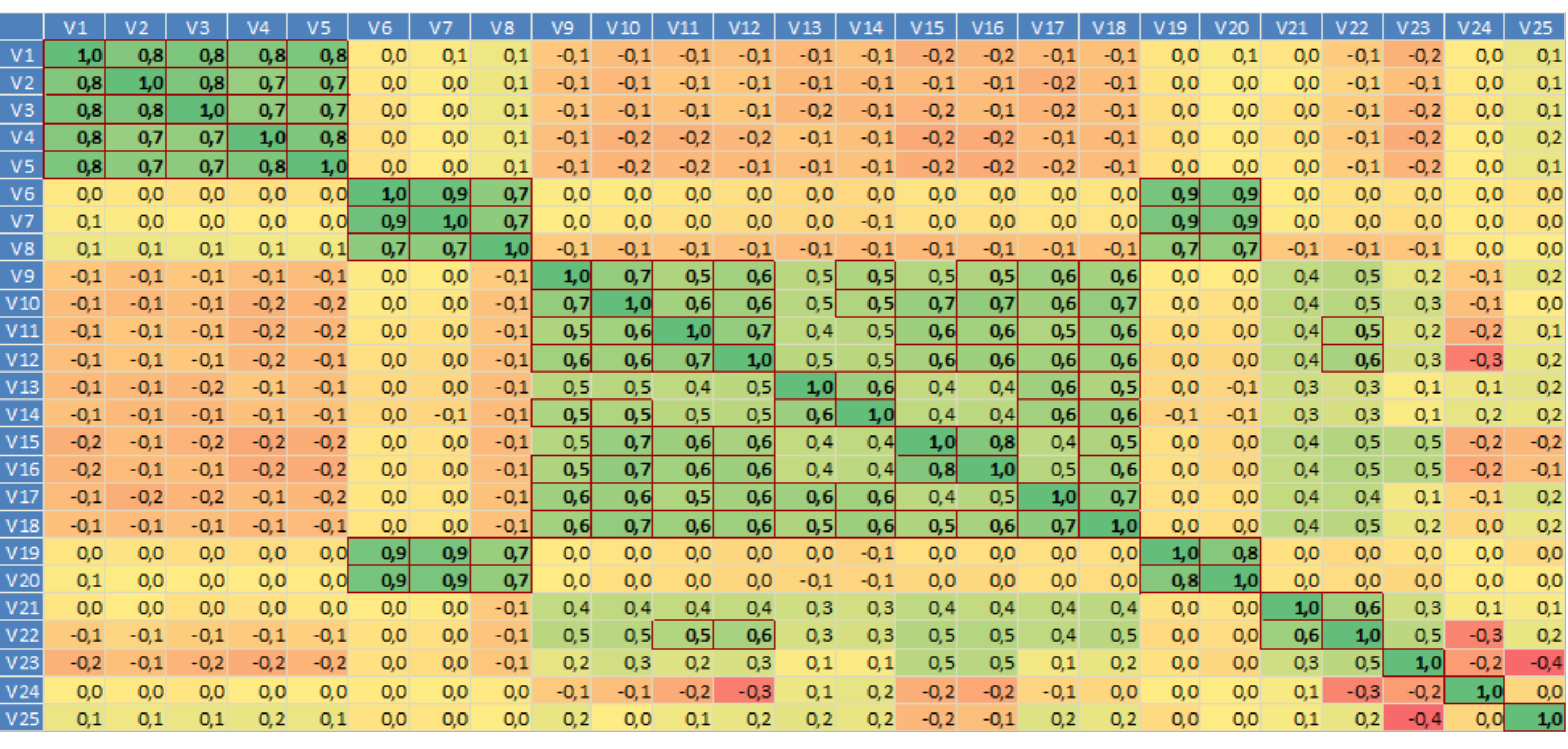}
\label{kendall2}
}
\caption{}
\end{figure}

Here we will present the linear correlations or Pearson (Figures \ref{pearson1} and \ref{pearson2}) and  Kendall correlations (Figures \ref{kendall1} and \ref{kendall2}) for the grid of 15x15. The variables used will be discussed in the next sections.

Analogously to the Pearson correlations, the next figures shows the Kendall correlations

We remark that after analyzing correlations, groups of correlated variables were observed, but given the data mining method used (random forest) was not imperative to eliminate or grouping variables and we decided to include all of them in modeling.

\section{Long-term static model: index for slow movements}
\
\par

In the next section we will see the construction of the long-term static model that will allow us to build risk maps.

\smallskip

\subsection{Choice of the model}
\
\par

Predictive long-term static model is based on a particular event. And the goal is to calculate the risk zone of likely future movements from the available data. Two ways to address the problem of developing a long-term model using \textit{data mining} techniques, such as:

\begin{itemize}
\item Spatial Autocorrelation Techniques
\item Cluster and Outlier Analysis
\item Variances based techniques
\item Regressions
\end{itemize}

The first approach is for the use of space data mining, whose main advantage is that the spatial correlation is contained in the method. Its disadvantage is the poor ability of temporal correlation.

The second approach is the use of classical data mining that has a lot of methods and techniques and easy to use database. However, it has the disadvantage of creating spatial correlations at the base.

\subsubsection{Methodology}
\
\par

As already mentioned in the construction of the analytical database, the following adjustments were made:
\begin{itemize}
\item In a first step, a grid was used with cells of 50x50 meters.
\item Radial neural networks where used to fill in missing data.
\item Inclusion of \textit{relevant neighbors}.
\item Choosing a time window of one year of data.
\end{itemize}

The variables used in the long-term model are listed in Table \ref {tab_var}, few neighbors are considered, the number of months considered for historical data collection until distance influences the RBF interpolation.

\begin{table}[htp]
\caption{Variables to consider in the model}
\begin{center}
\resizebox{0.7\textwidth}{!} {
\begin{tabular}{|l|c|c|c|c|}
\hline
Sensor& Point& Neighbors & History & Interpolation RBF\\
\hline
\hline
INSAR & Yes & 4 & -9 months & 75 m \\
\hline
DEM  & Yes & 4 & -12, -9, -6, -3 months & -- \\
\hline
Geology & Yes & -- & -- & -- \\
\hline
Prisms & Yes & -- & -12, -9, -6, -3 months & 50 m \\
\hline
Piezometers & Yes & -- & -- & 50 m \\
\hline
\end{tabular}}
\end{center}
\label{tab_var}
\end{table}

\subsubsection{Results}
\
\par

\begin{figure}[h]
\centering
\subfloat[Summary]{\includegraphics[width=0.5\textwidth,height=4cm]{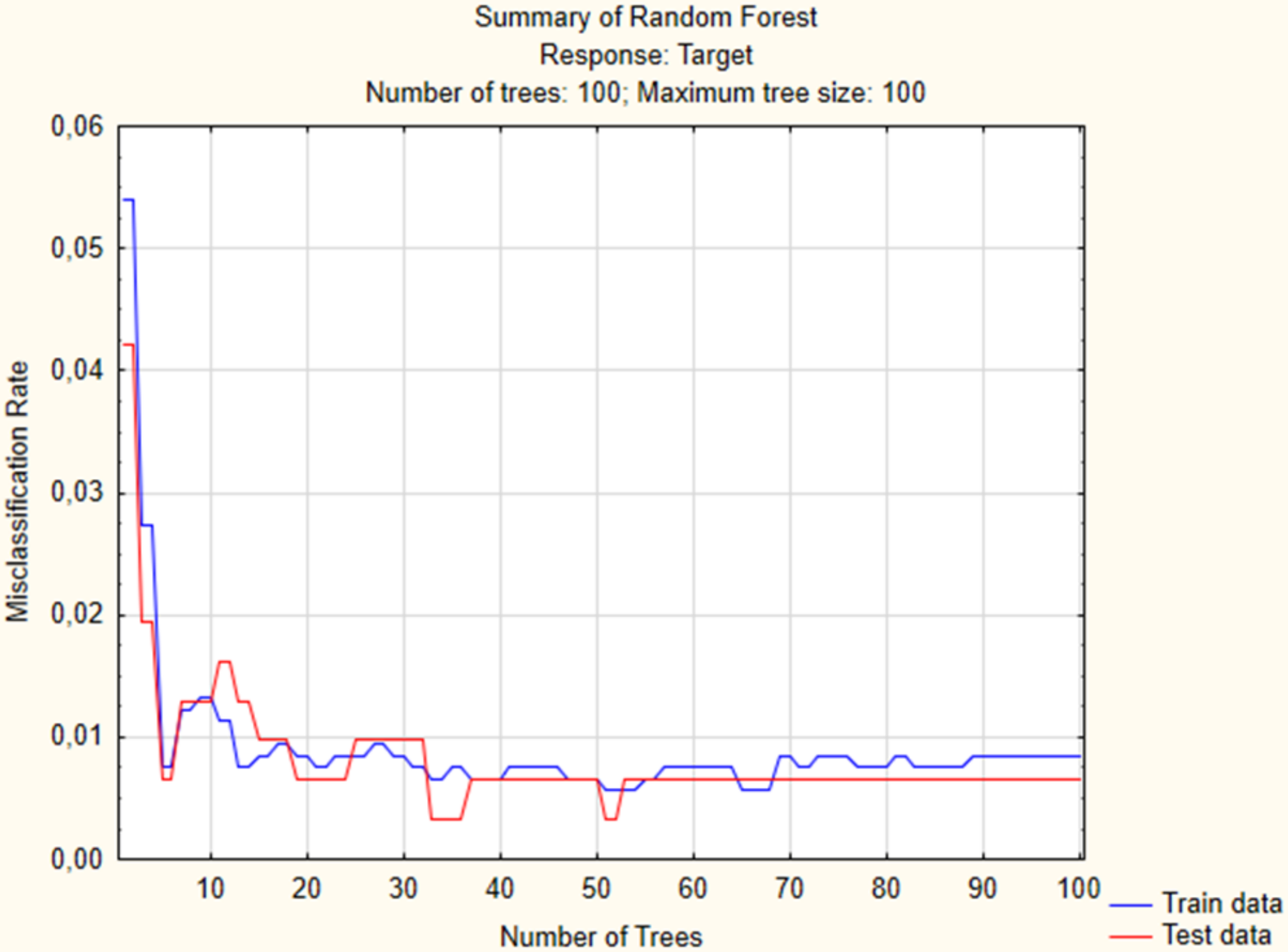}} \vspace{0.01 \linewidth}
\subfloat[Treel 2]{\includegraphics[width=0.35\textwidth]{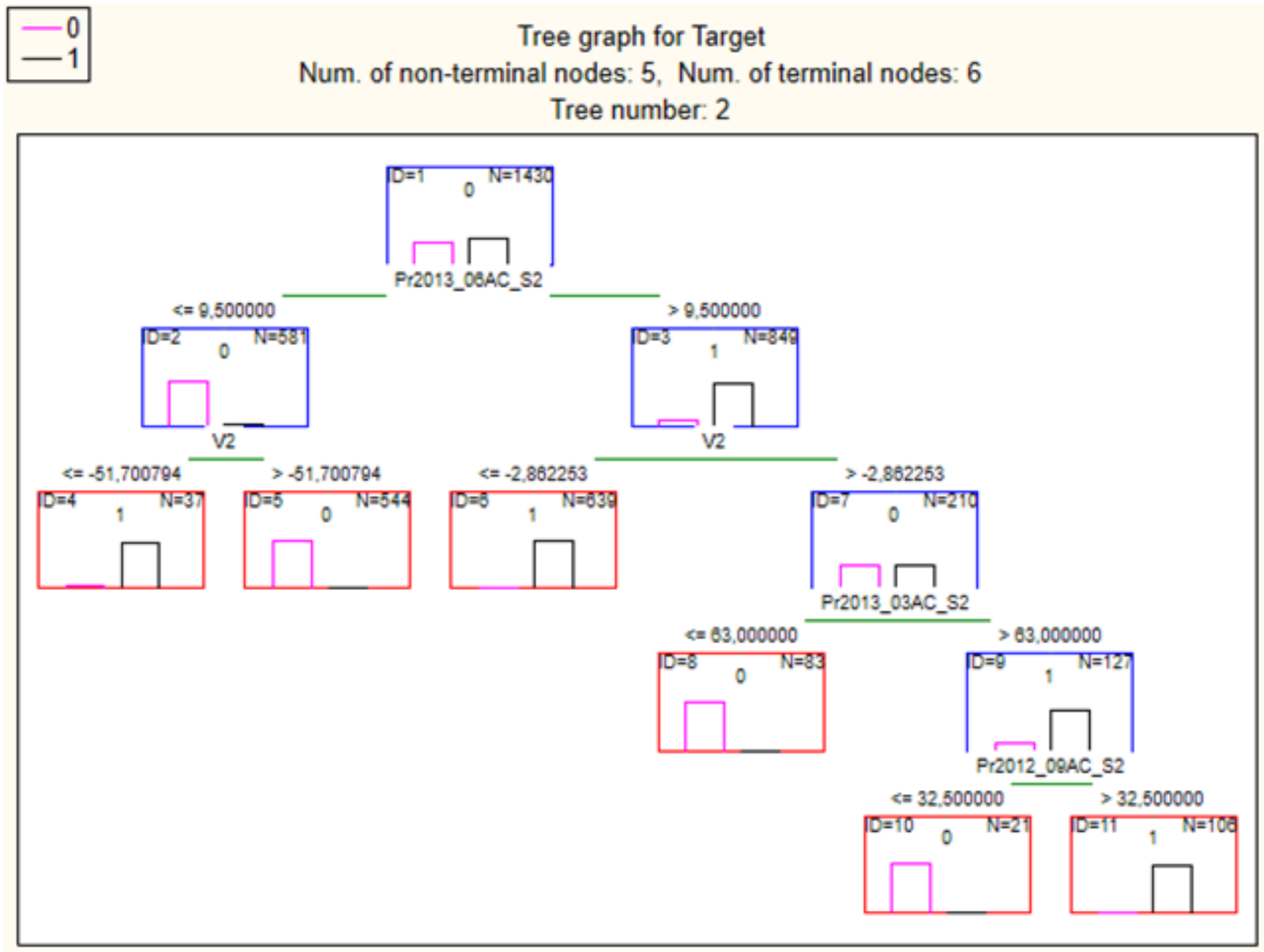}}
\subfloat[Tree 100]{\includegraphics[width=0.35\textwidth]{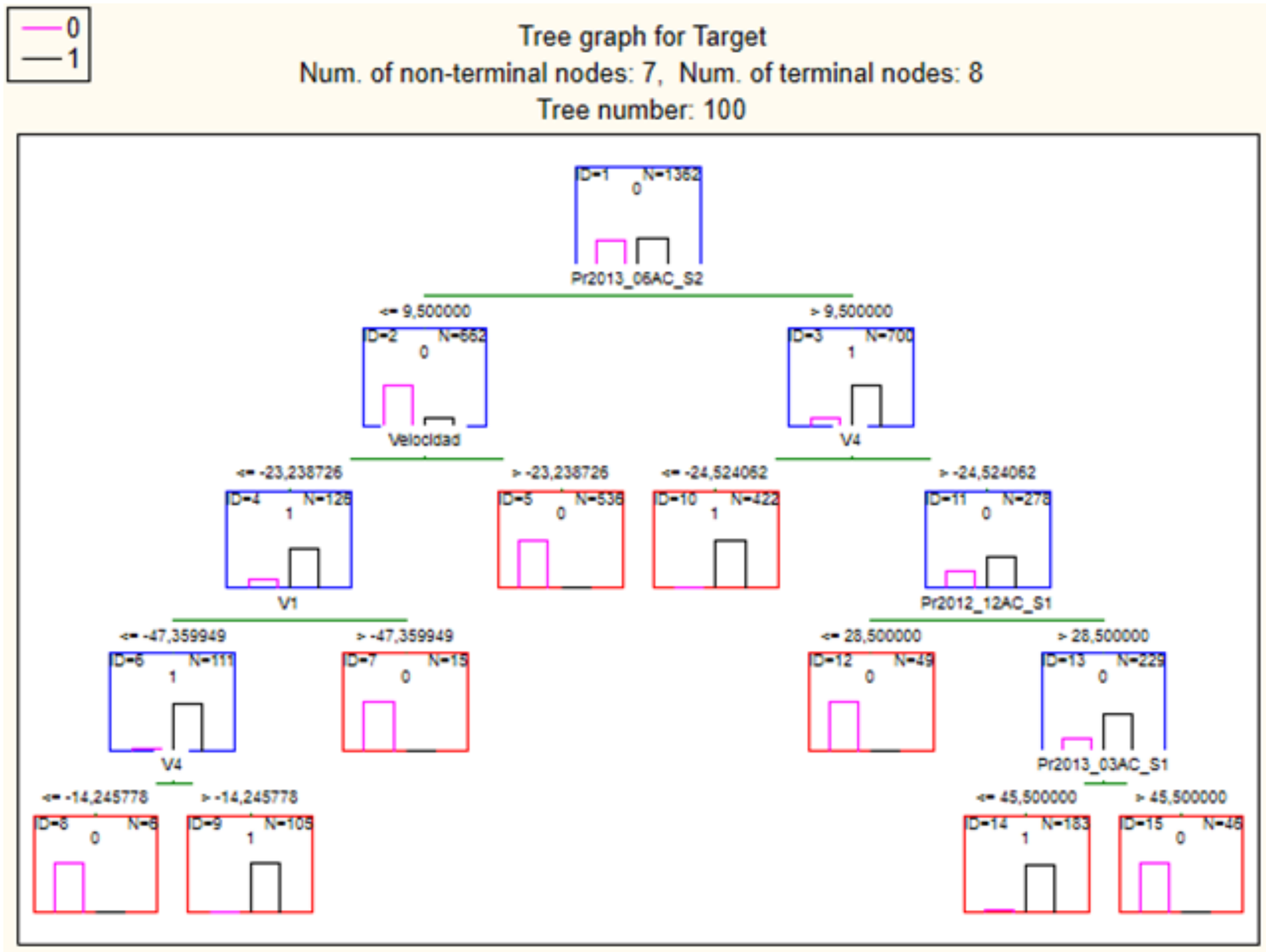}}
\caption{Result of the  (6 months). 
Summary of the 100 trees and graphs of trees  2 and 100
} 
\label{ran_for}
\end{figure}

After testing the previous six models, the best result was obtained using \textit{Random Forest}. 100 decision trees were created. Figure \ref{ran_for} show the results for 100 trees and the graphs of trees 2 and 100.

\begin{figure}[h]
  \subfloat[Results with 3 months of data]{\includegraphics[width=0.45\textwidth]{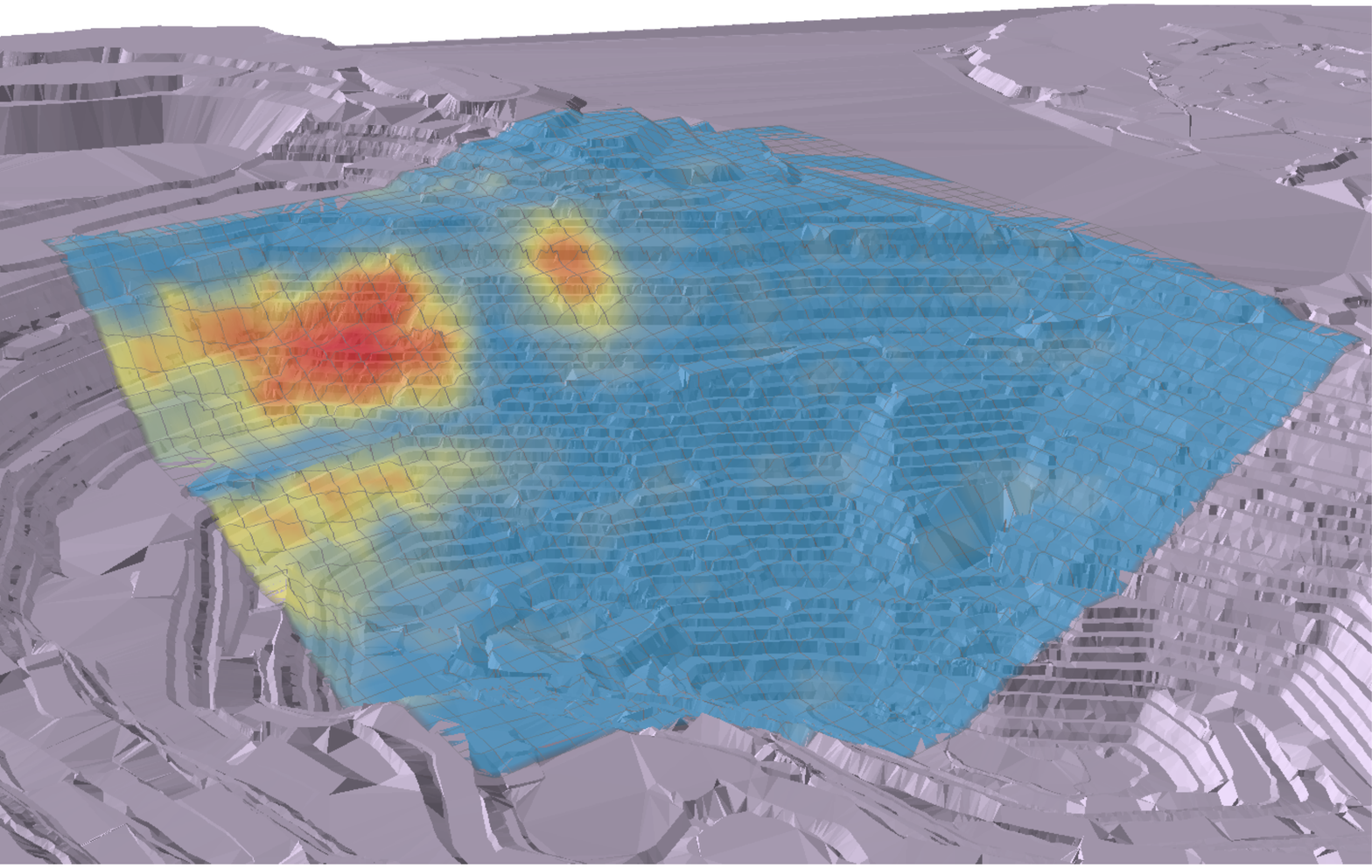}
 \label{Resultado3}}
 \subfloat[Results with 6 months of data]{
  \includegraphics[width=0.45\textwidth]{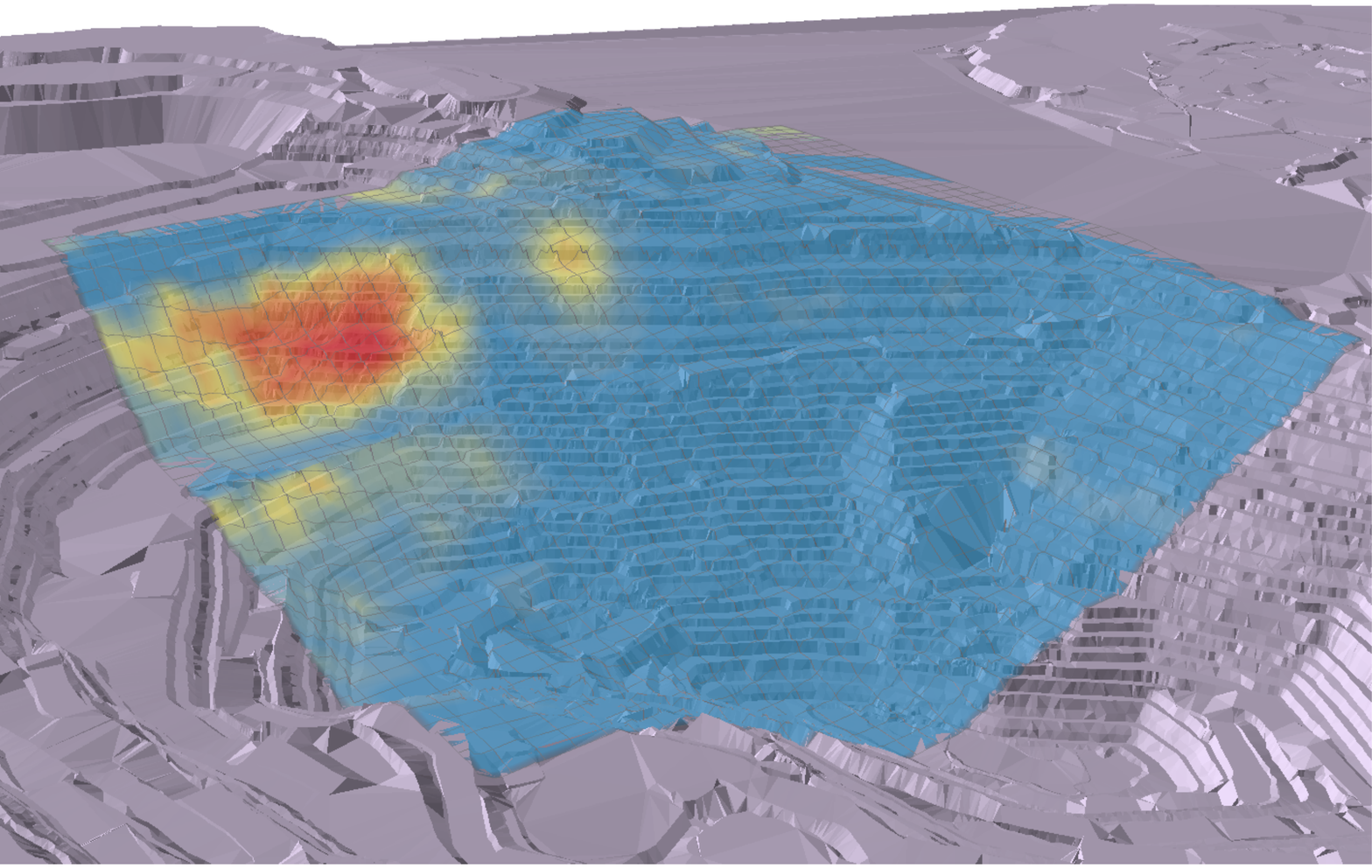}}
 \label{Resultado4}
 \caption{Long term results}
\end{figure}

In Figures \ref{Resultado3} and \ref{Resultado4} model predictions are shown. The model was run with data of 3 and 6 months before the event respectively, that is, with a capacity of anticipation of 3 and 6 months. The main used variables are 
INSAR data, DEMs, which are directly related to the mining operation and Geology.

The results of the Random Forest  showed that the relevant attributes (variables) correspond to:
\begin{itemize}
\item Speeds of lower blocks
\item Production accumulated in lower blocks
\item Mineralogy and lithology
\end{itemize}

\subsubsection{Conclusions}
\
\par

With this first static model developed to predict future movements, it can be concluded that:
\begin{itemize}
\item It is possible to build a tool for generating risk maps for movements of the pit.
\item The developed model can anticipate 6 months at the beginning of the movement.
\item The most important variables in the model correspond to the speed of neighboring (gradient), INSAR average speed, and the mining operation which is represented by the difference of the DEM.
\end{itemize}
	
Models were created at different resolutions, covering 15x15 meters 30x30 meters and 45x45 meters, which are easily associable banks as a measure of area (15m), because the method of filling variables through Gaussian explodes with the amount of training samples, as it involves matrix inversion, the method when running especially the pit is limited to a resolution of 30x30 meters, independent of a higher resolution is used in the model.

\smallskip

\section{Short time indicator: Index for fast movements}
\
\par

The short-term model proposed in this chapter is based on an index based on data from the Ground Based Radar (GBR).
Indices or indicators serve to show or indicate particular situations, as well as to quantify processes or conceptual models. In this regard, the generation of a dimensionless index, facilitates the classification or observation of the phenomenon and decision making. In other words your understanding. The main features or desired properties of such indicators are:
\begin{itemize}
\item dimensionless parameter. It allows scalability and clear definition of thresholds. Allows easy understanding and easy viewing, alarm generation or classifications.
\item Values Dimension. A clear characterization of the range of this index.
\item Robustness. Low sensitivity to noise in the data.
\end{itemize}

In this sense, the study has focused on the behavior of GBR data from the sensor, based on such indicators, which allow a marked reduction in the complexity of the phenomenon to study.

\subsection{Simulation of the pit movement}
\
\par

\begin{figure}[h]
 \centering
    \includegraphics[width=0.7\textwidth,height=3.5cm]{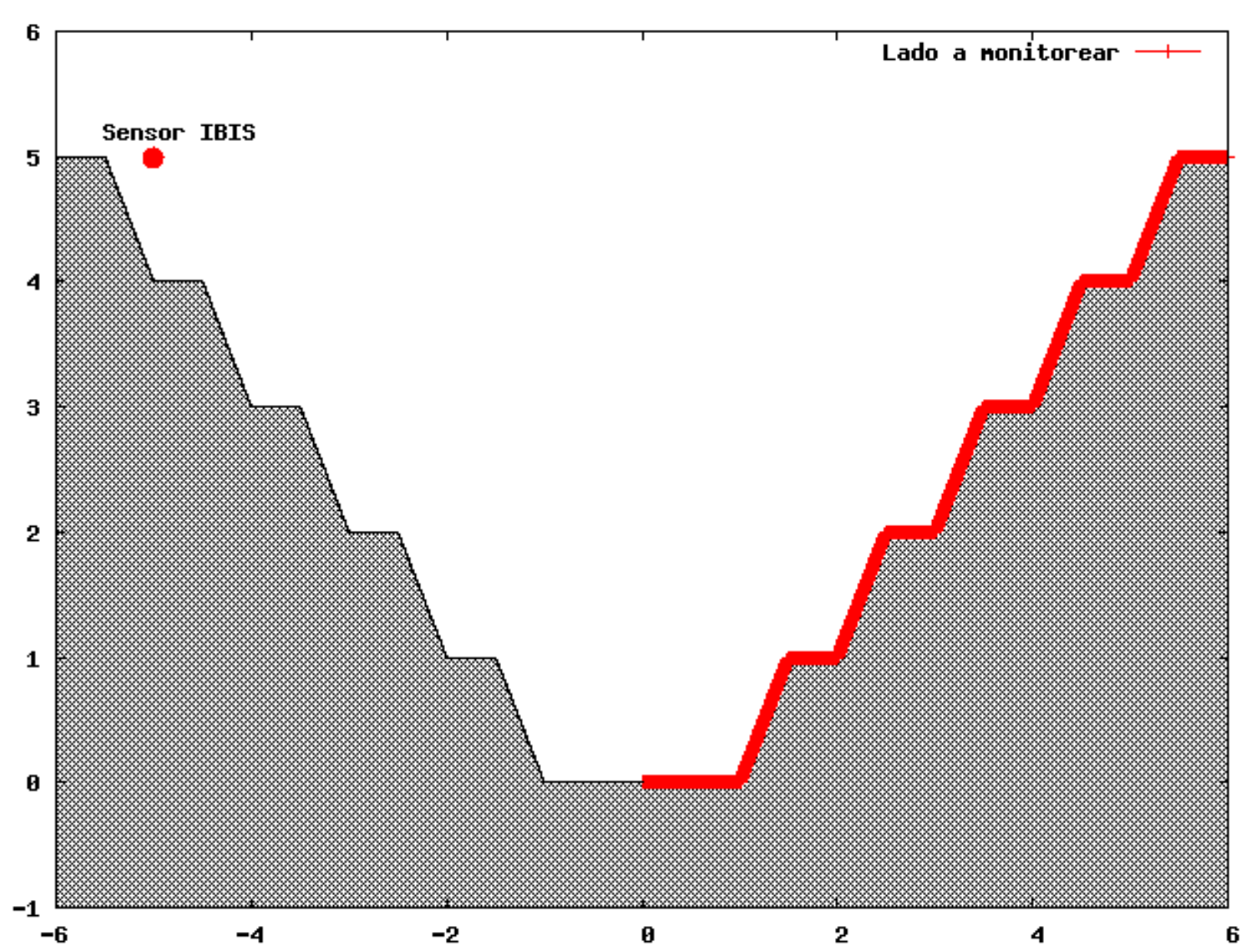}
 \caption{Scheme  of the mine and location of the sensor}
 \label{Esquema_Mina}
\end{figure}

The Ground Based Radar (GBR) is able to record movements or deformations of the pit in the same radar direction measurement. That is, it is possible to measure specific changes that move away or approach it in a large area of the pit, depending on the distance from the sensor to the wall to be measured and the angles of incidence of the measurement.
In a first stage, a study of movement or deformation of the pit was performed. For which a 2D pit model shown in Figure \ref{Esquema_Mina} was created. This diagram shows a section of the pit with its banks, representing the position of the GBR  and highlighted in red color, the surface being monitored. The objective is to understand the dynamics of movement and generating recorded by the sensor, especially the positive values. The values represented in both the $ $ X and $ Y $ axis correspond to artificial units of length to facilitate calculations and understanding of the phenomenon. The zero on the axis $ X $ marks the midpoint of the base of the pit.

\begin{figure}[h]
 \centering
 \subfloat[Simulated movements corresponding to vectors  $\vec{d_1}$  and $\vec{d_2}$]{   \includegraphics[width=0.5\textwidth]{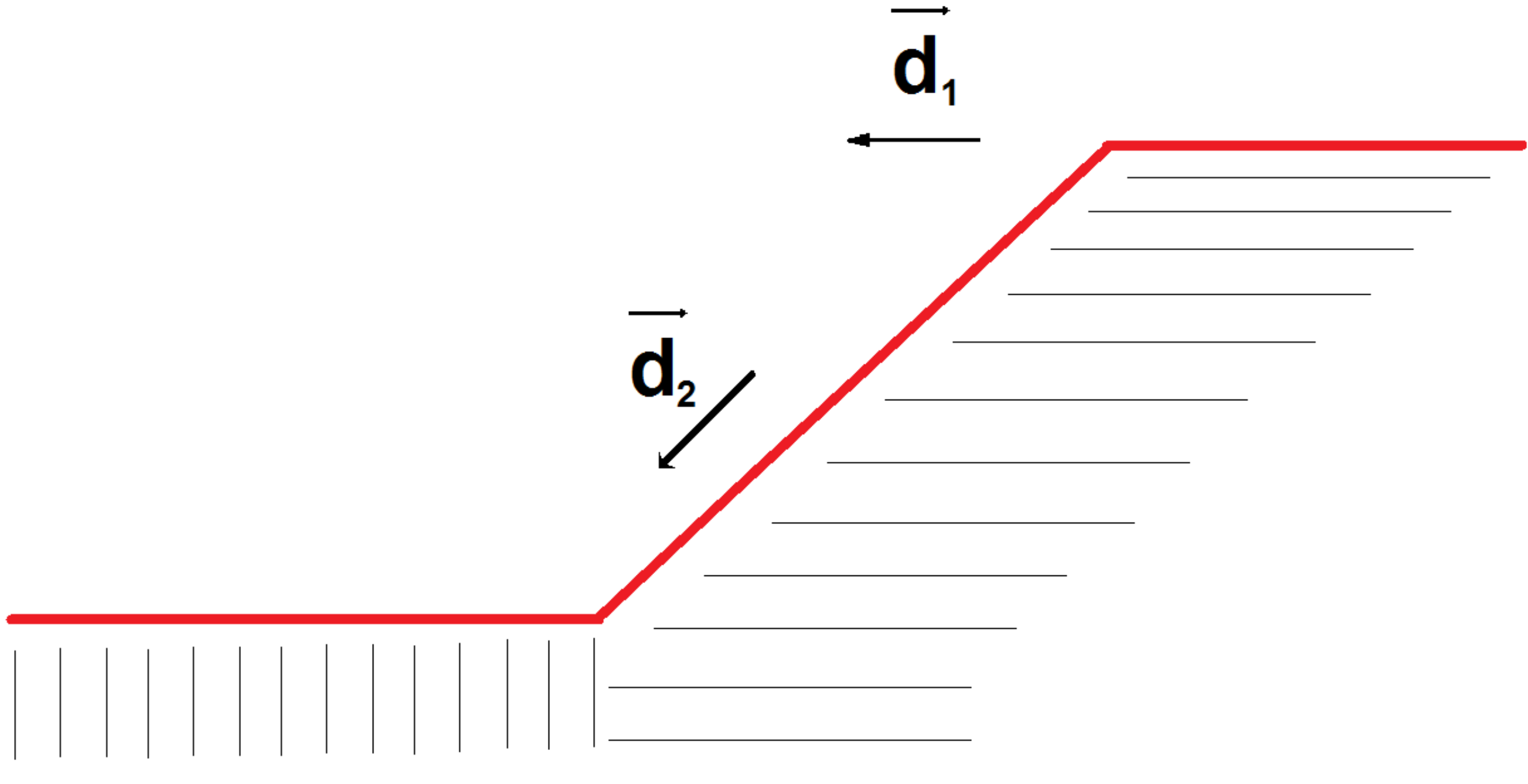}
 \label{vectores}}
 \subfloat[Considered displacements]{\includegraphics[width=0.5\textwidth,height=4.0cm]{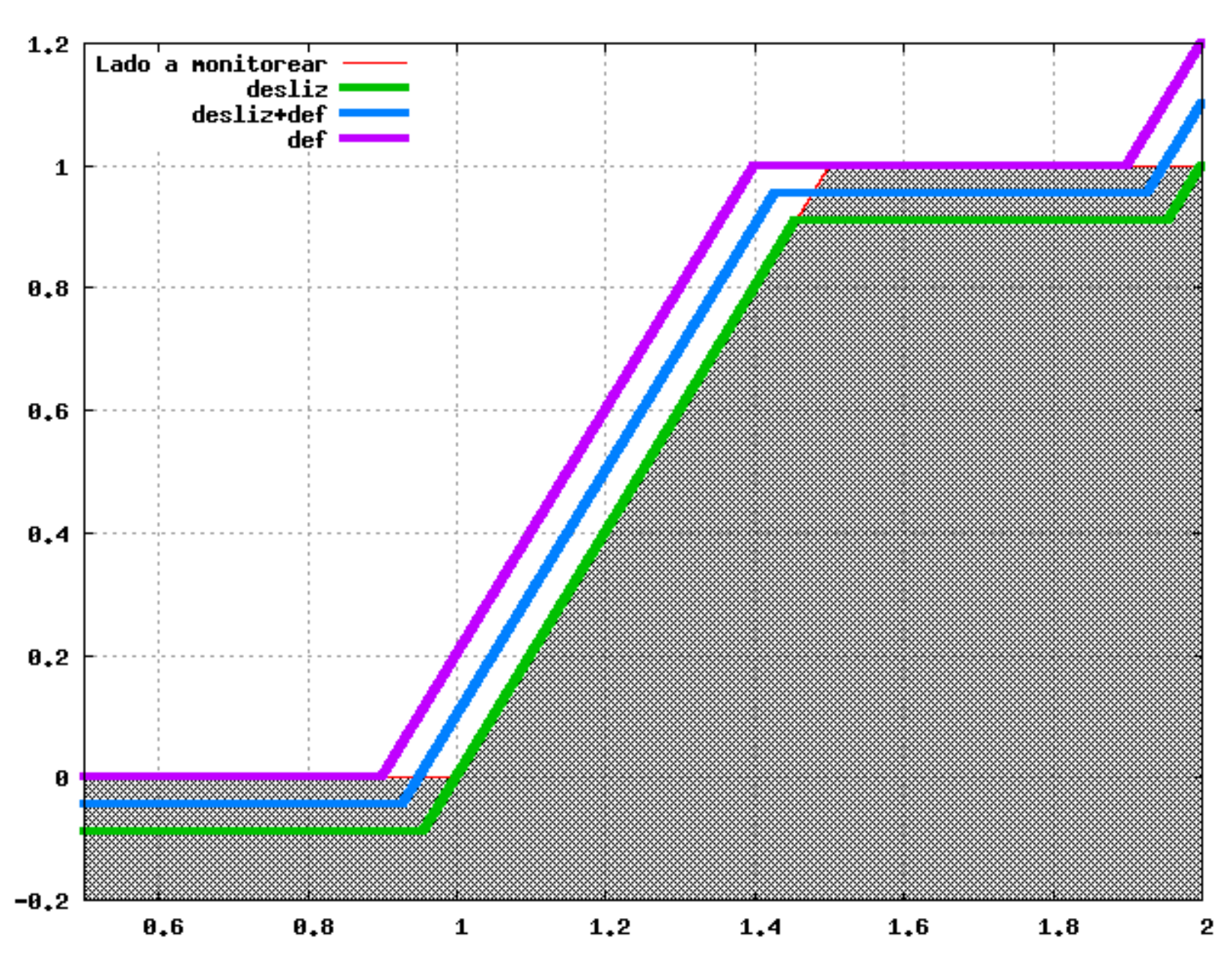}
 \label{movimientos1}}
 \caption{}
\end{figure}

The movement of the  wall motion and the measurements of the GBR sensor were simulates. As shown in Figure \ref{movimientos1}, three different movements were considered:
\begin{itemize}
\item Slip: Movement where the entire surface moves in the direction of the slope.
\item Deformation: Movement of the pit where the surface moves in the direction of the sensor.
\item Slip + Deformation. In this case the two movements occur, i.e., the pit  moves down towards the sensor.
\end{itemize}

The simulated deformations corresponds to an instantaneous change of the wall and records of the displacements are made on the profile of the wall. Figure \ref{movimientos1} represents the simulated movements  in a portion of the wall. The graph represents on the axis $ X $ the distance from the sensor to the measured point. The axis $ Y $ gives the magnitude of displacement recorded by the sensor. The movements performed have two components, one direction horizontal to the sensor, represented by the vector: $ \vec{d_1} = \binom{-1}{0} $, and another in the direction of the slope of the pit  $\vec{d_2} = \binom{-0.5}{- 1}.$ Figure \ref{vectores} represents these two movements.

Thus, we have the following:
\begin{itemize}
\item Deformation: It is a movement where the wall moves $ 0.1 $  units only in the horizontal direction, i.e. towards $ \vec{d_1}$.
\begin{equation}
D_{deformation} = 0.1*\frac{ \vec{d_1}}{|| \vec{d_1}||}
\end{equation}

\item Slip:  It is a movement where the entire surface moves $ 0.1 $ units in the direction of the slope, i.e. towards  $\vec{d_2}$.
\begin{equation}
D_{slip} = 0.1*\frac{ \vec{d_2}}{|| \vec{d_2}||}
\end{equation}

\item Slip + Deformation. In this case the two movements occur, i.e., the pit moves $ 0.05 $ units  in the horizontal direction and $ 0.05 $ units  in the direction of the slope at a time.
\begin{equation}
D_{slip + deform} =   0.05*\frac{ \vec{d_1}}{|| \vec{d_1}||} + 0.05*\frac{ \vec{d_2}}{|| \vec{d_2}||}
\end{equation}
\end {itemize}

\begin{figure}[h]
 \centering
    \includegraphics[width=0.5\textwidth, height=4.0cm]{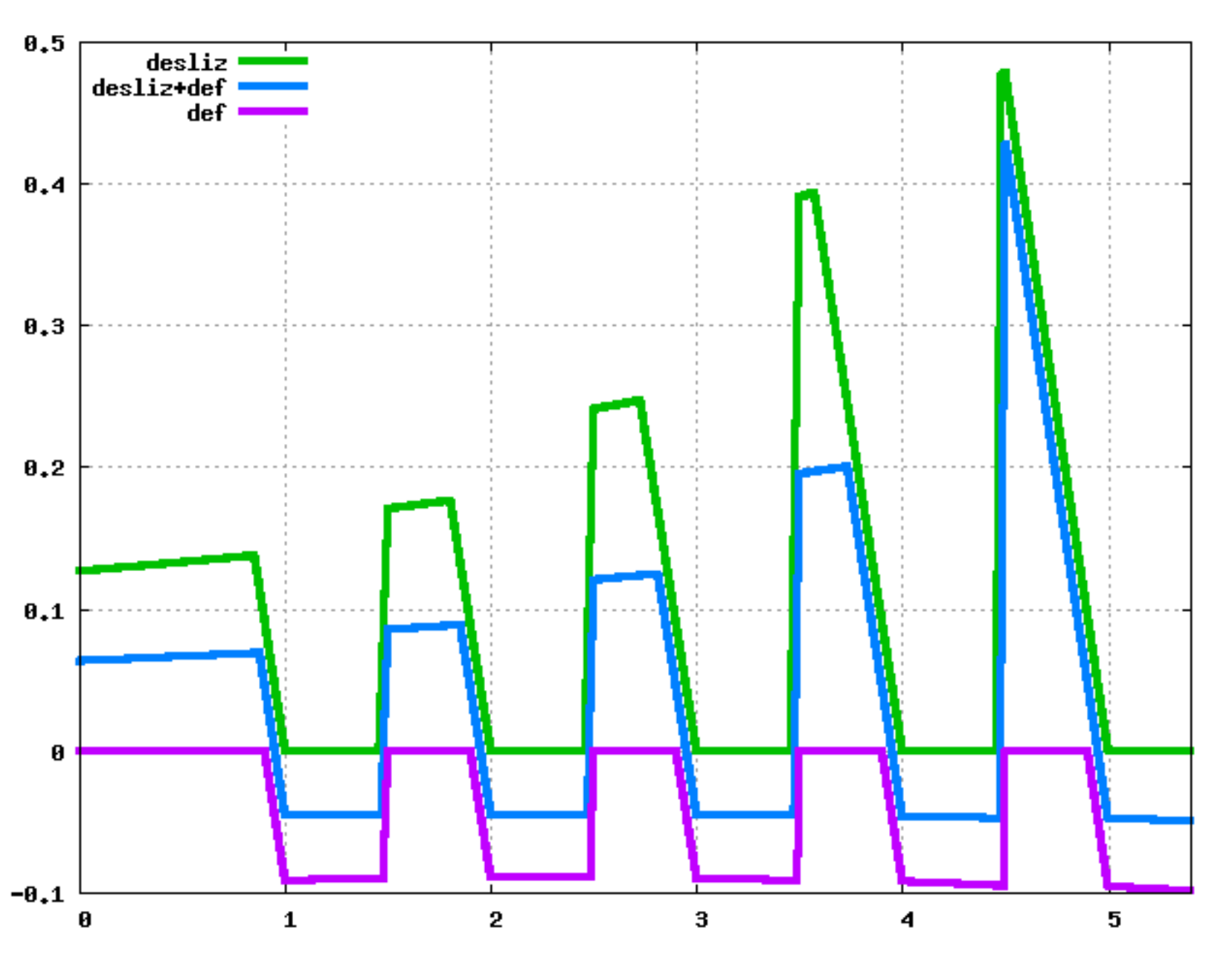}
 \caption{Datos obtenidos por el sensor IBIS, correspondientes a los 3 desplazamientos }
 \label{movimientos2}
\end{figure}

From these two movements the GRB data can be generated. This is represented in Figure \ref{movimientos2}, where each color corresponds to each simulated movement across the wall.
We remark that the $Deformation$ generates only negative and zero distances. On the other hand,
the movement $ Slip $ generates only positive measurements, an the combined movement  $Slip +   Deformation$  causes measurements alternating positive and negative values.
Looking at the  results obtained with the simulation model, together with  the observations and analysis of the real data recorded by the sensor, one can conclude that the predominant movement recorded by the GRB  is the $deformation$. That is, the predominant movement of the wall is in the direction of the sensor, and the $slip$ movement does not clearly seen in the real data.
We can also conclude that to take into account all data at full spatial and temporal resolution ensures that you can see very localized phenomena, it would be impossible to capture without having all the information generated. Moreover, the great difficulty of handling a large amount of data, determines that for efficient data management is required to reduce the dimensionality of the problem, reducing the amount of data to look through sampling or, using parameters that may characterize the amount of reduced data.

\subsection{Movement Index}
\
\par

According to the preliminary analysis of the data, we had built a movement indicator for the pit wall, which could anticipate a movement thereof.
In what follows, given a time instant $t>0$, the GBR sensor records the movements of a particular area, we denote this scene deformation by $ x_t $, which is a data vector. Thus the data recorded can be see as a time series of an Euclidian space $\mathbb{R}^n$, that is $t\to x_t\in \mathbb{R}^n,$ where $n$ is the dimension of the space or the number of recorded points.

In order to evaluate the  behavior of $ x_t $,  it is necessary to establish a comparison with the data prior to $ t $, thus  we consider for an arbitrary time $ t $ fixed,  and constants  $ \delta, \tau >0 $, 
$$ \overline{x}_{t, \delta, \tau} = \frac{1}{\tau} \sum\limits_{h\in [t- \tau- \delta, t- \delta]} x_ {h}, $$
 the average movement of data in the interval $[t -\tau -\delta, t-\delta ]$.
 
We want to represent the variation of $ x_t $ with respect the average  $ \overline {x} _ {t, \delta, \tau} $, thus since  $ x_t $ and $ \overline {x} _ {t, \delta, \tau} $ are vectors in $ \mathbb {R}^n$, we consider the cosine of the angle between these two vectors as an indicator of change, therefore we define the
\emph{Index Movement"} as:
\begin{equation} 
\label{InMov}
Im(t,\delta,\tau) \ = \ \displaystyle\frac{<x_t,\overline{x}_{t,\delta,\tau}>}{\|x_t\|\|\overline{x}_{t,\delta,\tau}\|},
\end{equation}
We note that this index  compares the directions  and not the magnitudes of the vectors. See Figure.

\begin{figure}[h]
 \centering
    \includegraphics[width=0.3\textwidth]{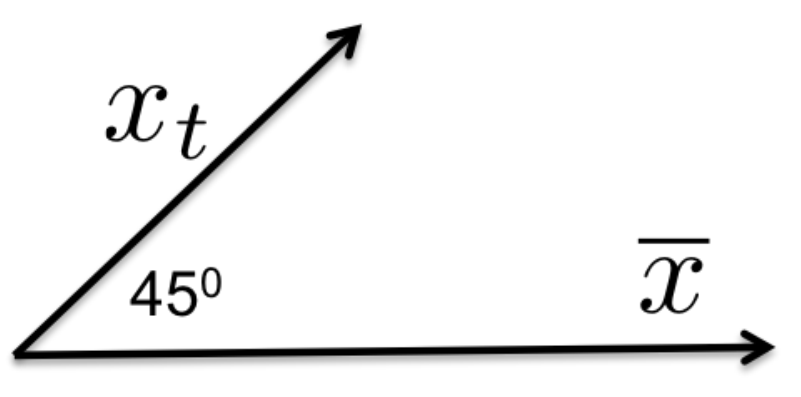}
 \caption{Representation of the Movement Index}
 \label{Esquema_Ejemplo02}
\end{figure}

\begin{figure}[h]
 \centering
    \includegraphics[width=0.5\textwidth]{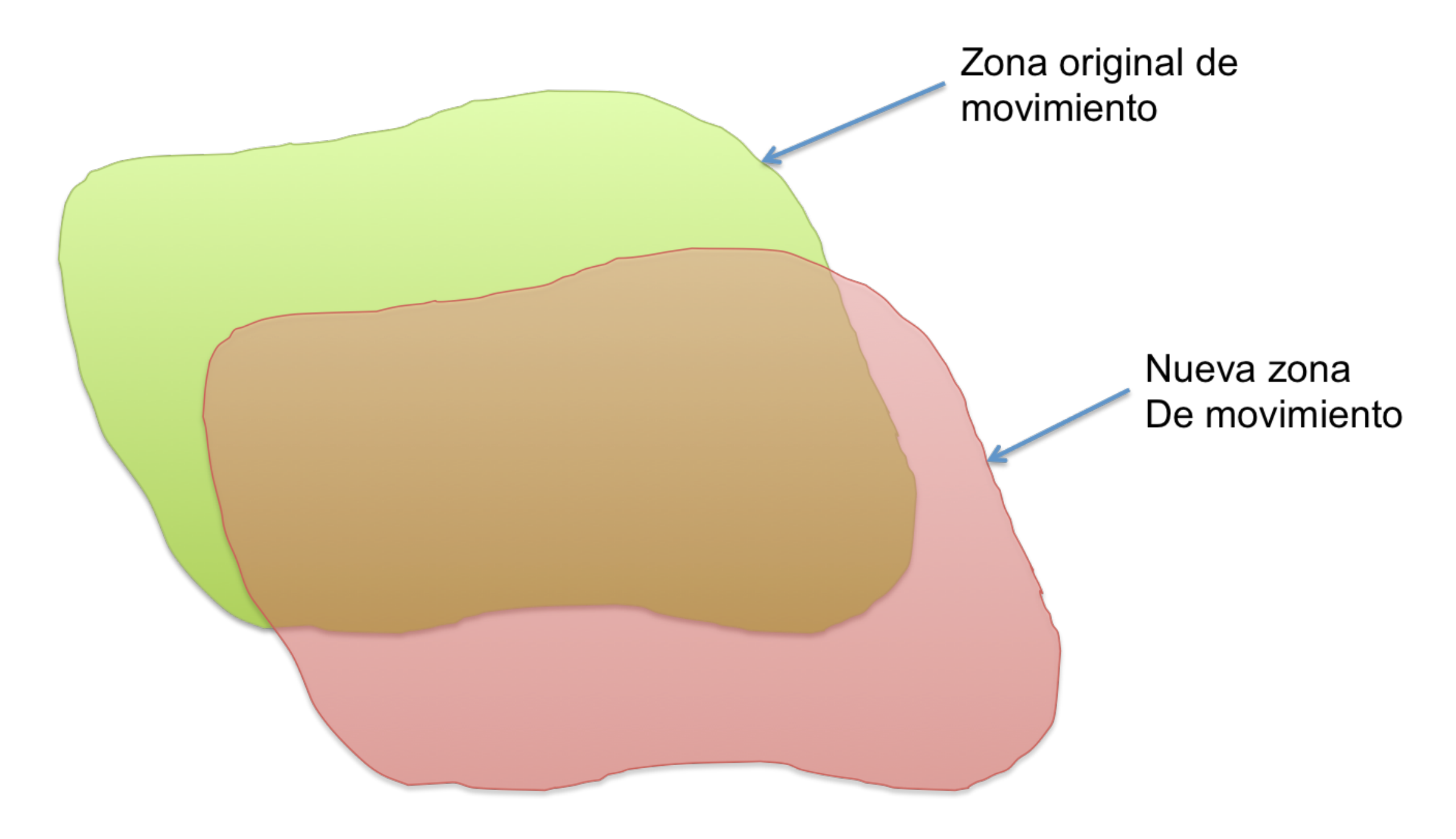}
 \caption{Scheme of the captured movement by the index \eqref{InMov}}
\label{Esquema_IndMov}
\end{figure}

From the above, the Movement Index, can describe changes in areas that are moving, regardless of the magnitude of these movements. For example if we have a case where there is an area that is moving (Figure \ref{Esquema_IndMov} original area of movement) and in a moment of time begins to move another area (Figure \ref{Esquema_IndMov} new movement area ), independent of the value of movement in these areas, the index  \eqref{InMov} will capture it.

\subsection{Sensitivity of the Index}
\
\par

\begin{figure}[htp]
 \centering
    \includegraphics[width=0.5\textwidth]{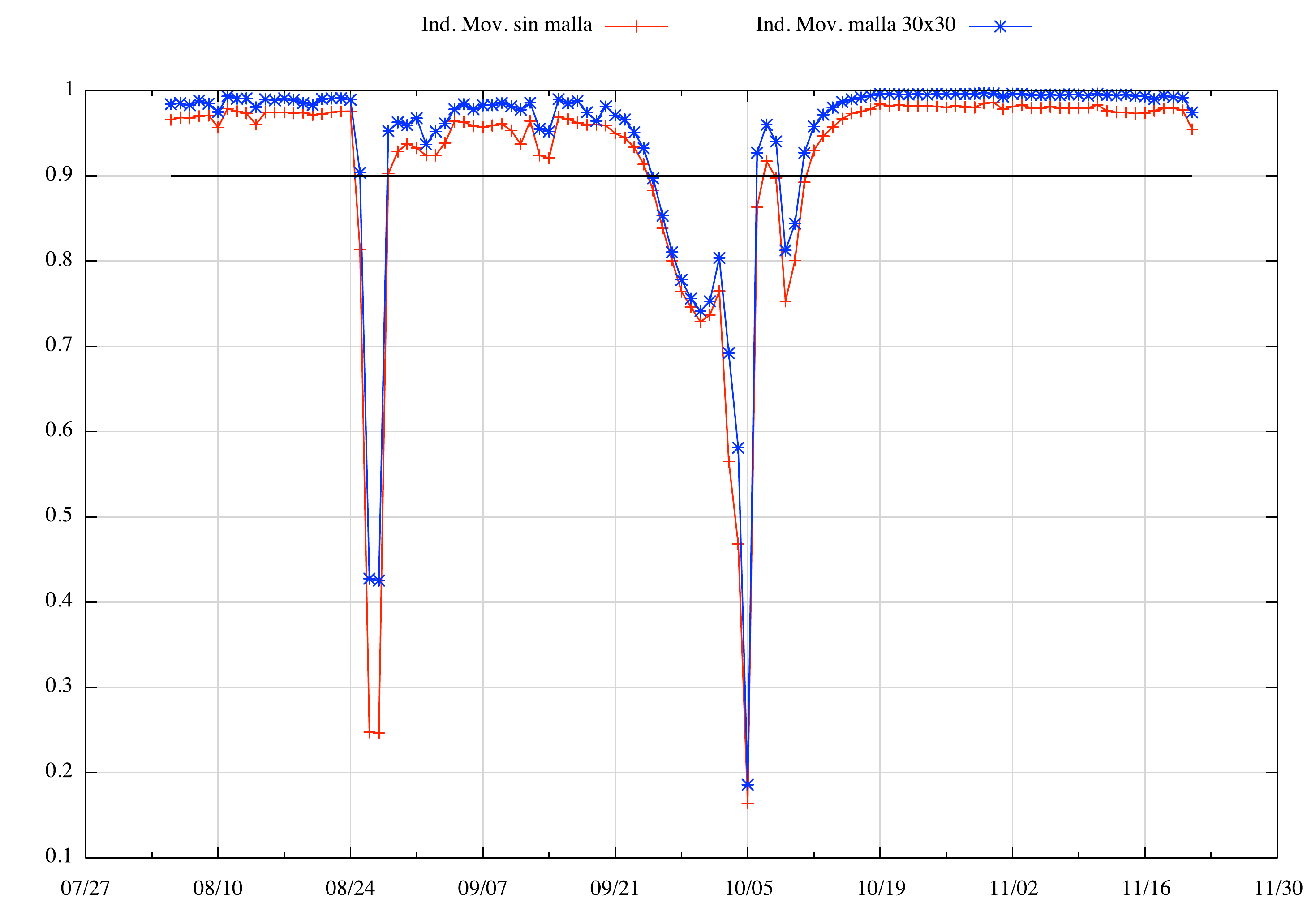}
 \caption{Index on a mesh of $ 30 \times 30 $ and on the original data for the sensor}
\label{comMalla}
\end{figure}

\begin{figure}[htp]
 \centering
\subfloat{ \includegraphics[width=0.45\textwidth]{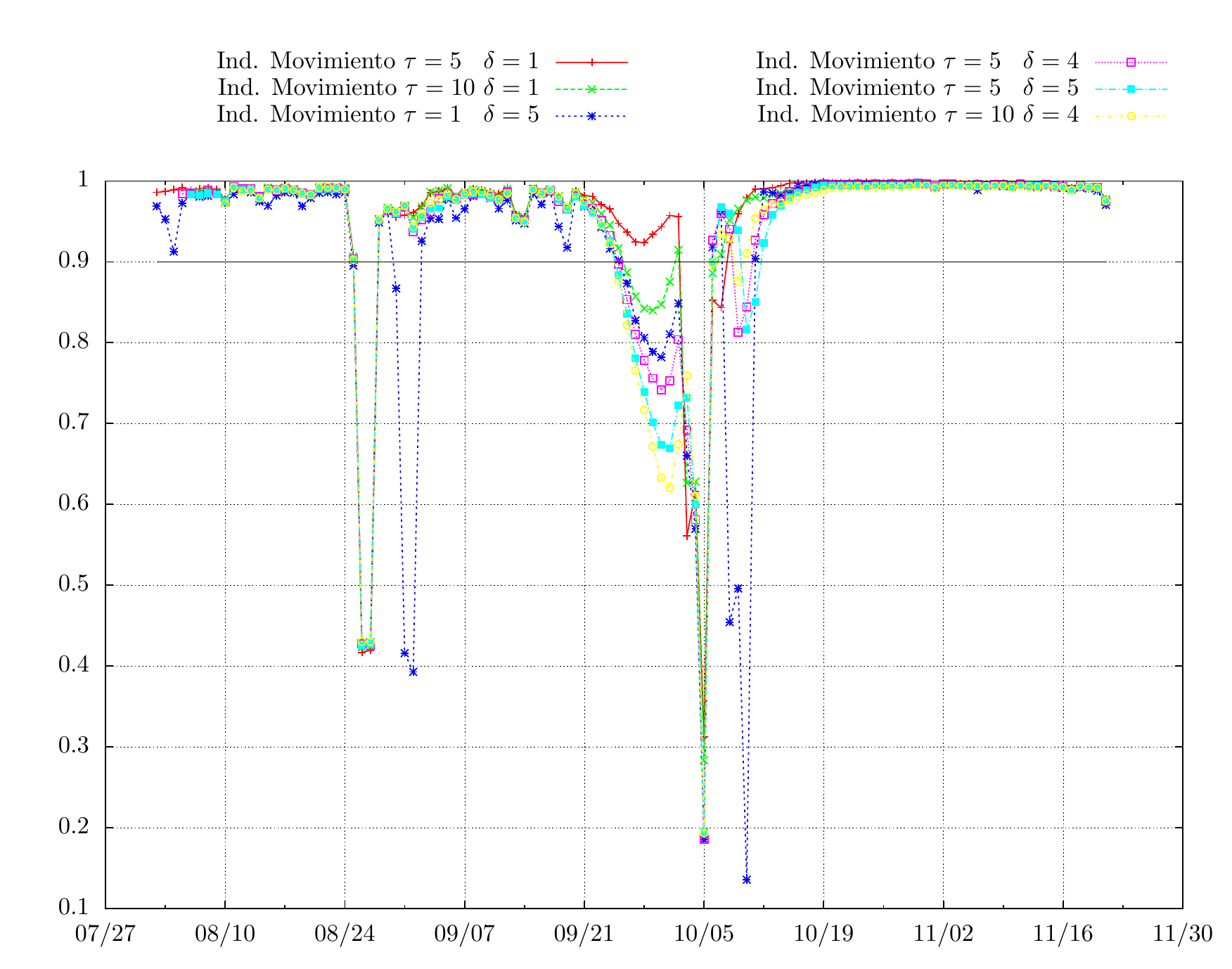}}
\subfloat{ \includegraphics[width=0.45\textwidth]{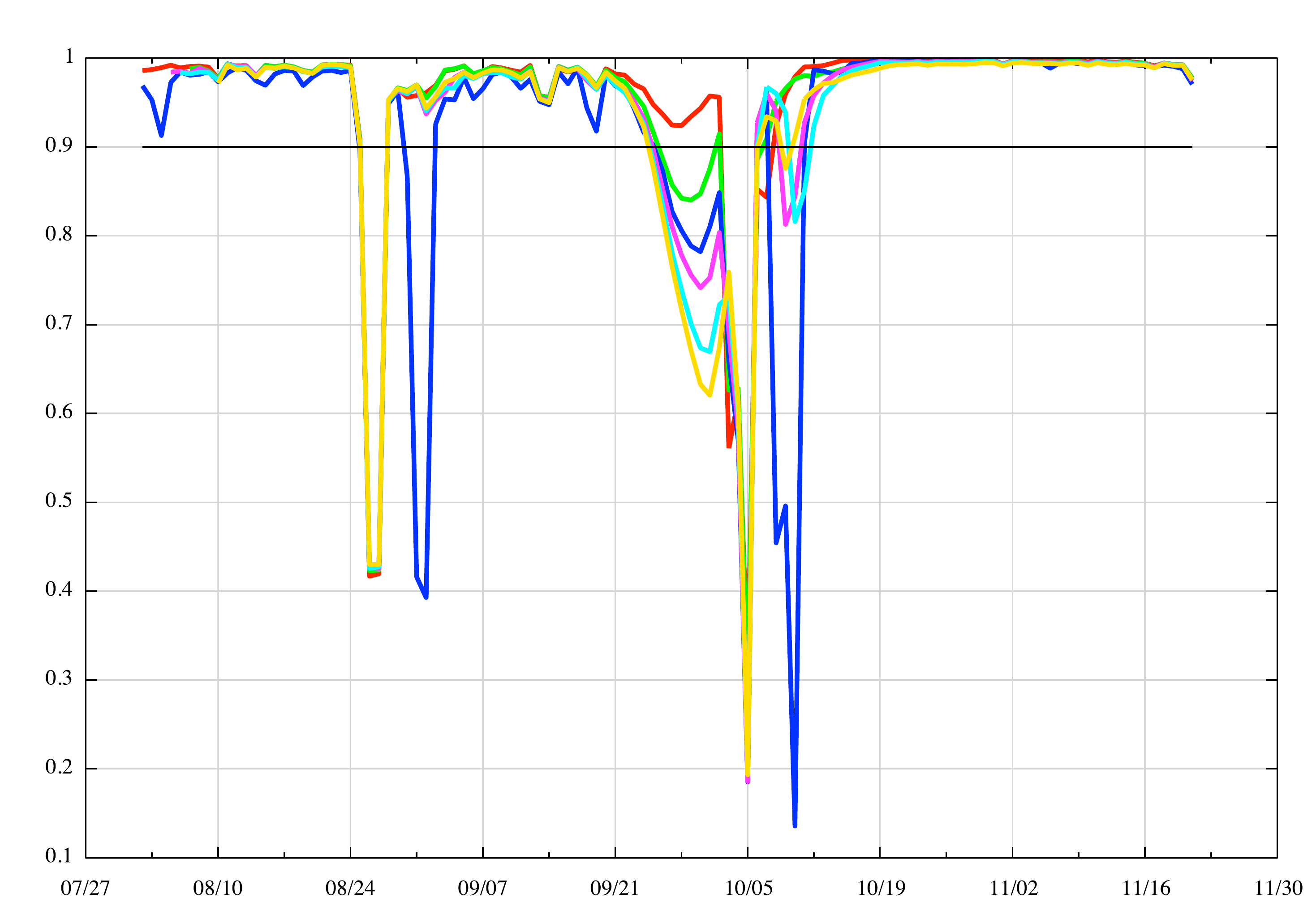}}
 \caption{Sensitivity of the index on the parameters $\delta$ and $\tau$}
\label{SensiIndiceMov}
\end{figure}

Since daily fluctuations in the GBR data reported, it is considered a daily average, thus the daily variation  is eliminated, moreover  we consider a mesh of $ 30 \times 30 $ meters, covering the desired area. 
We have observed that the index values with the original data or with a mesh of $ 30\times 30 $ meters do not change significantly, see Figure \ref {comMalla}.
In  Figure \ref{comMalla}, we consider the values $\delta=4$ and $\tau=5$.
We note that the reduction in information and increase in computation time is enough to consider as an great advantage the use of a mesh of $ 30 \times 30$ meters. We can observe the sensitivity of the index on the parameters $ \ delta $ and $ \ tau $ in Figure \ref{SensiIndiceMov}.
 Moreover, from Figure \ref{SensiIndiceMov} we can conclude that in the case of $\delta$ and $\tau$ have similar values, the index has the best performance, on the other hand, when  one of them is very large relative to the other, the index or it becomes very sensitive to noise or very little perceptive of the changes of the behavior.  Therefore, all future registrations herein we consider $ \delta = 4 $ and $ \tau = 5 $.

\begin{figure}[htp]
\tiny
 \centering
  \subfloat[\tiny Studied Zone]{
 \includegraphics[width=0.25\textwidth,height=0.25\textwidth]{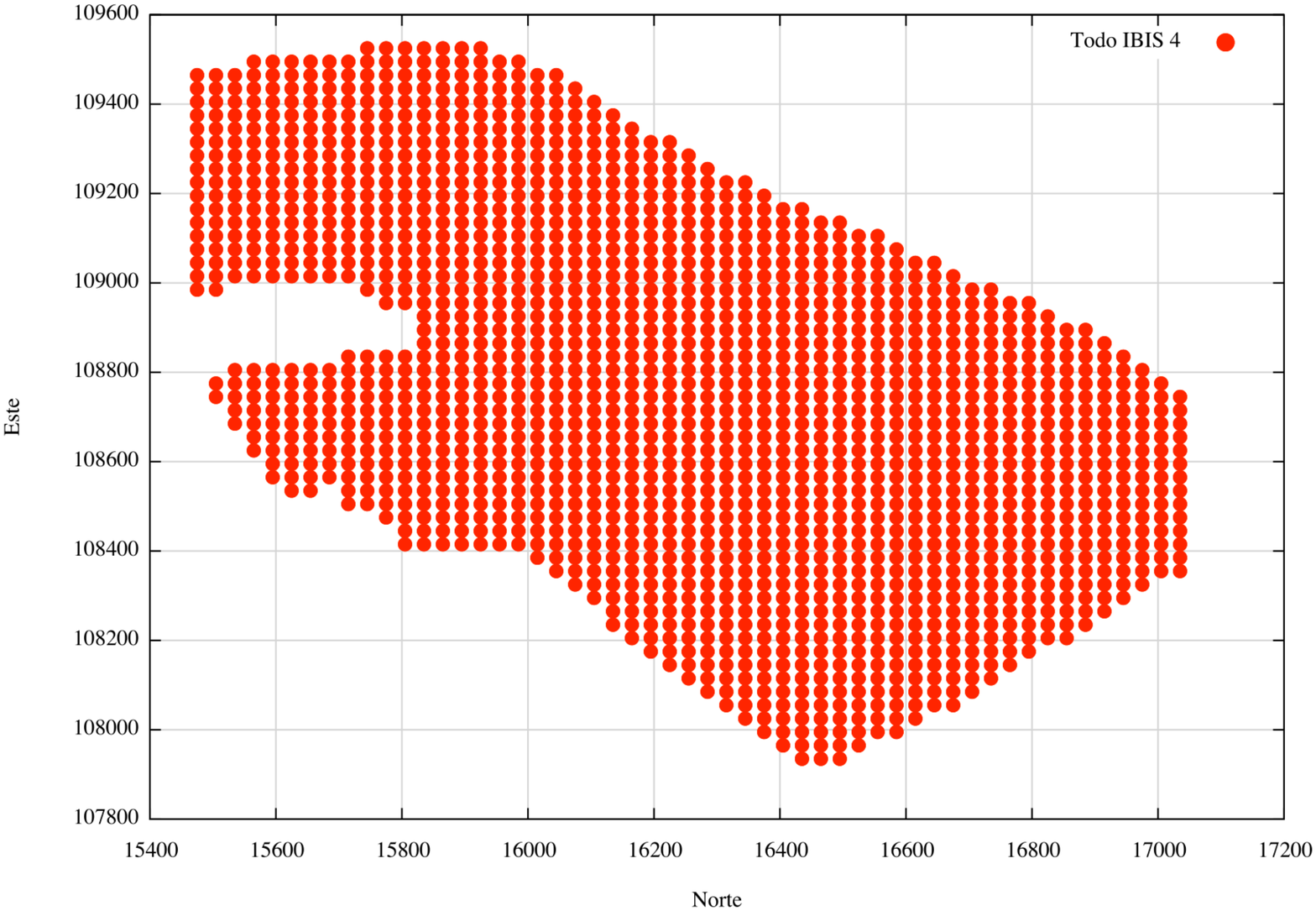}}
  \subfloat[\tiny Zone 1]{
    \includegraphics[width=0.25\textwidth,height=0.25\textwidth]{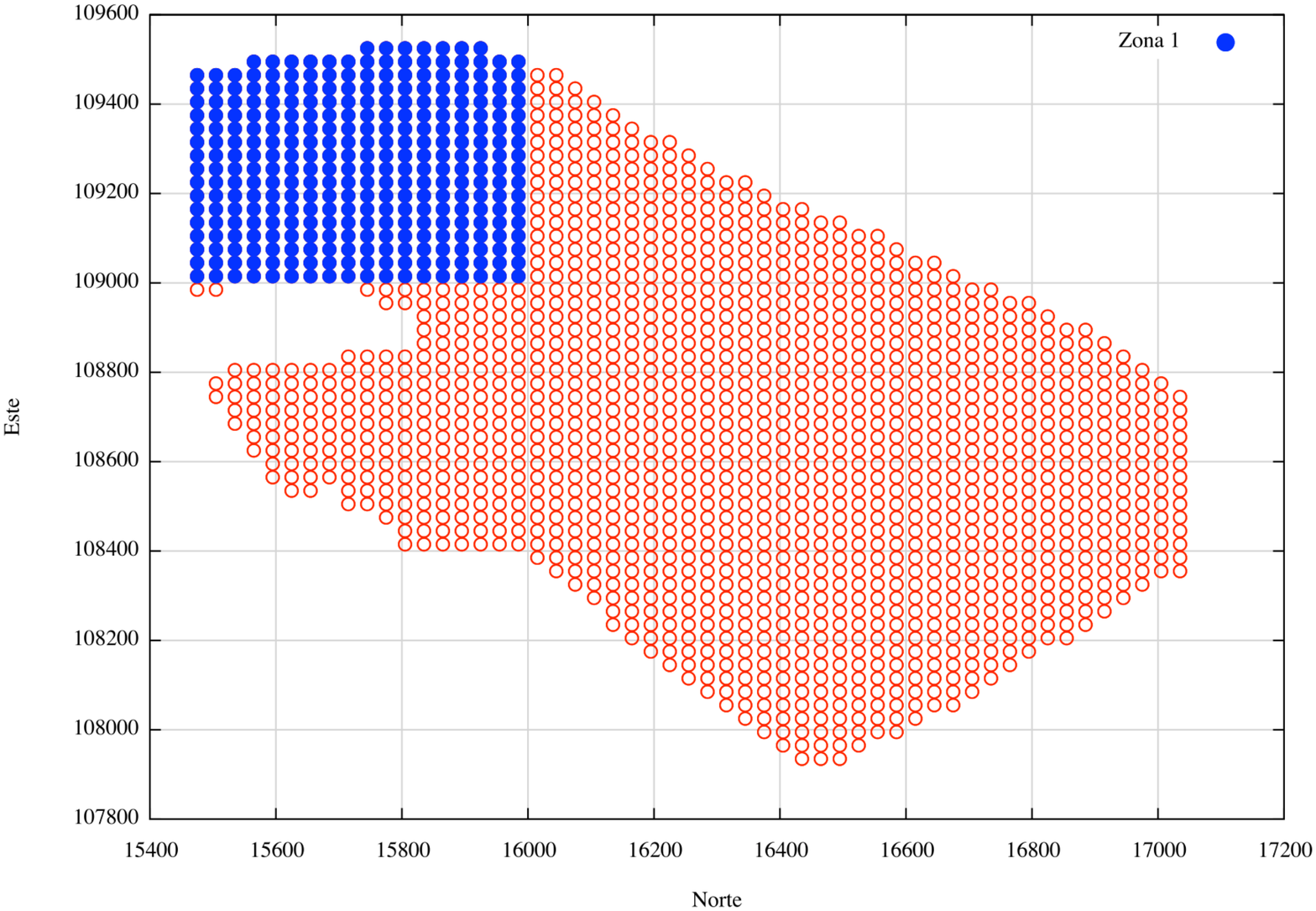}}
  \subfloat[\tiny Zone 2]{
    \includegraphics[width=0.25\textwidth,height=0.25\textwidth]{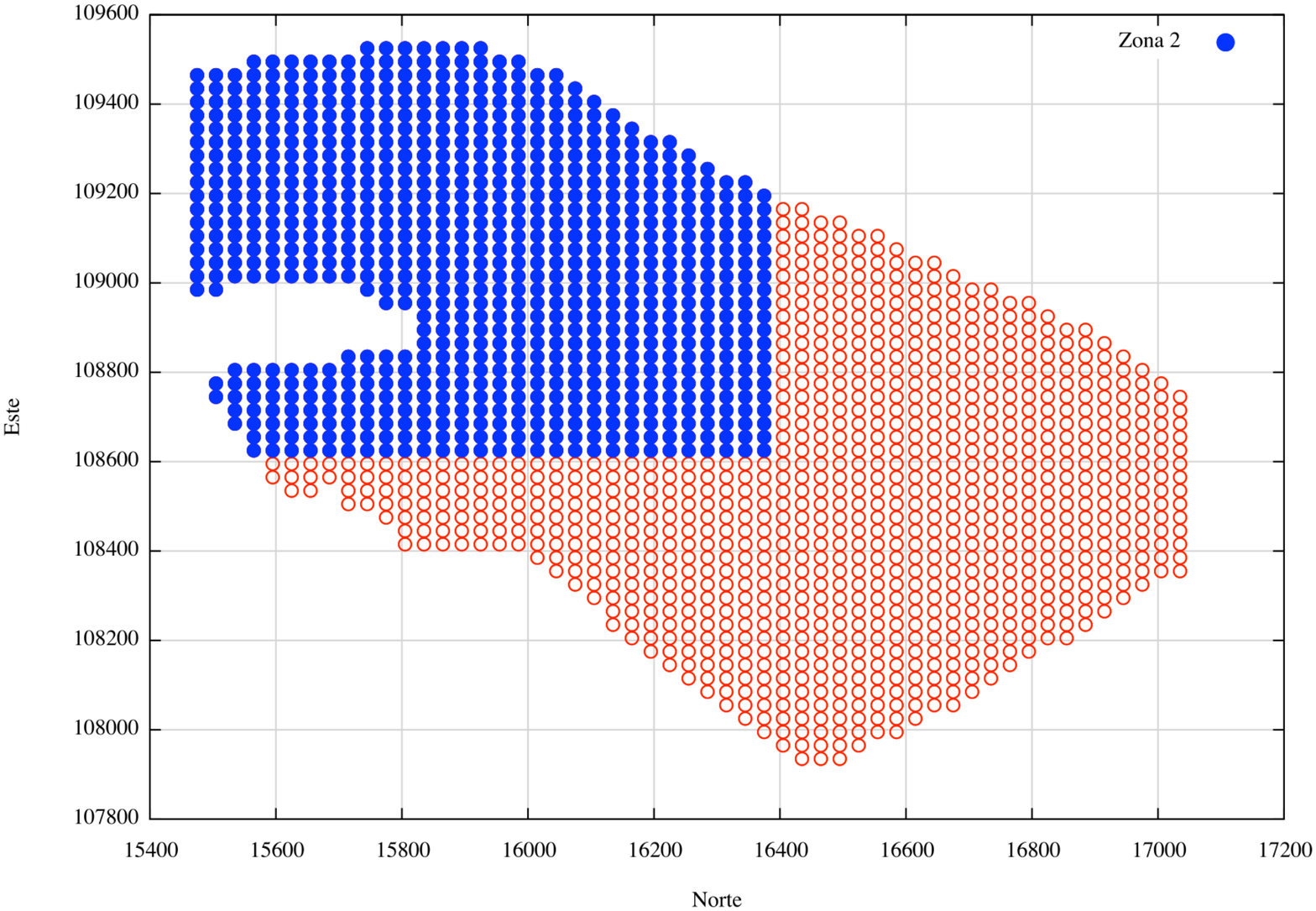}}
     \subfloat[\tiny Zone 3]{
    \includegraphics[width=0.25\textwidth,height=0.25\textwidth]{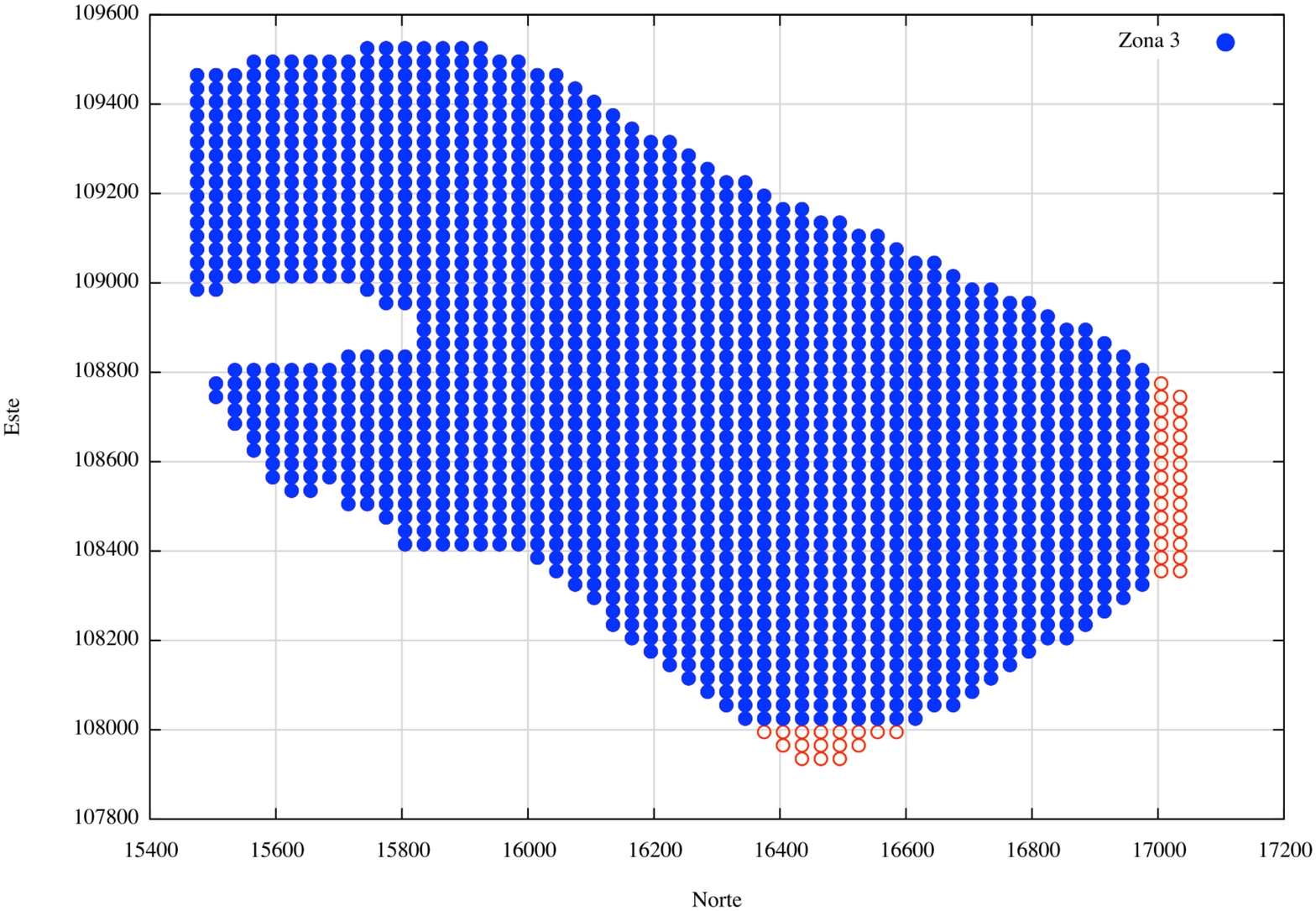}}\\
  \subfloat[\tiny Zone 4]{
    \includegraphics[width=0.25\textwidth,height=0.25\textwidth]{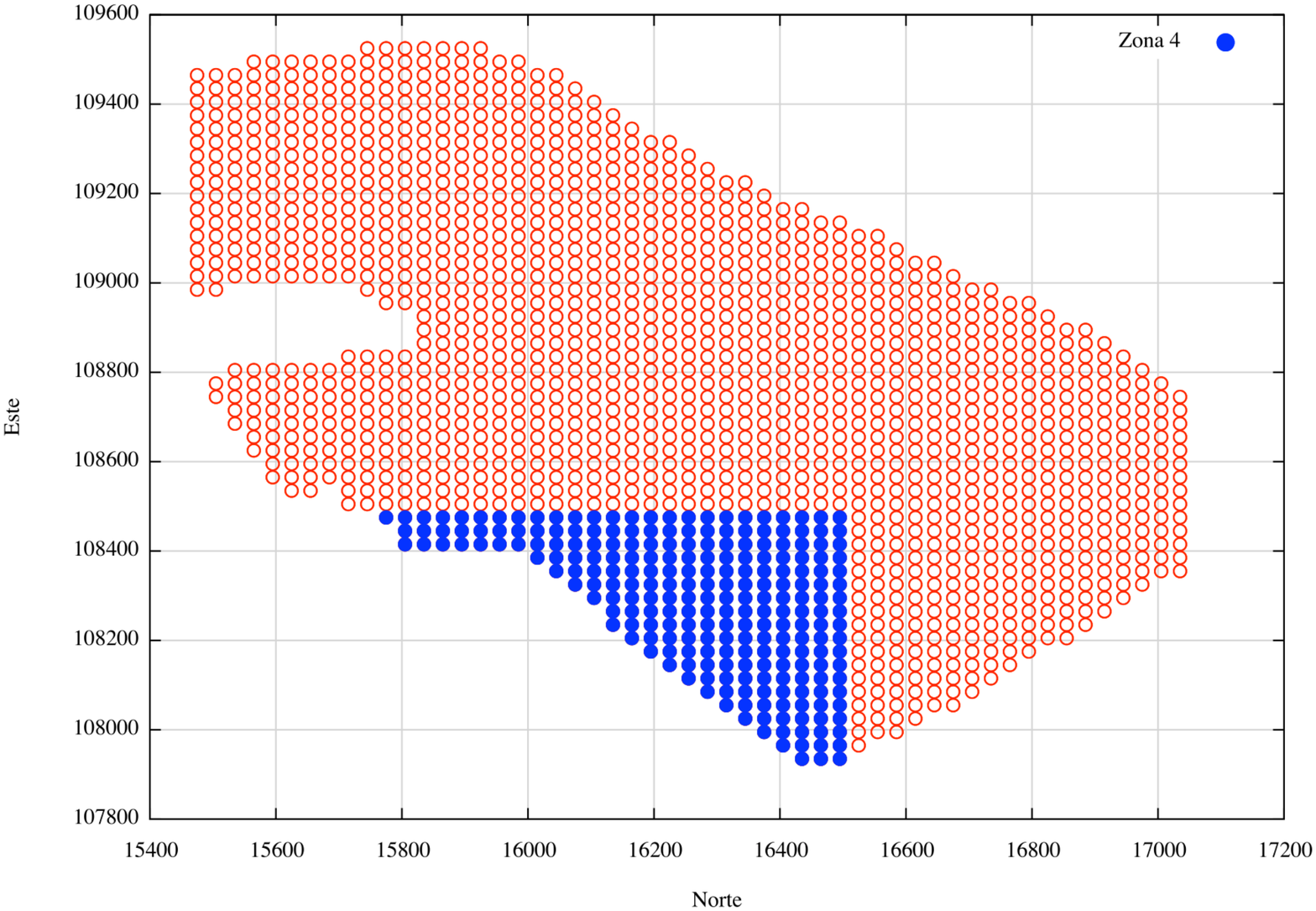}}
    \subfloat[\tiny Zone 5]{
    \includegraphics[width=0.25\textwidth,height=0.25\textwidth]{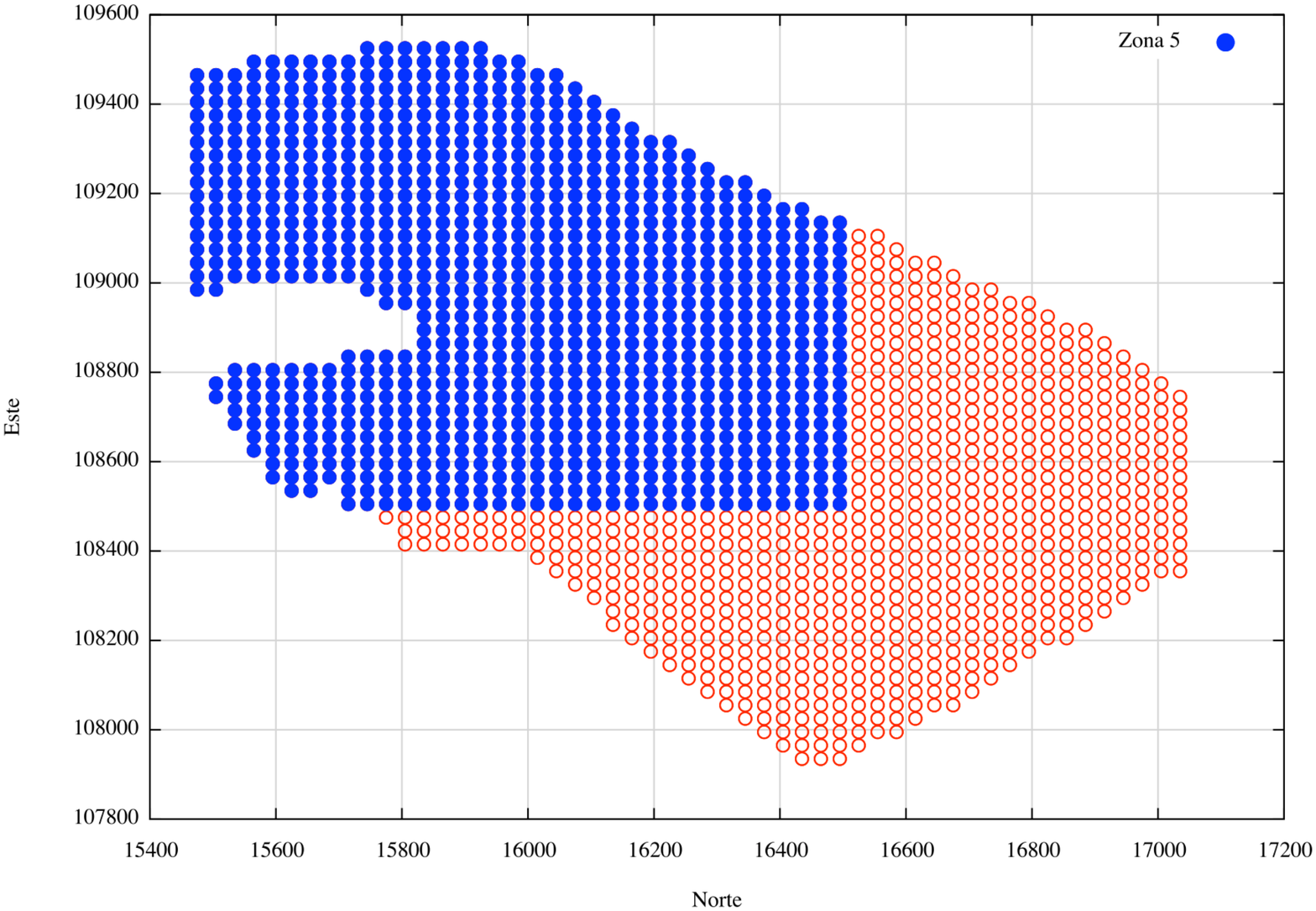}}
    \subfloat[\tiny Zone 6]{
    \includegraphics[width=0.25\textwidth,height=0.25\textwidth]{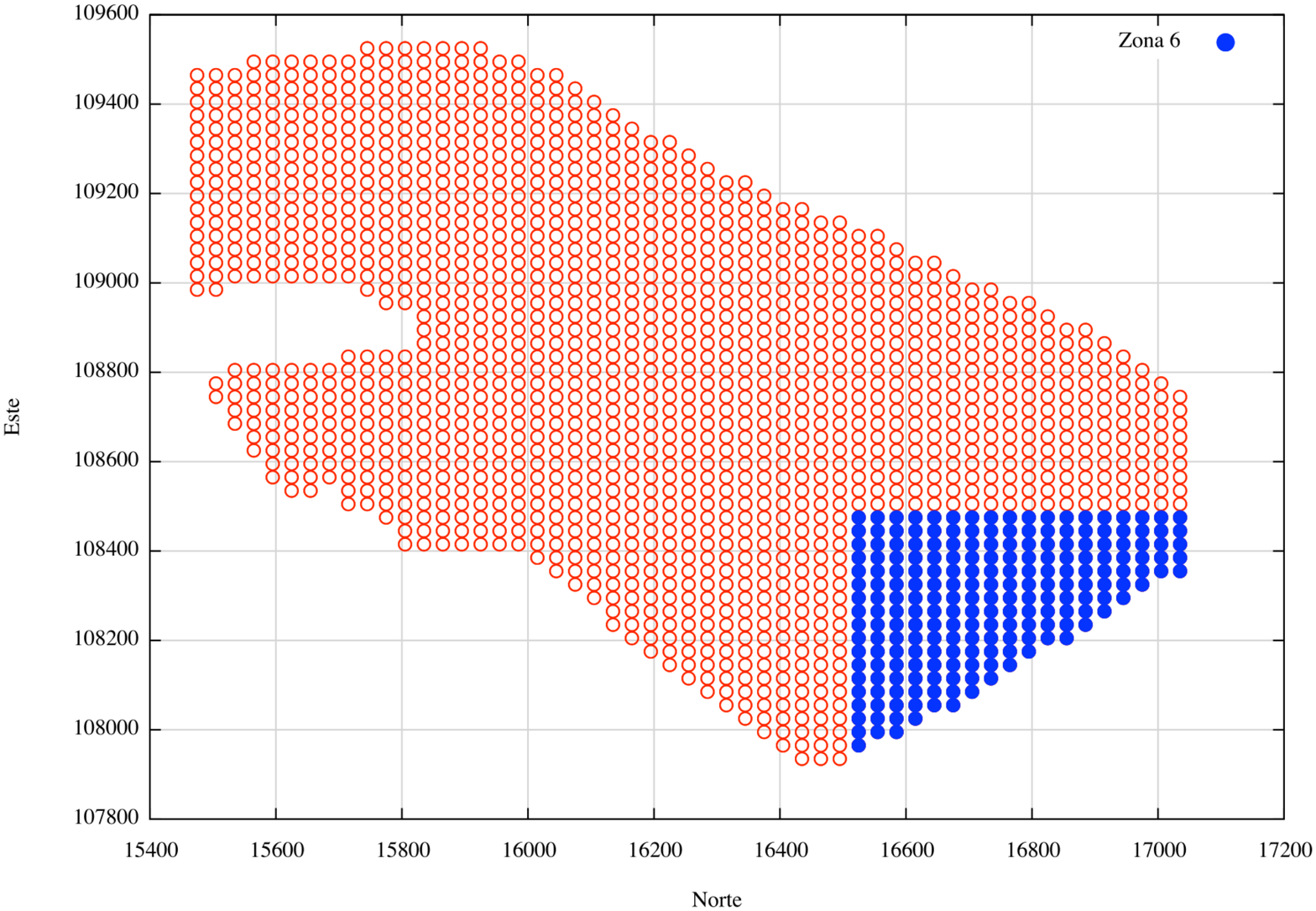}}
    \subfloat[\tiny Zone 7]{
    \includegraphics[width=0.25\textwidth,height=0.25\textwidth]{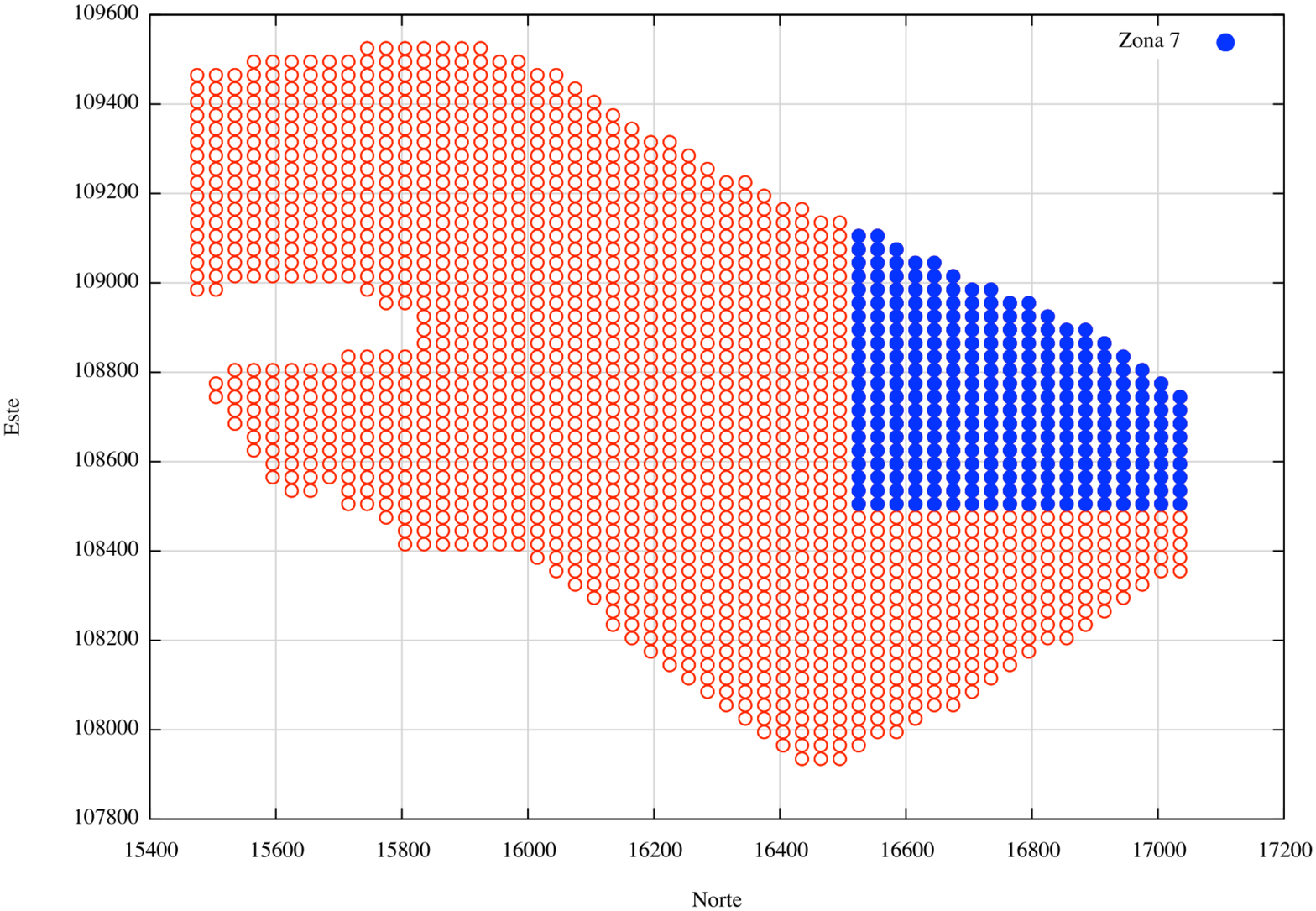}}
 \caption{Studied Zones by the GBR sensor}
 \label{ZonasIBIS4}
\end{figure}

\begin{figure}[htp]
 \centering
\subfloat[Index on zones  1,2 and 3 compared with the entire zone] {  \includegraphics[width=0.5\textwidth]{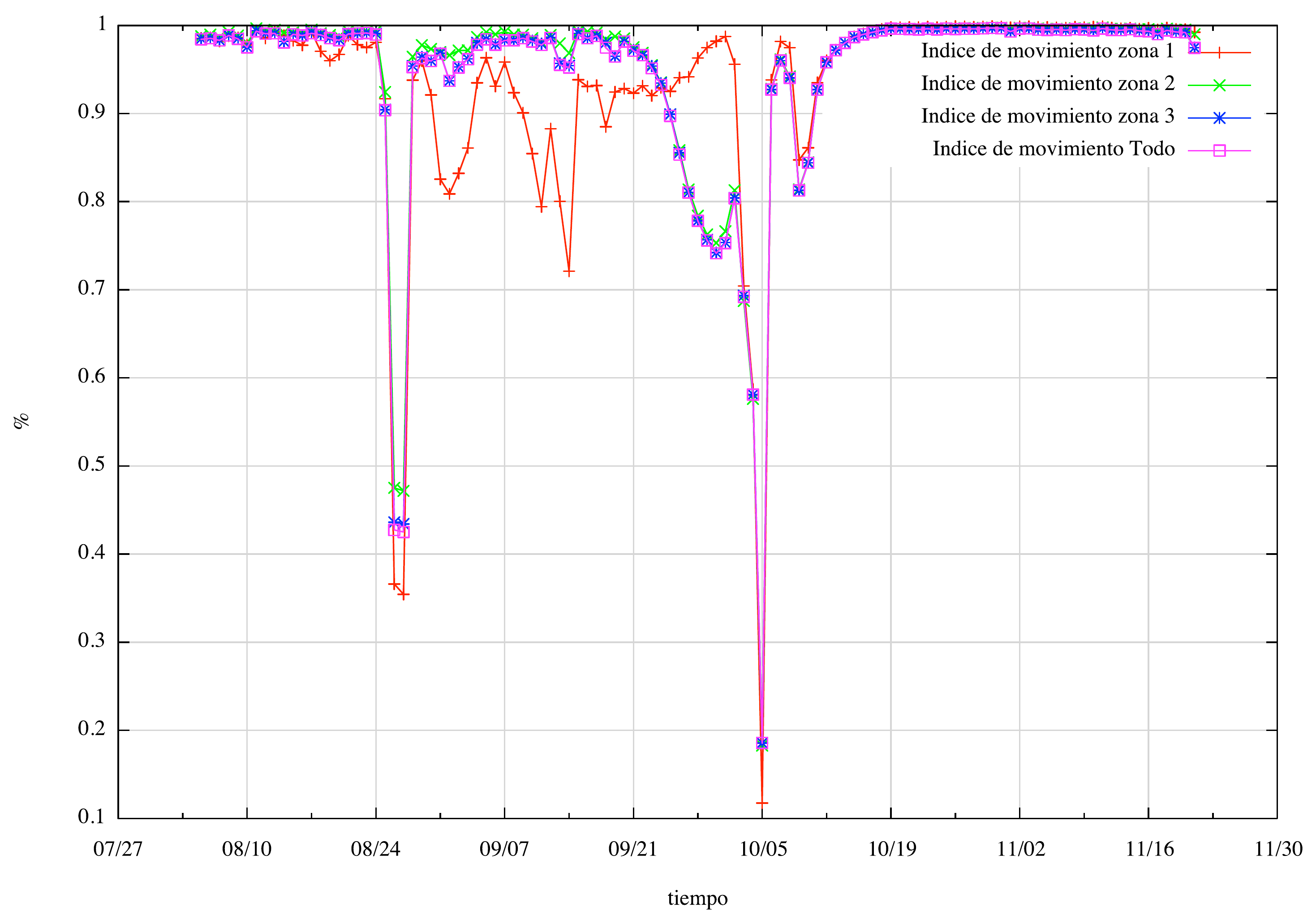}
\label{Ind_zona123_IBIS4}
} \\
\subfloat[Index on zones  4 and 5 compared with the entire zone]{    \includegraphics[width=0.5\textwidth]{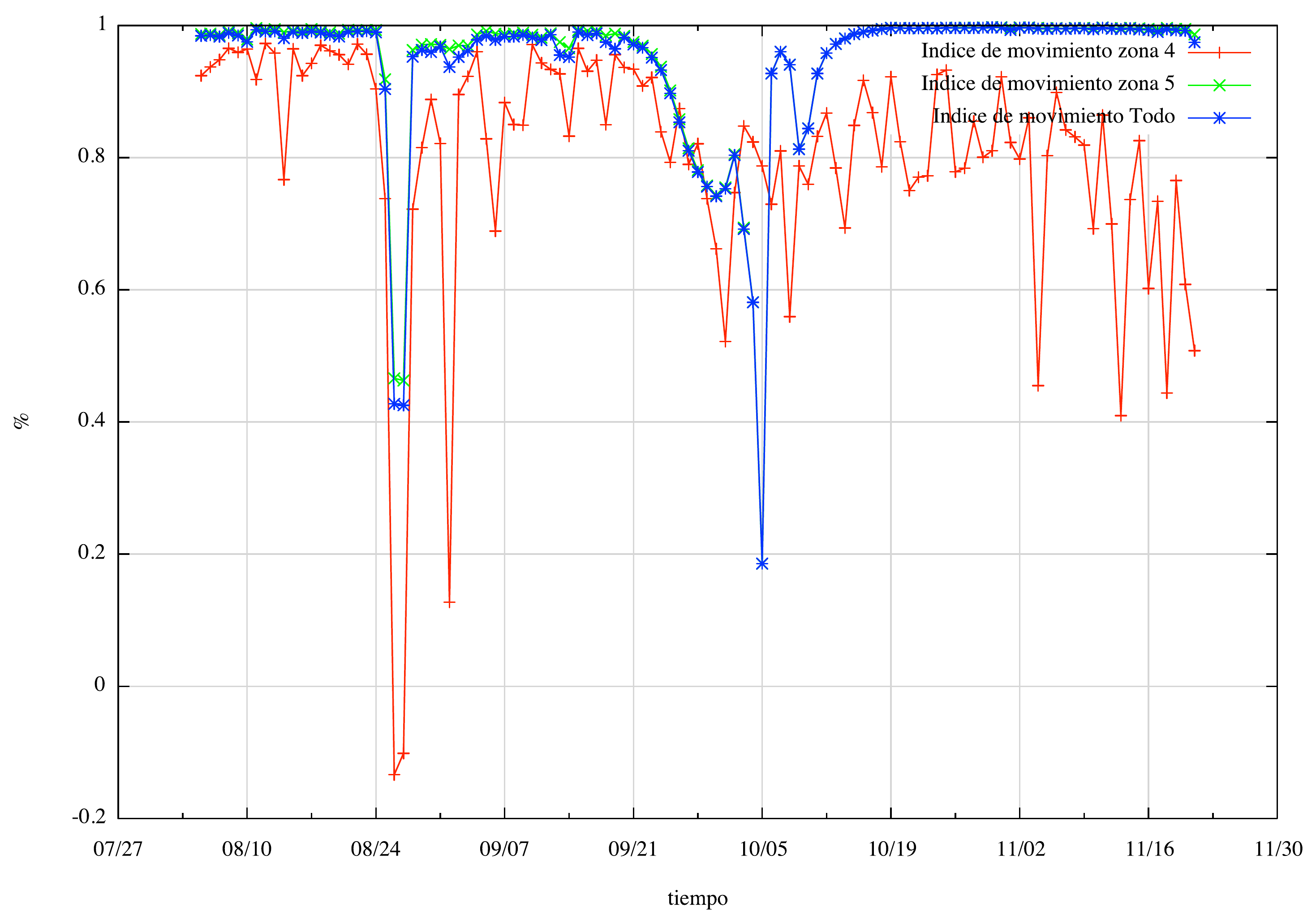}
\label{Ind_zona45_IBIS4}}
\subfloat[Index on zones  6 and 7 compared with the entire zone]{\includegraphics[width=0.5\textwidth]{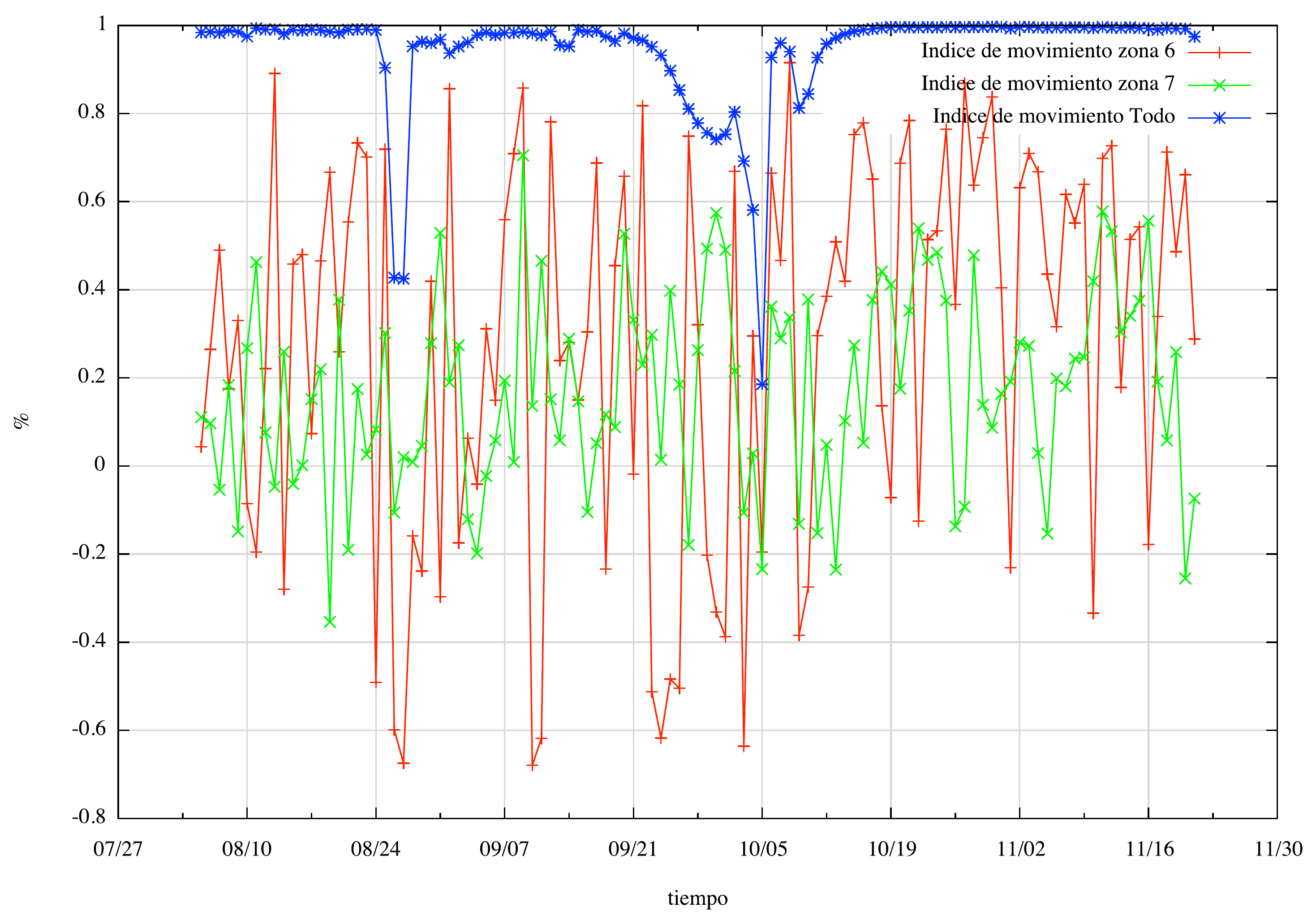}
\label{Ind_zona67_IBIS4}
}
\caption{}
\end{figure}

With respect to the spatial sensitivity of the index, that is,  the influence of the area considered by the GBR data on the index values, we consider the index 
in  seven sub-areas of the region recorded by the sensor, the Figure \ref{ZonasIBIS4}  we can see the areas concerned. 
Figure \ref{Ind_zona123_IBIS4} show us the index on the zones 1,2 and 3, compared with the index over the entire zone.
We can also see the index obtained in zones 4 and 5 compared to the entire project index in Figure \ref{Ind_zona45_IBIS4}. Finally in Figure \ref{Ind_zona67_IBIS4}, we have the index zone 6 and 7 compared to the index of the entire project. 

In conclusion, the index turns out to be robust under geographical changes. Moreover, if the zone is not moving, the index values represent noise and it lose credibility. It is also advisable do not to use the index in small areas, because these could be all moving and therefore the indicator may not register changes.
\begin{figure}[htp]
 \centering
  \subfloat[All  projects]{
 \includegraphics[width=0.33\textwidth]{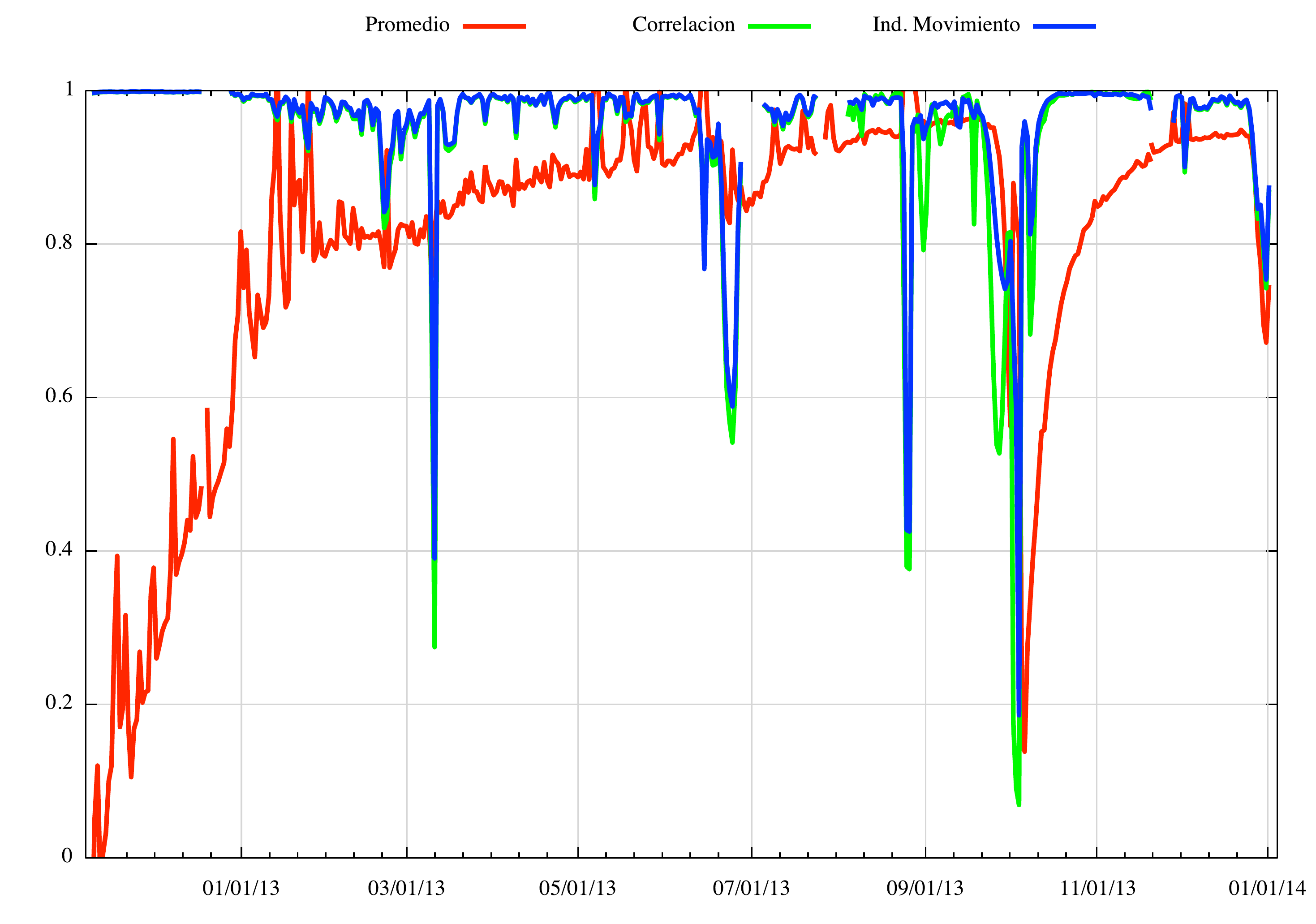}}
  \subfloat[Project 1]{
    \includegraphics[width=0.33\textwidth]{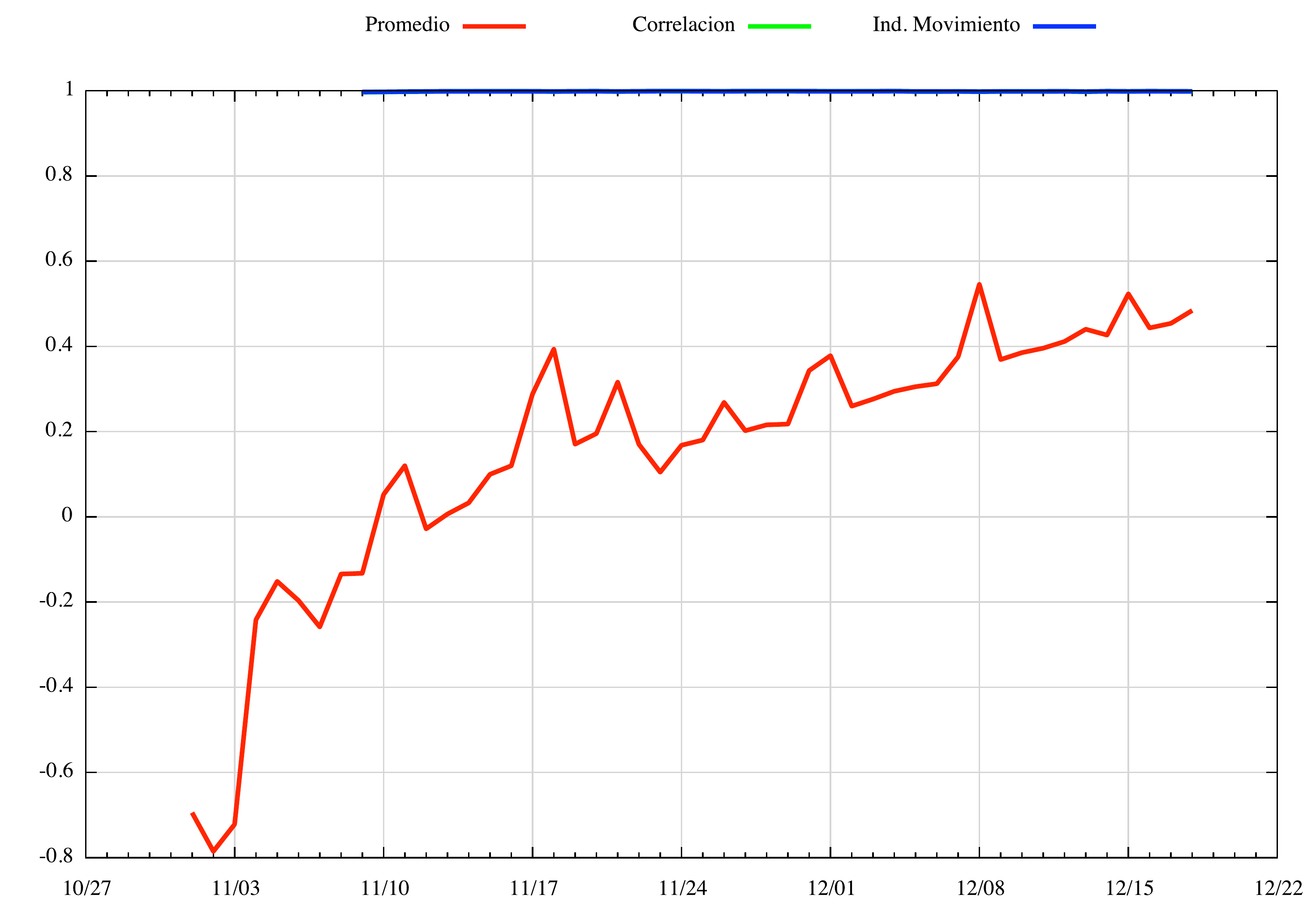}}
  \subfloat[Project  2]{
    \includegraphics[width=0.33\textwidth]{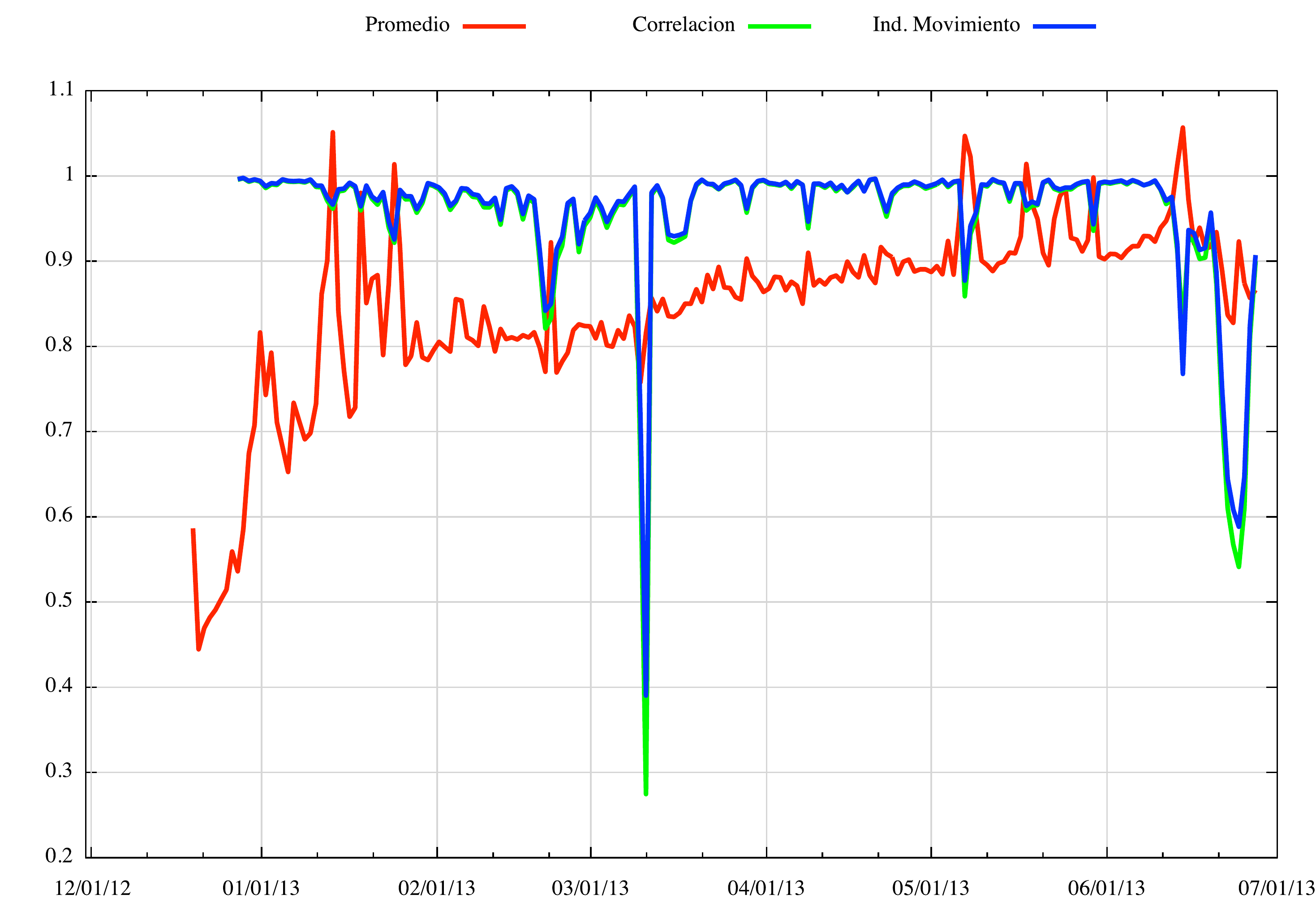}}\\
     \subfloat[Project  3]{
    \includegraphics[width=0.33\textwidth]{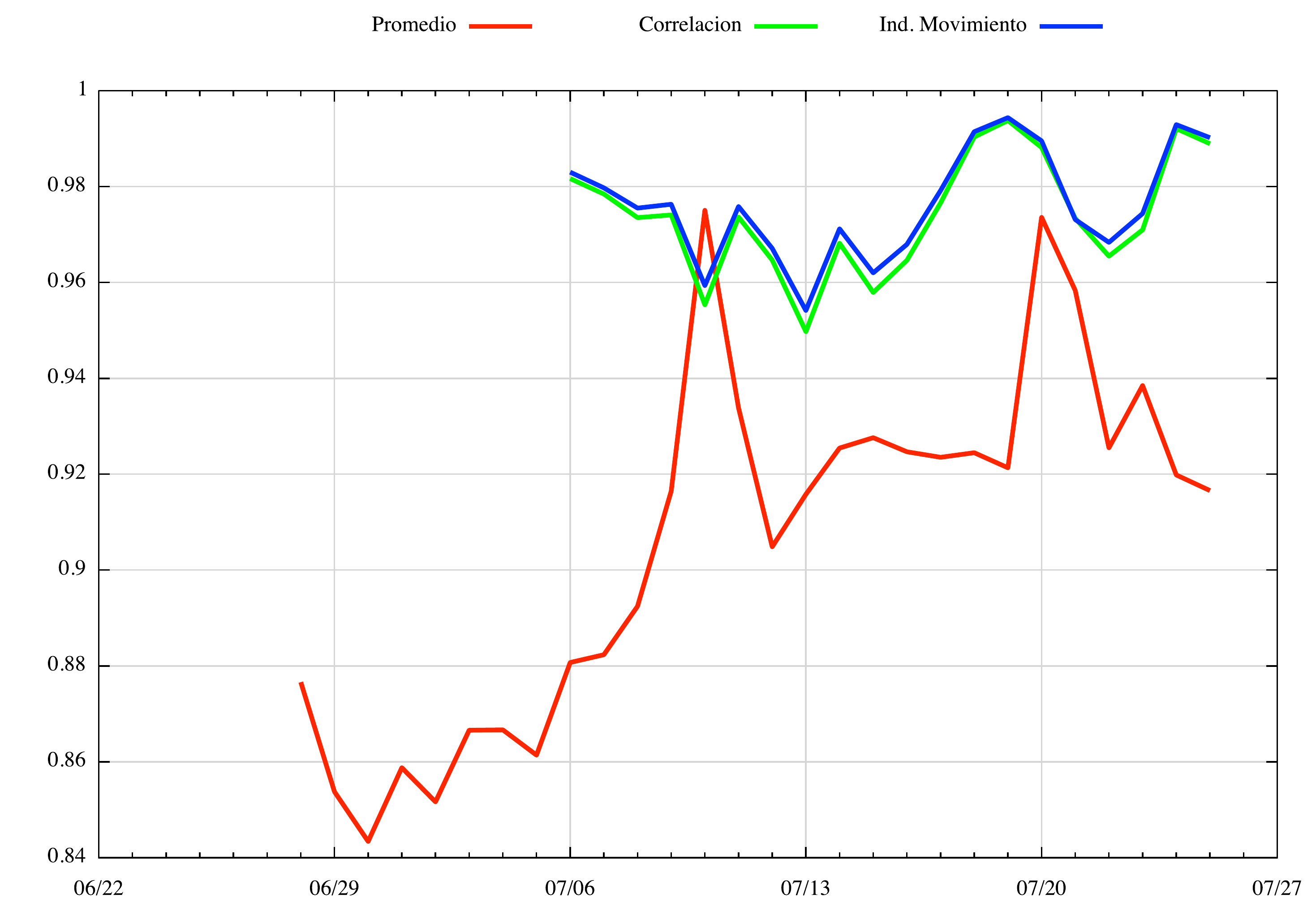}}
  \subfloat[Project  4]{
    \includegraphics[width=0.33\textwidth]{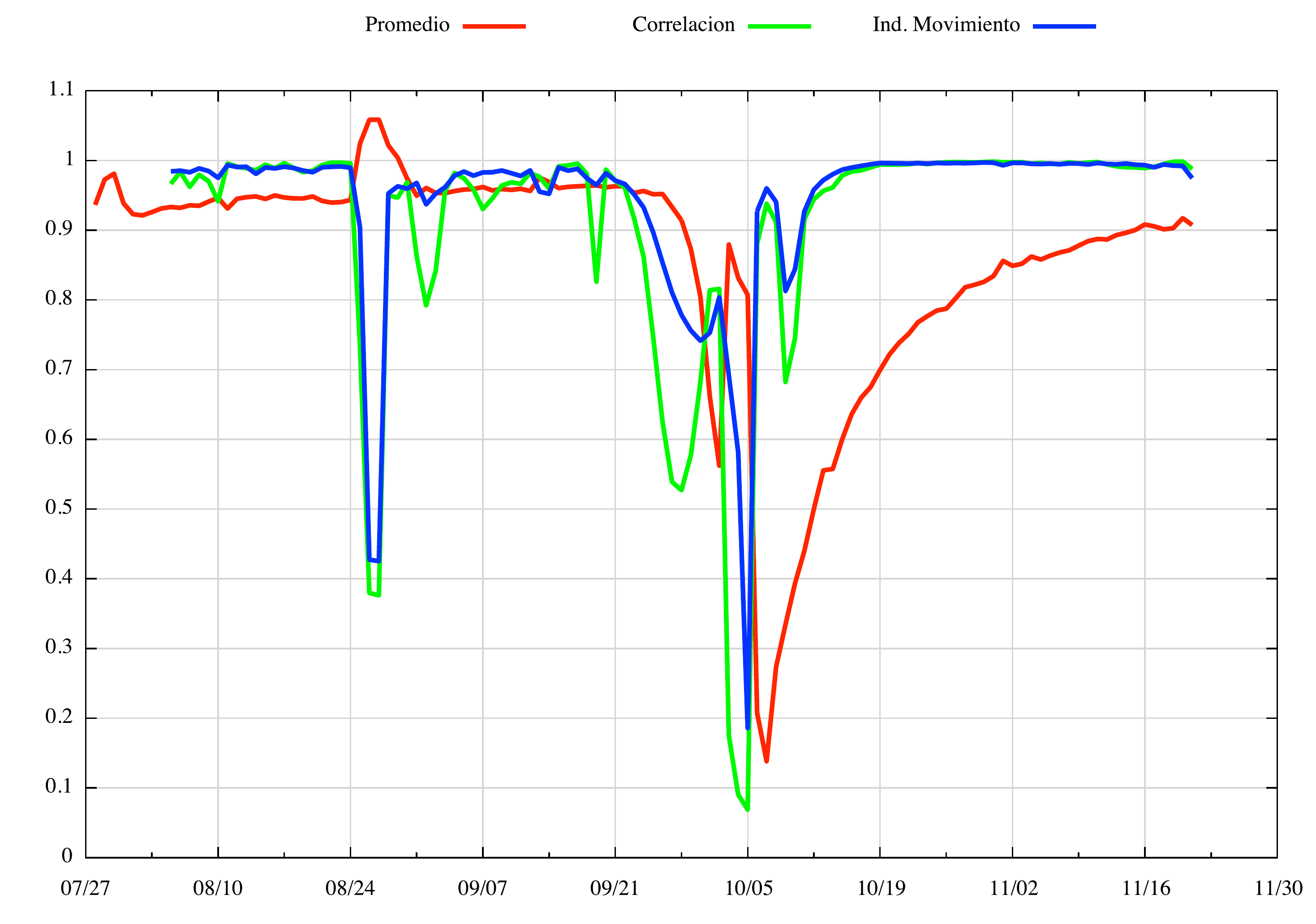}}
    \subfloat[Project  5]{
    \includegraphics[width=0.33\textwidth]{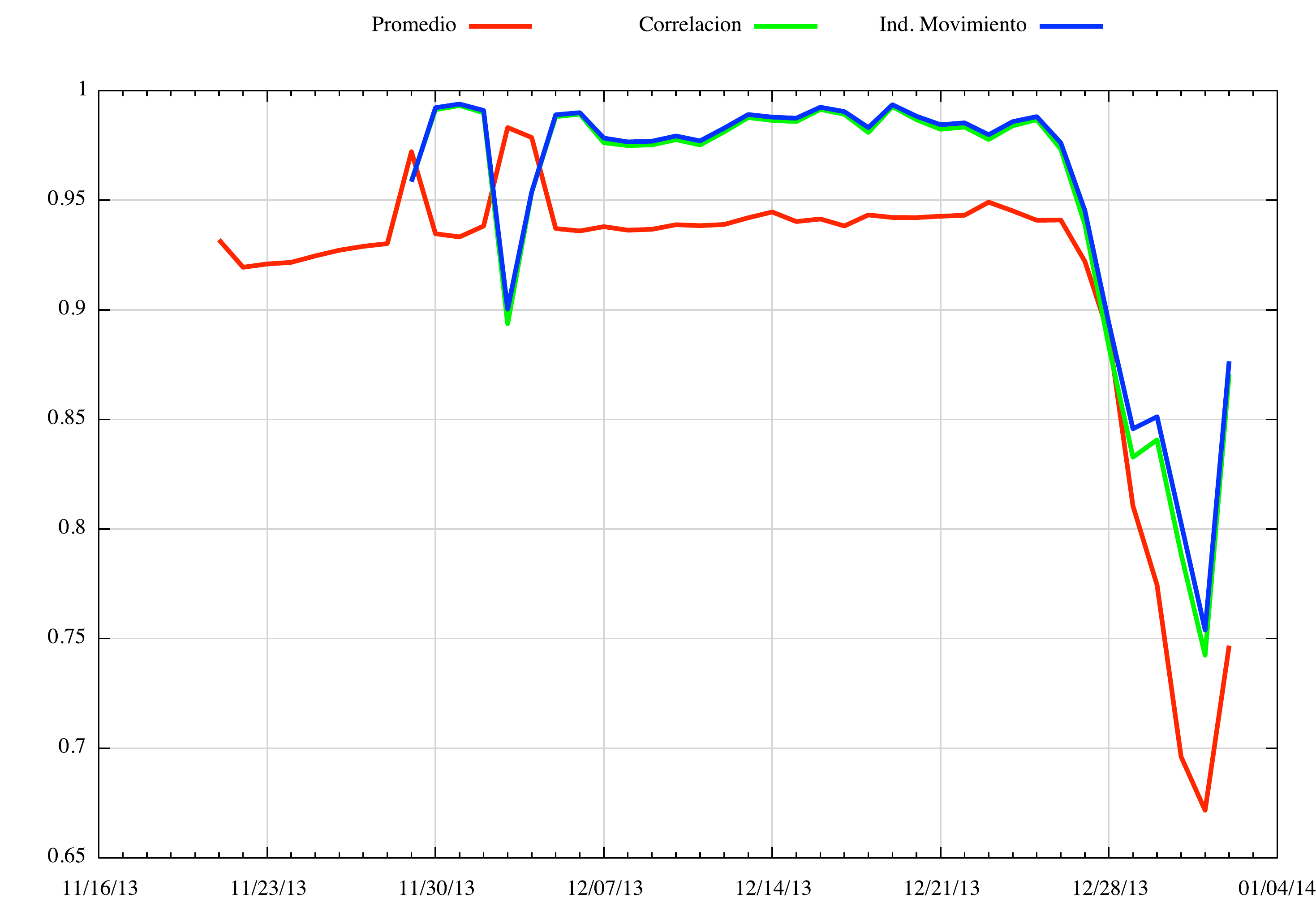}}
 \caption{Comparison of the index with the correlation and the average of the data}
 \label{ComparaTodo}
\end{figure}

\subsection{Movement Zone}
\
\par 

For a better interpretation of the index, see Figure  and due to its geometric representation \ref{Esquema_Ejemplo02}, we can decompose $\frac{x_t}{\|x_t\|}$ on its projection over  $\overline{x}_{t,\delta,\tau},$ and the perpendicular movement, that is, 
$$\frac{x_t}{\|x_t\|}=\lambda\overline{x}_{t,\delta,\tau}+v(t,\delta,\tau).$$
In this way, we can define the movement zone $v(t,\delta,\tau)$, such that:
\begin{equation}\label{vecmov}
v(t,\delta,\tau)=\frac{x_t}{\|x_t\|}-Im(t,\delta,\tau)\frac{\overline{x}_{t,\delta,\tau}}{\|\overline{x}_{t,\delta,\tau}\|}
\end{equation}
We remark that the movement zone represents  the direction in which the vector $ x_t $  has mainly changed  about it  past, represented by $ \overline {x} _ {t, \delta, \tau} $.
\begin{figure}[htp]
 \centering
  \subfloat[19-09]{
    \includegraphics[width=0.33\textwidth]{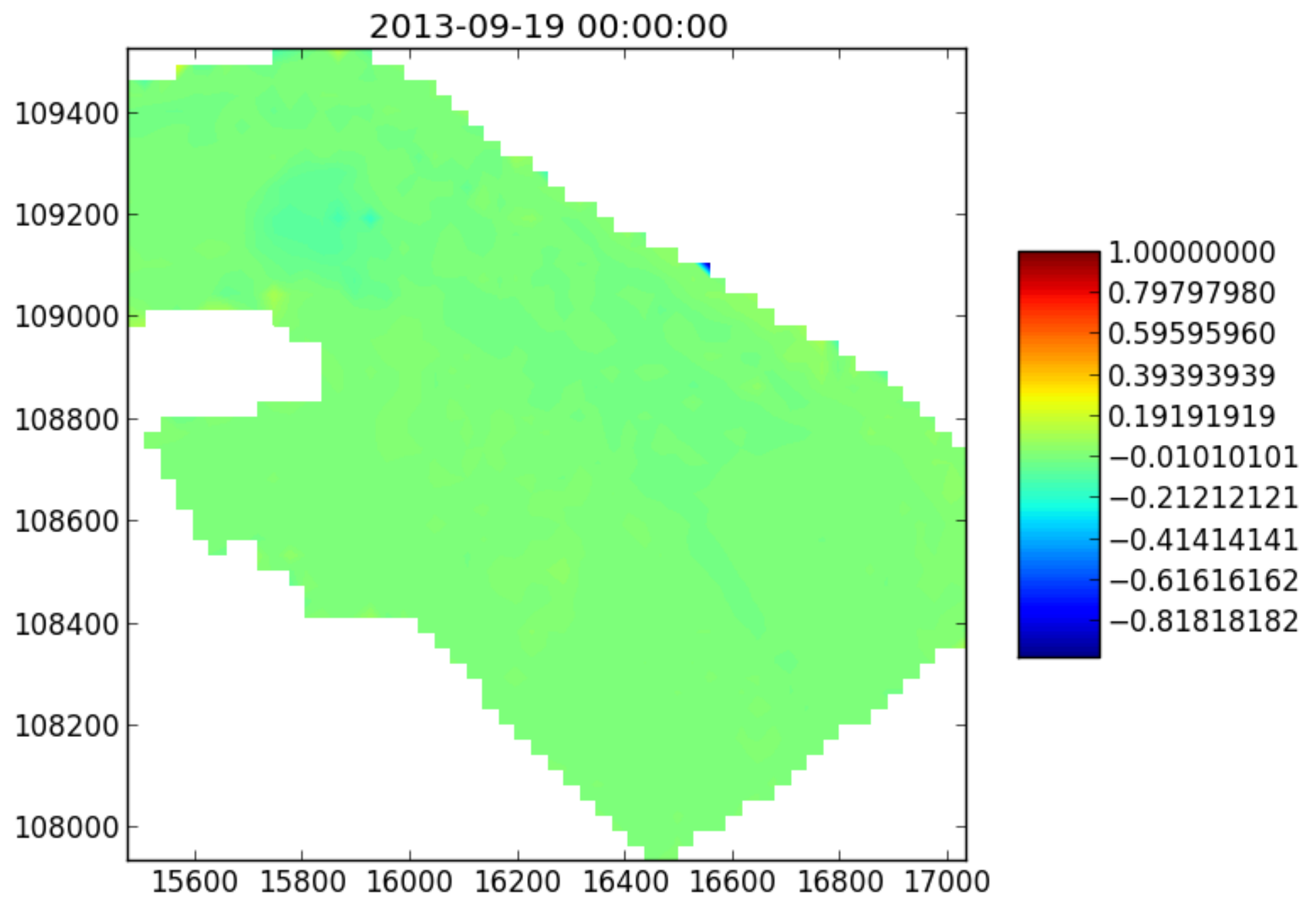}}
  \subfloat[21-09]{
    \includegraphics[width=0.33\textwidth]{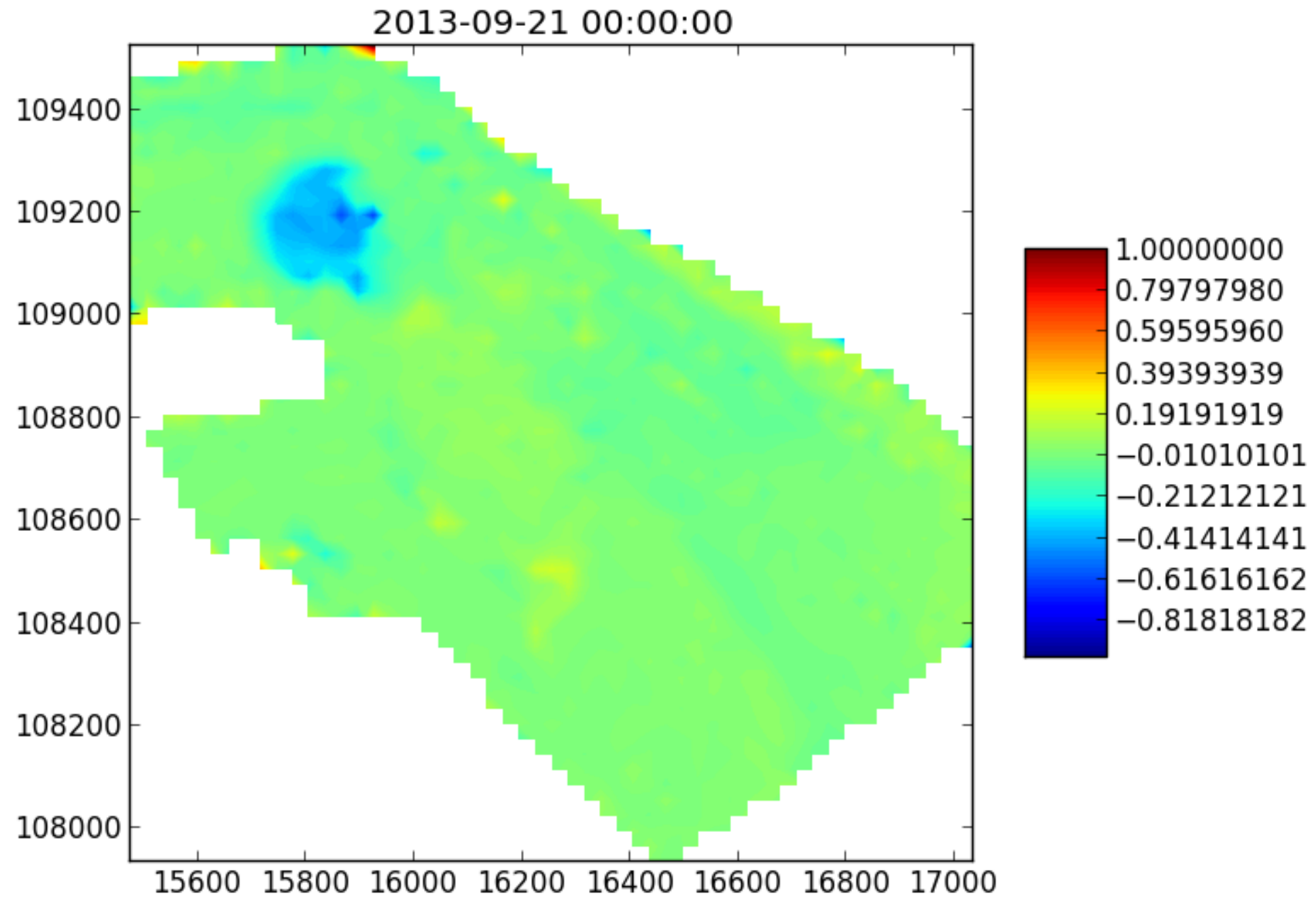}}
    \subfloat[23-09]{
    \includegraphics[width=0.33\textwidth]{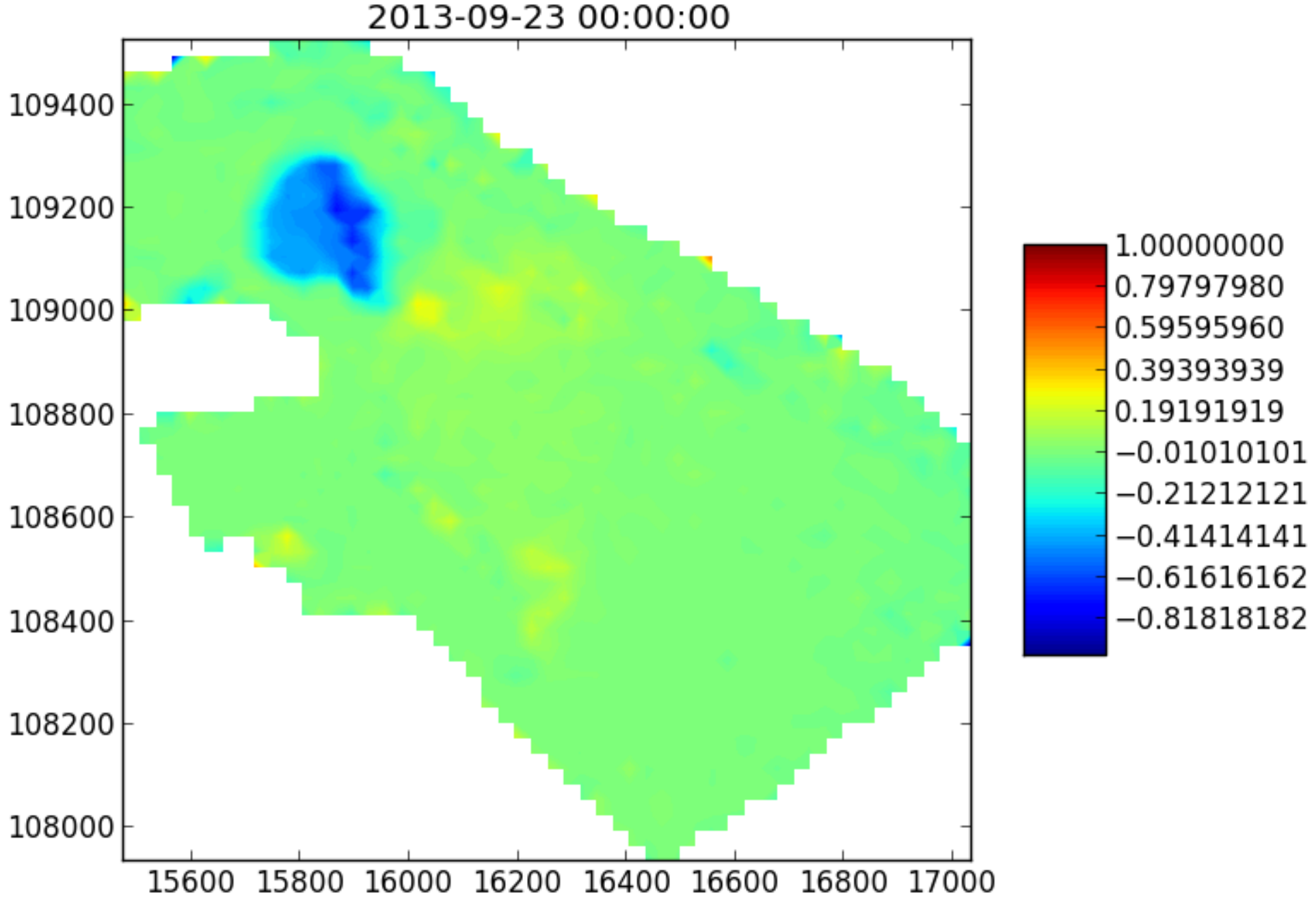}}\\
     \subfloat[25-09]{
    \includegraphics[width=0.33\textwidth]{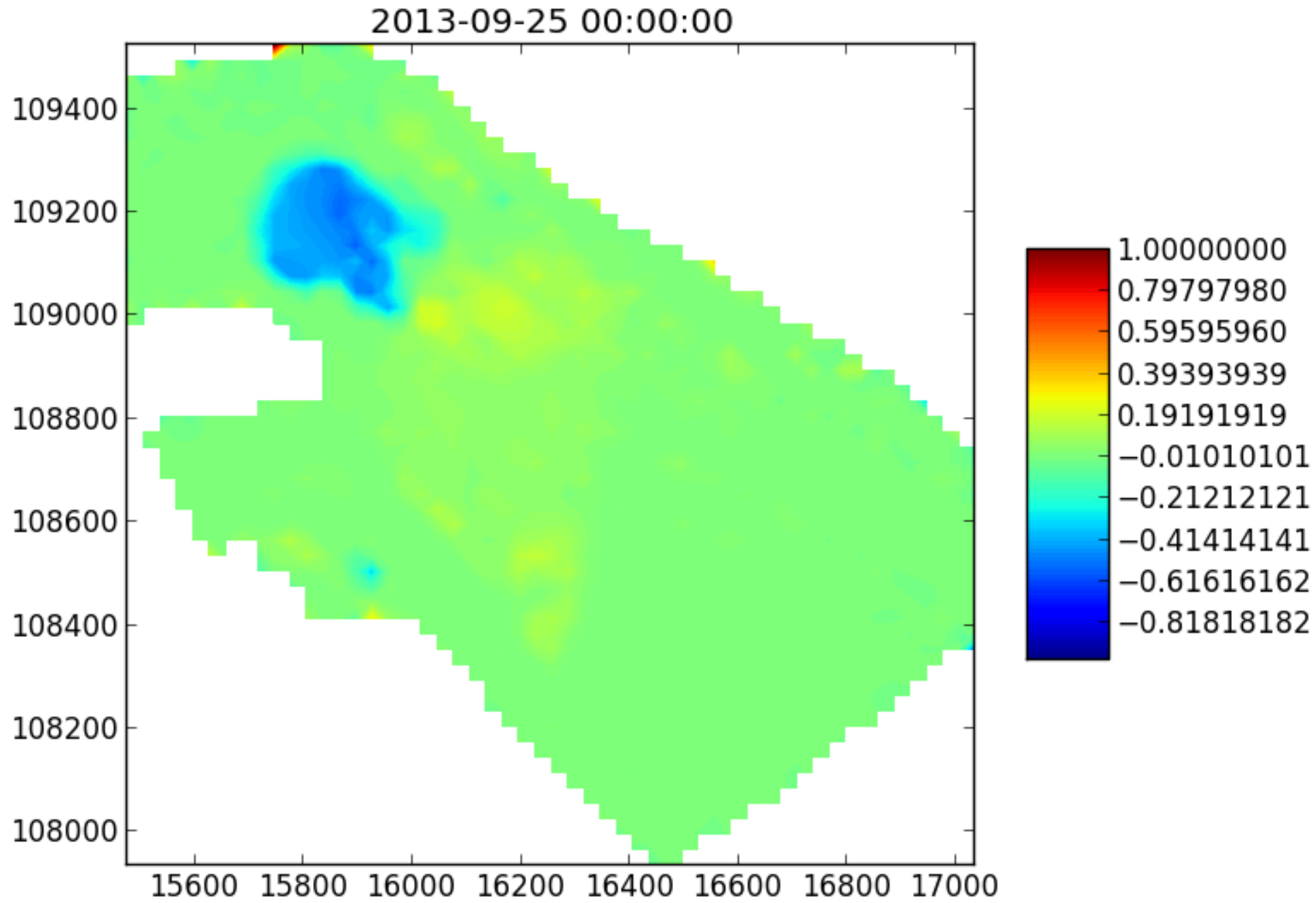}}
  \subfloat[27-09]{
    \includegraphics[width=0.33\textwidth]{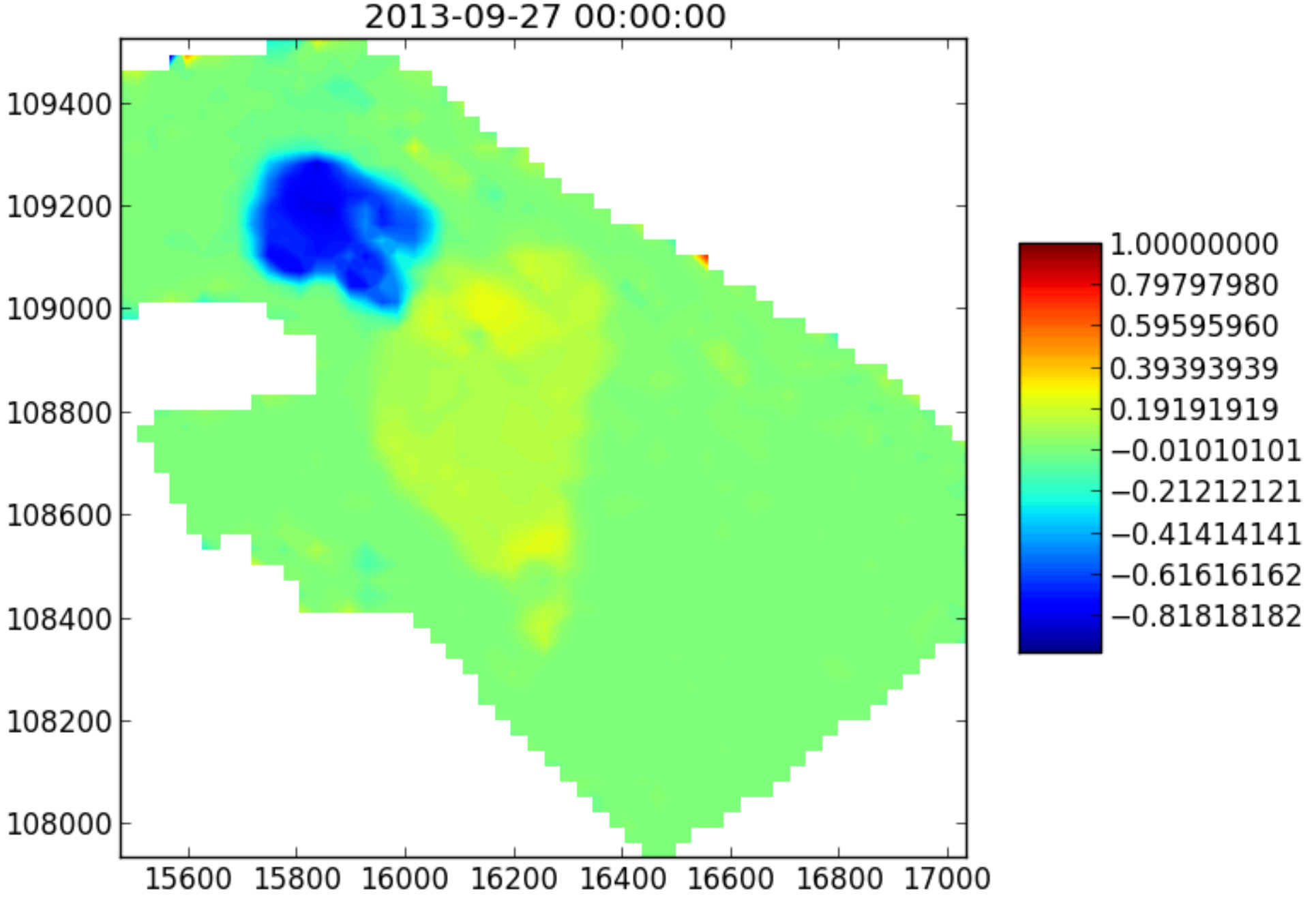}}
    \subfloat[29-09]{
    \includegraphics[width=0.33\textwidth]{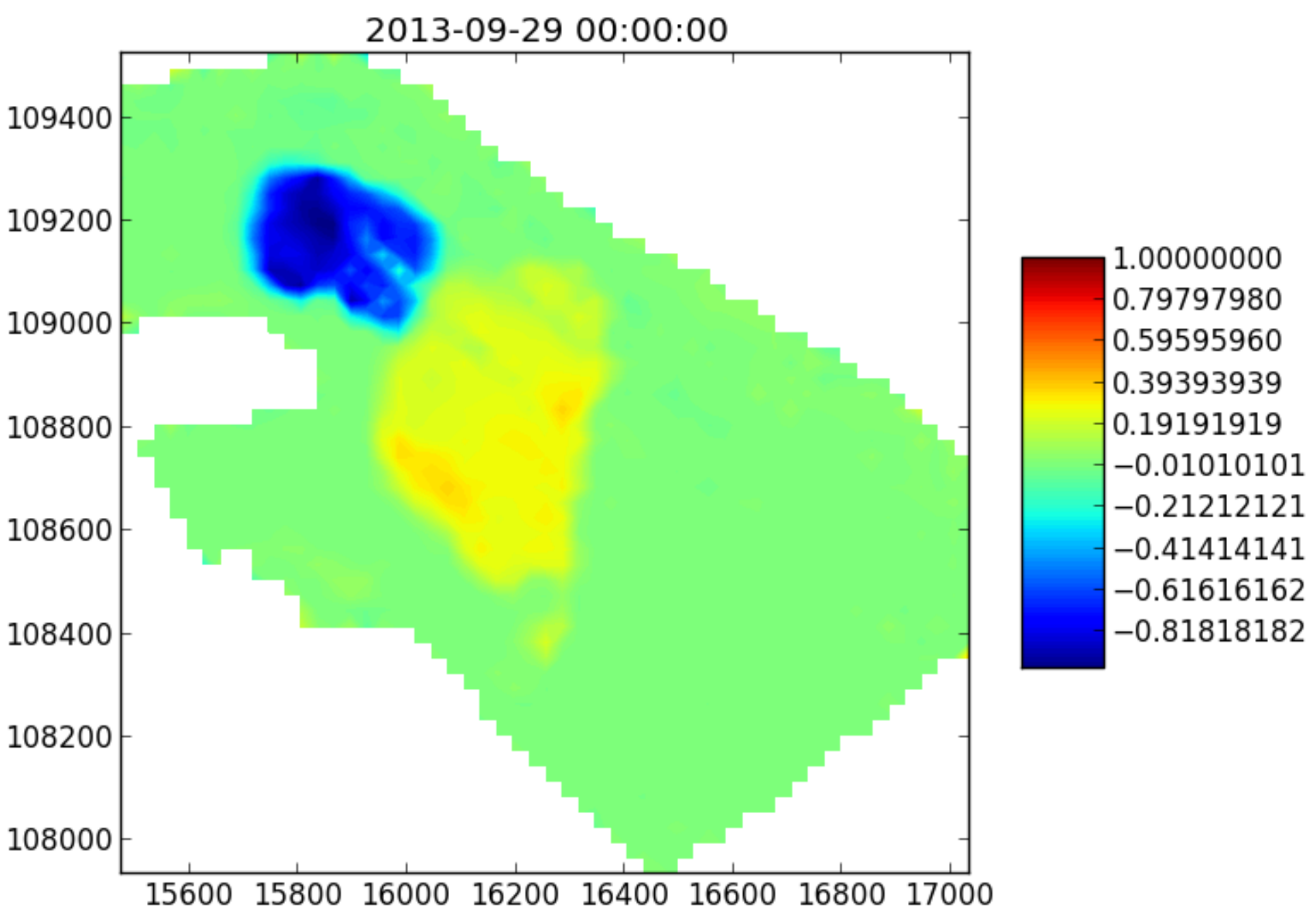}}\\
    \subfloat[01-10]{
    \includegraphics[width=0.33\textwidth]{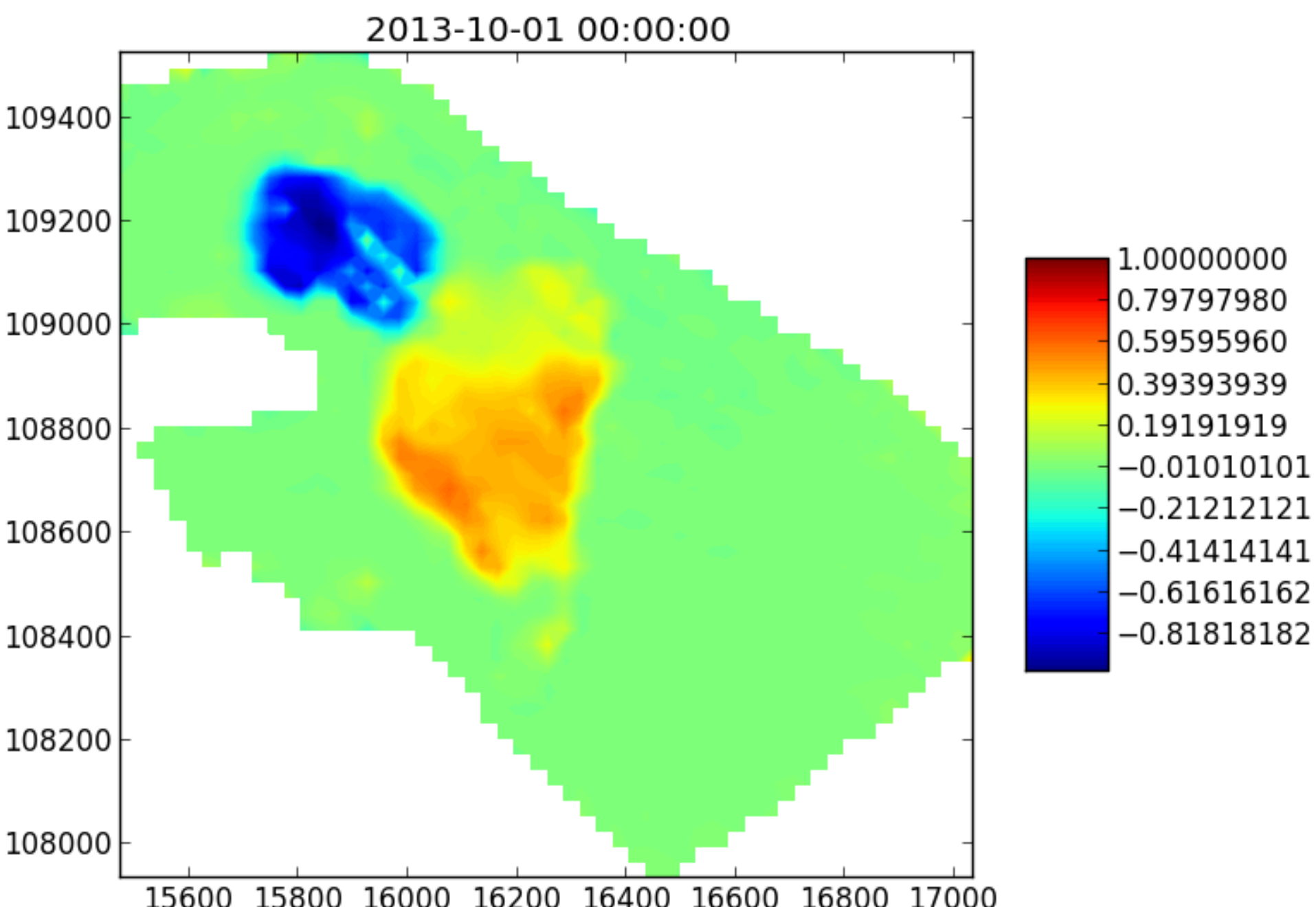}}
  \subfloat[03-10]{
    \includegraphics[width=0.33\textwidth]{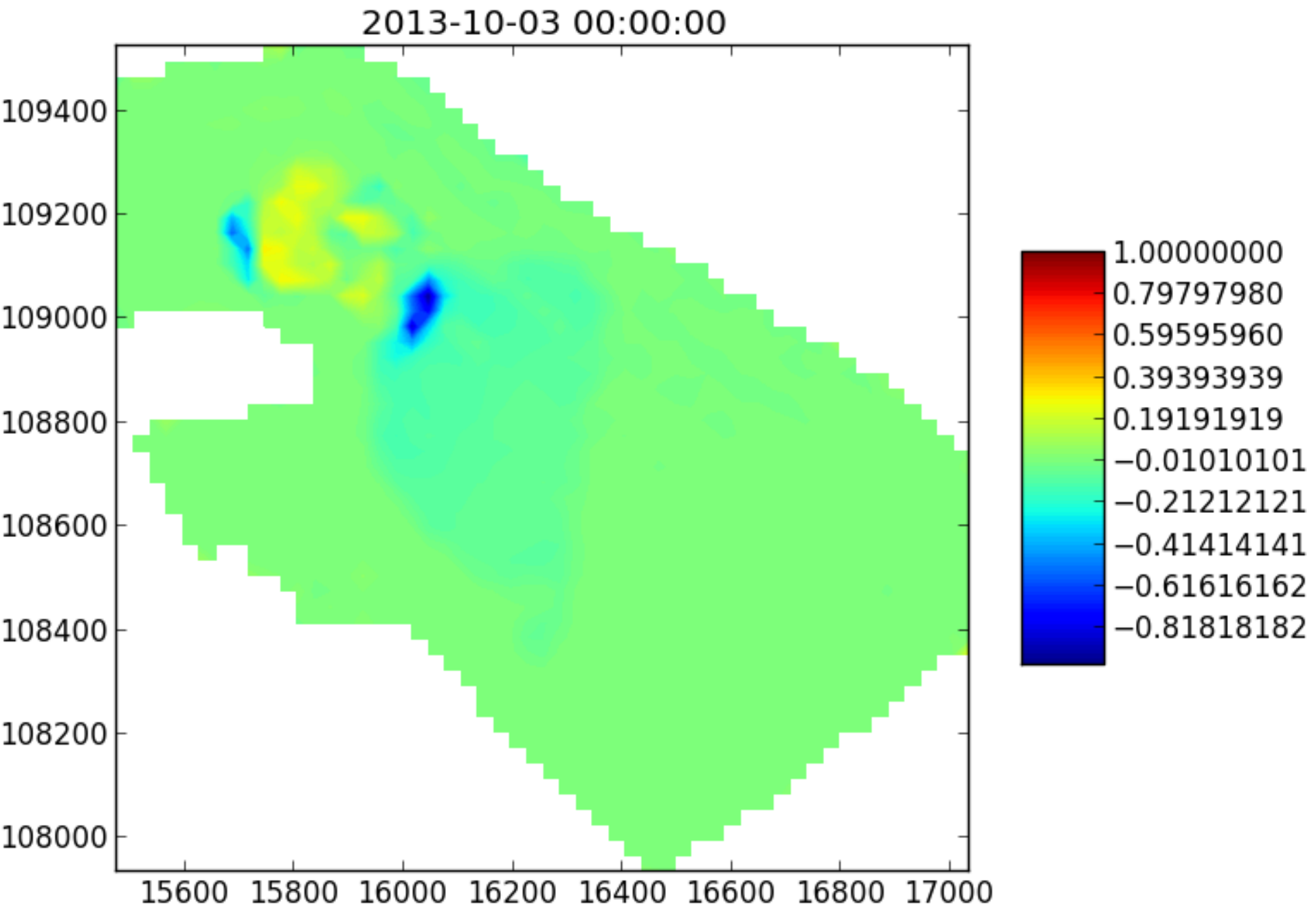}}
    \subfloat[05-10]{
    \includegraphics[width=0.33\textwidth]{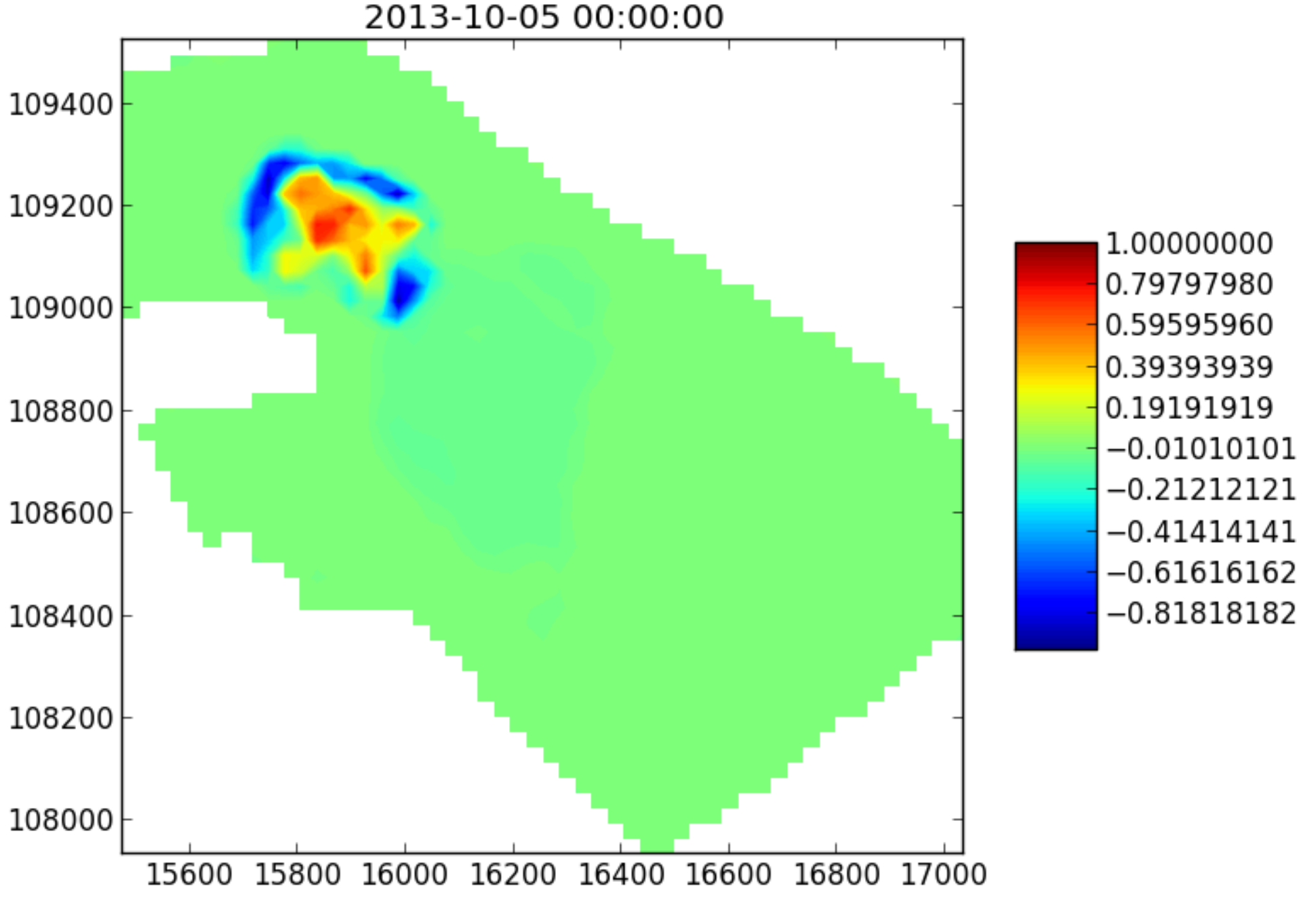}}
 \caption{View of the movement area around the event registered by the GBR on the daily mesh and $ 30 \times 30 $ meters}
 \label{VectorMovimiento}
\end{figure}

In Figure \ref{VectorMovimiento} we can observe the movement area every two days from September 19 until 5 October. Moreover, with respect to the colors, blue indicates an approach to sensor and yellow zone indicates that the sensor is going far away.

\subsection{Conclusions}
\
\par

 This index give us the geometric change suffered by the GBR data. It makes a comparison between the region which is moving today which was moving in the past. This indicator is dimensionless and takes values between -1 to 1, where one represents that the current movement is identical to the movement of the past, whereas zero corresponds to the current movement has no relation to the previous movement, and the value -1 , represents a movement in the opposite direction.

The main characteristics of the index are:
\begin{itemize}
\item Dimensionless and bounded range.
\item Insensitive  to increased magnitude of the phenomenon, that is, if the region moving in the past is the same as is currently moves but with a different intensity, the indicator take values very close to one. Whereas if the magnitude is similar but the moving region is different, the index  take values far away from one.
\item Under suitable approximations and  ideal conditions on data (without noise and high differentiability of the data), we can establish a relationship between the indicator and the time derivative of the data.
\item The index  has low sensitivity to noise in the information.
\item The index allows to identify the phenomenon before the average of variance (in the data studied was obtained 6 days before).
\item The index has a high correlation with the index of temporal correlation, but the latter tends to exacerbate certain phenomena, which the index turns out to be more robust.
\item This indicator is sensitive to the variation of the ratio between positive and negative data for each time instant.
\end{itemize}

\newpage

\section{acknowledgements} 
\
\par

This research was supported by Gesecology Ltda. Part of the work was also partially supported by Grant CORFO-INNOVA  2013-24066 and PFB03-CMM UChile. We remark also that the results of this research belongs to gesecoloy Ltda and it is under consideration for an Industrial Patent.

\end{document}